\documentclass{aa}

\usepackage[varg]{txfonts} 
\usepackage{graphicx} 
\usepackage{natbib}
\bibpunct{(}{)}{;}{a}{}{,} 
\usepackage{hyperref}
\usepackage{booktabs}
\usepackage[usestackEOL]{stackengine} 
\usepackage{siunitx}

\usepackage{orcidlink}

\newcommand*{\fullref}[1]{\hyperref[{#1}]{\autoref*{#1} \nameref*{#1}}}

\def\kobe/{\texttt{KOBE}}
\def\kobeshadows/{\texttt{\kobe/-Shadows}}
\def\kobetransits/{\texttt{\kobe/-Transits}}
\def\kobevetter/{\texttt{\kobe/-Vetter}}
\def\similar/{similar}
\def\mixed/{mixed}
\def\antiordered/{anti-ordered}
\def\ordered/{ordered}
\def\pip/{peas in a pod}
\def\figwidth{8cm}

\defcitealias{Mishra2022b}{Paper II}
\def\papertwo/{\citetalias{Mishra2022b}}

\newcommand{\edit}[1]{#1}
\newcommand{\change}[1]{#1}
 
 \newcommand{\lang}[1]{#1}

\graphicspath{{./images/}}

\def\mearth{M_\oplus}
\def\rearth{R_\oplus}

\def\msun{M_\odot}

\def\mj{M_\text{J}}

\def\mcore{M_{\rm core}}

\def\sigmas0{\Sigma_\mathrm{s,0}}

\def\snr/{S/N}

\def\periodictce/{\text{\textit{p}TCE}}

\def\nplanet{n}

\def\cvtext/{coefficient of variation}
\def\cstext/{coefficient of similarity}
\def\cs{C_{S}}
\def\cv{C_{V}}

\hypersetup{pdfauthor={Lokesh Mishra}, 
        pdftitle=Framework for architecture of exoplanetary systems I,  
        colorlinks=true, 
        urlcolor=blue, 
        linkcolor=blue,  
        citecolor=blue}

\begin{document}

\title{Framework for the architecture of exoplanetary systems}
\subtitle{I. Four classes of planetary system architecture\thanks{Catalogue of observed planetary systems used in this work is available online at \url{https://cdsarc.cds.unistra.fr/cgi-bin/qcat?J/A+A/}.}}

\author{Lokesh Mishra\inst{\ref{unibe},\ref{unige}}\orcidlink{0000-0002-1256-7261}
                \and Yann Alibert\inst{\ref{unibe}}\orcidlink{0000-0002-4644-8818}
        \and St\'ephane Udry\inst{\ref{unige}} \orcidlink{0000-0001-7576-6236}
        \and Christoph Mordasini\inst{\ref{unibe}}\orcidlink{0000-0002-1013-2811}
}

\authorrunning{L. Mishra et al.}
\titlerunning{Architecture Framework I -- Four classes of planetary system architecture}

\institute{
        Institute of Physics, University of Bern, Gesellschaftsstrasse 6, 3012 Bern, Switzerland\label{unibe}
        \\\email{\hyperref{mailto:exomishra@gmail.com}{}{}{exomishra@gmail.com}}
        \and
        Geneva Observatory, University of Geneva, Chemin Pegasi 51b, 1290 Versoix, Switzerland\label{unige}       
}

\date{Received 10 04 2022; accepted 05 12 2022}


\abstract{We present a novel, model-independent framework for studying the architecture of an exoplanetary system at the system level. This framework allows us to characterise, quantify, and classify the architecture of an individual planetary system. Our aim in this endeavour is to generate a uniform systematic method to study the arrangement and distribution of various planetary quantities within a single planetary system. We propose that the space of planetary system architectures be partitioned into four classes: similar, mixed, anti-ordered, and ordered. A central aim of this paper is to introduce these four architecture
classes. We applied our framework to observed and synthetic multi-planetary systems, thereby studying their architectures of mass, radius, density, core mass, and the core water mass fraction. We explored the relationships between a system's (mass) architecture and other properties.
Our \lang{work} suggests that: (a) similar architectures are the most common outcome of planet formation; (b) internal structure and composition of planets shows a strong link with their system architecture; (c) most systems inherit their mass architecture from their core mass architecture; (d) most planets that started inside the ice line and formed in-situ are found in systems with a similar architecture; and (e) most anti-ordered systems are expected to be rich in wet planets, while most observed mass ordered systems are expected to have many dry planets. We find, in good agreement with theory, that observations are generally biased towards the discovery of systems whose density architectures are similar, mixed, or anti-ordered. 
This study probes novel questions and new parameter spaces for understanding theory and observations. Future studies may utilise our framework to not only constrain the knowledge \lang{of} individual planets, but also the multi-faceted architecture of an entire planetary system. We also speculate on the role of system architectures in hosting habitable worlds.}



        \maketitle
        
        \section{Introduction}
        \label{sec:introduction}

        Over the last 25 years, our knowledge of exoplanetary astrophysics has improved dramatically. While the first decade was marked by sensational discoveries of individual exoplanets \citep[e.g.][]{2003Natur.422..143V, 2004A&A...426L..19S, 2005A&A...444L..15B, 2007A&A...469L..43U,2008Sci...322.1345K, 2009Natur.462..891C, 2010Natur.465.1049S}, we are now in an age of population-level exoplanetary statistics \citep[for a recent review, see][]{2021arXiv210302127Z}.
        We now know that (statistically) almost every star hosts a planet and one in two Solar-like stars host a rocky planet in their habitable zone \citep{Hsu2019,Bryson2021}. Moreover, many exoplanet-hosting stars have multiple planets orbiting them. 
        

    The arrangement of multiple planets and the collective distribution of their physical properties around host star(s) characterises the architecture of a planetary system \citep{Mishra2021}. Exoplanets in some multi-planetary systems are thought to behave \lang{like} `\pip/' \citep{Lissauer2011, Ciardi2013, Millholland2017, Weiss2018}. The \pip/ trend consists of the following correlations: size, whereby adjacent exoplanets are either similar or ordered in size (i.e. the outer planet is larger); mass, whereby adjacent exoplanets are either similar or ordered in mass; spacing, whereby for a system with three or more planets, the spacing between an adjacent pair of exoplanets is similar to the spacing between the next consecutive pair; packing, whereby smaller planets tend to be packed together closely and larger planets are in wider orbital configurations.

    While the statistical method used by \cite{Weiss2018} has been debated \citep{Zhu2019, Murchikova2020, Weiss2019}, support for the astrophysical nature of the \pip/ correlations (as opposed to \lang{emerging} from detection biases) has emerged from theoretical studies and numerical simulations \citep{Adams2019, Adams2020, He2019, He2020, Mulders2020}. In particular, \citet{Mishra2021} reproduced the observations from \citet{Weiss2018} using a model of planet formation and evolution (the Bern Model \cite{Emsenhuber2021A, Emsenhuber2021B}) and a model for the detection biases of a Kepler-like transit survey (using KOBE). We showed that when nature's underlying exoplanetary population (consisting of detected and undetected exoplanets) resembles \pip/, then a population of transiting exoplanets will have correlations that are consistent with those found by \cite{Weiss2018}. In addition, \cite{Mishra2021} suggested that the four trends are not independent of each other. The size correlations seem to emerge from the mass correlations, while the mass and packing trends could combine to give rise to the spacing trend. The \pip/ trends are amenable to a unification.

        Most of the current studies on this topic utilise statistical correlation coefficients at the population level, that is, the correlation is measured for adjacent planetary pairs from several planetary systems. While useful in terms of testing the existence (or otherwise) of architecture trends, these coefficients may have limited utility for analysing the architecture of a single planetary system. Being statistical in nature, a reliable estimate of these coefficients requires large datasets - which seems difficult for a single system. Although there are some planetary system-level studies \citep[][discussed in Sect. \ref{subsec:comparemetrics}]{Kipping2018,Alibert2019, Mishra2019, Gilbert2020, Bashi2021}, the current literature lacks a prescription for  \edit{uniformly assessing the multi-faceted architectures of several quantities (e.g. mass architecture, radius architecture, or eccentricity architecture)} for a single planetary system.   

        We seek a framework that allows us to characterise the architecture of an individual planetary system. Our motivations for developing such a framework arise from questions related to: formation, such as the extent to which \lang{a system's} architecture \lang{is} shaped by initial conditions (i.e. the environment in and around the star and protoplanetary disk formation regions) \citep{Jin2014, Safsten2020}; evolution, the role of physical processes such as orbital migration or giant impacts in shaping the final architecture of planetary systems \citep{Mulders2020}; identification, \lang{which} particular stars host planets that resemble \pip/, and, in particular, whether the planets in systems like TOI-178 \citep{2021A&A...649A..26L}, Trappist-1 \citep{2021PSJ.....2....1A}, or 55 Cancri \citep{Bourrier2018} show mass/size similarities; other architectures, \lang{we know} that there are many planetary systems that do not follow the peas in a pod architecture (e.g. the Solar System). Overall, it is not obvious how the architecture \lang{of any} individual planetary system should be uniformly assessed. 

        In this series of papers, we propose a framework for examining the architecture of planetary systems at the system level. The philosophy behind system level analysis is to consider the entire planetary system as a single unit of a physical system. This framework allows us to not only quantify, compare, and investigate a system's architecture, \lang{but also offers} some unexpected benefits. As it turns out, the framework allows for a \edit{conceptually} intuitive partitioning of the space of possible architectures. We label the four classes of planetary system architectures as: similar, ordered, anti-ordered, and mixed. In this way, our work extends the trends initiated by the notion of \pip/ architecture. Furthermore, we verify the unification of the \pip/ correlations proposed in \cite{Mishra2021}. \lang{We find that}, Similar architectures are the most common type of planetary system architectures and their high occurrence explains why the intra-system radius uniformity was already observable from the first four months of Kepler data \citep{Lissauer2011}.
        
        Our framework \lang{engenders} novel questions. For instance, \lang{if nature produces distinct classes of architecture in multi-planetary systems, then what is the frequency or occurrence rates of these architecture classes? How does the occurrence of an architecture class depend on stellar and protoplanetary disk environment? How does the architecture of a system evolve over time? What is the role of stellar evolution, protoplanetary disk interactions, and planet formation in shaping the final architecture? How is a planet's internal composition related to the system's architecture? Or does the ability of a planet to host life depends on the architecture of the planetary system?} In this series of papers, we explore  these questions. Although the number of multi-planetary systems is low today, this may change in the next few decades. Thanks to large survey missions such as PLATO \citep{Rauer2014}, GAIA \citep{2016A&A...595A...1G}, TESS \citep{2015JATIS...1a4003R}, LIFE \citep{2021arXiv210107500Q}, and others, the growing number of known multi-planetary systems will allow for a better understanding to emerge. We hope our work encourages observers to dedicate more observation time to detecting planets within a known planetary system, that is, in \lang{finding} multi-planetary systems. 
        

        The architecture classification scheme proposed in this paper is a model-independent framework. \change{To demonstrate our classification framework and explore its consequences, we applied our framework to simulated planetary systems. To illustrate our framework on real systems, we also applied our framework to observed exoplanetary systems}. We emphasise that while the results emerging from the application of our framework on these datasets may suffer from some limitations (arising from theoretical modelling or detection biases for observed systems); however, the concept of our architecture classification scheme, being model-independent, does not share these limitations.  
        In this paper, we present the catalogues of planetary systems  we apply our framework to in Sect. \ref{sec:catalogues}, along with a newly curated catalogue of observed exoplanetary systems and simulated planetary systems, using the Bern Model. We introduce our framework in Sect. \ref{sec:framework}. In Sect. \ref{sec:architecturetypes}, the characteristics of the architecture classes are discussed. We explore the link between the internal composition of planets and the system architecture class in Sect. \ref{sec:internalcomposition}. Then, in Sect. \ref{sec:discussion}, we speculate on how habitability could depend on the architecture of planetary systems. Our conclusions are given in Sect. \ref{sec:conclusion}.
        
        In a companion paper, we investigate the formation pathways, i.e. the role of initial conditions and physical processes in shaping the final architecture (\cite{Mishra2022b} referred to as \papertwo/). Our work demonstrates that the processes of planet formation and evolution are imprinted on the entire system-level architecture. We find that  protoplanetary disks with low solid-mass 
        \lang{give rise to} planetary systems endowed with a mass similarity. On the other hand, massive disks and high metallicity often lead to mass Ordered, Anti-Ordered, or Mixed system architectures. Planet-planet and planet-disk interactions play a decisive role in shaping these three architectures.

\section{Catalogues}
\label{sec:catalogues}

\subsection{Theoretical dataset: Bern Model}
\label{subsec:bernmodel}

\begin{figure}
        \resizebox{\hsize}{!}{\includegraphics{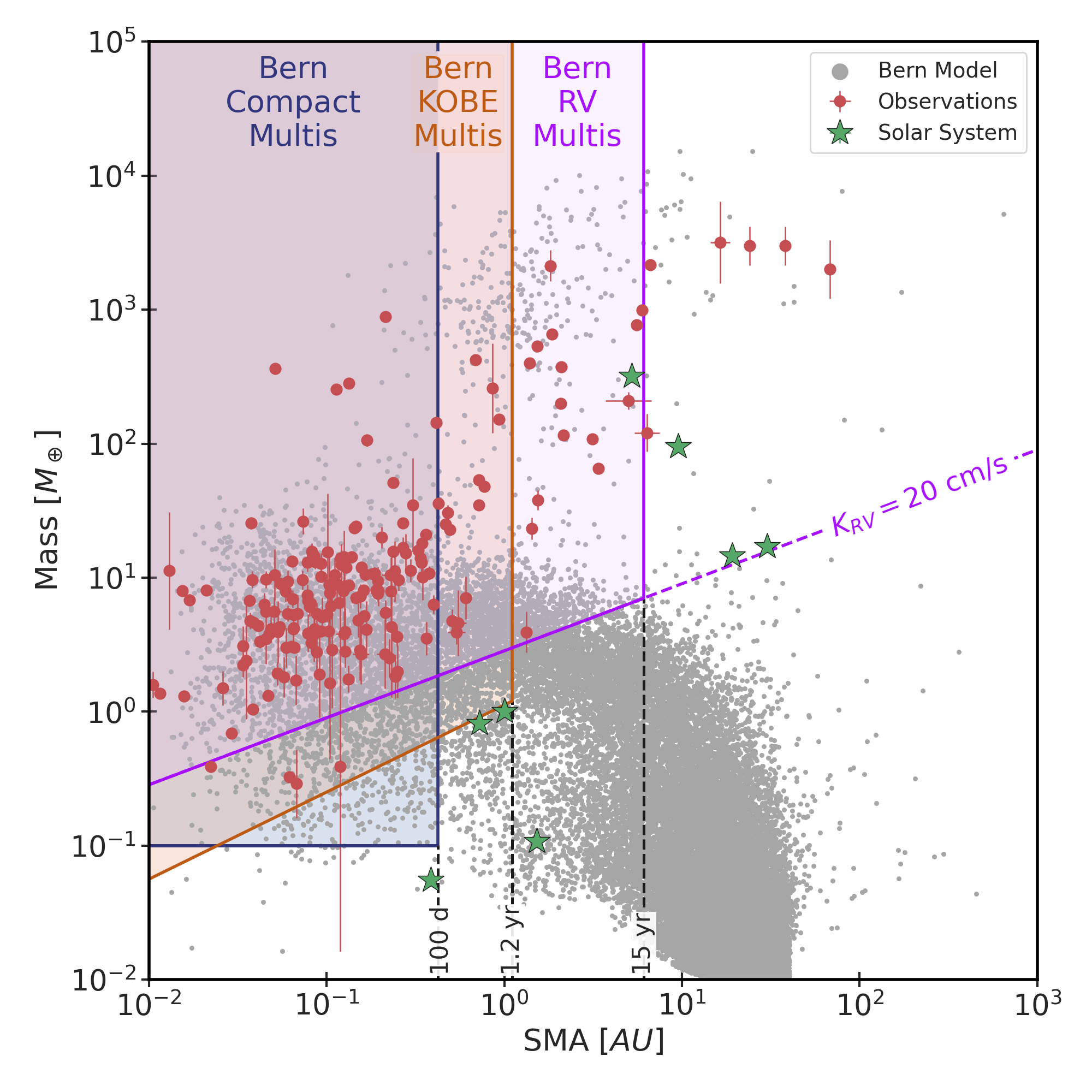}}
        \caption{Mass-distance diagram. This figure shows the masses and the distances of planets in all catalogues used in this study. Shaded regions show the parameter space spanned by synthetic planets observed via radial velocity surveys (Bern RV Multis), transit surveys (Bern KOBE Multis), and ongoing missions (Bern Compact Multis). The parameter space for Bern KOBE Multis has been mapped from its original radius-period plane.}
        \label{fig:masssma}
\end{figure} 

In this series of works, we demonstrate our architecture framework by analysing the architecture of synthetic planetary systems. These systems were numerically computed using the Generation III Bern \lang{M}odel of planet formation and evolution \citep{Emsenhuber2021A,Emsenhuber2021B} that \change{is based on the core-accretion paradigm of planet formation \citep{Pollack1996, Alibert2004, Alibert2005}}. The model follows the growth of protoplanetary embryos embedded in a protoplanetary disk of gas and solids around a solar-type star. A diverse range of physical processes are simultaneously occurring and coherently computed in this 1D star-disk-embryo system. These include: stellar and disk physics (evolution of and interaction between star and viscous disk, condensation of volatile and refractory species, etc.), planetary formation physics (accretion of planetesimals and gases, internal structure calculations, etc.), and additional physics (orbital and tidal migration, planet-planet \textit{N}-body interactions, planet-disk interactions, atmospheric escape, deuterium fusion, etc.). We describe these physical processes in Sect. \ref{sec:bernmodel} and a descriptive summary of these processes is provided in \citet[][in particular, Fig. 1 and Sections 2, 3, and Appendix A]{Mishra2021}. More details can also be found in \citep{Emsenhuber2021A,Emsenhuber2021B}. 

We synthesised 1000 planetary systems, each starting with 100 lunar mass protoplanetary embryos, wherein the following initial conditions were varied: mass of protoplanetary gas disk, photo-evaporation rate, dust-to-gas ratio, disk inner edge, and the starting location of embryos. In Fig. \ref{fig:masssma}, we show all synthetic planets on the mass-distance diagram. For each synthetic planetary system failed embryos, objects with mass less than $0.1 \mearth$, were removed from further analysis\footnote{\change{As long as the mass threshold for failed embryos is kept under $0.1 \mearth$, the results presented in this paper are not sensitive to the threshold limit. We removed these small objects since they (a) failed to grow as massive planets, (b) are insignificant to the dynamical evolution of the system, and (c) are currently unobservable in exoplanetary systems. All results arising from the Bern RV Multis, Bern KOBE Multis, and Bern Compact Multis are insensitive to these failed embryos.}}.  

Three observationally motivated catalogues were prepared from the synthetic dataset. This allowed us to facilitate a comparison of the architecture from observed planetary systems with the synthetic planetary systems and to make predictions. The parameter space \lang{spanned by the planets in these catalogues} is shown in Fig. \ref{fig:masssma}. These catalogues are as follows:

\textit{Bern RV Multis}: We assume a radial velocity (RV) survey which can find planets with periods $\leq 15\ \si{yr}$ and semi-amplitude $K_{RV} \geq 20\ \si{cm/s}$. These numbers are motivated by (a) long-running RV surveys such as the HARPS survey \citep{Mayor2003, Mayor2011} and the California Legacy Survey \citep{Rosenthal2021,Fulton2021}; (b)  current precision achieved by ESPRESSO \citep{Lillo-Box2021,Netto2021}; and (c) making predictions for future RV surveys. Such RV detectable synthetic planetary systems with four or more planets form the Bern RV Multis catalogue, \lang{which includes} $3\,828$ planets around $565$ stars.

\textit{Bern KOBE Multis}: We assume a Kepler-like transit survey which continuously observes $\num{2e5}$ stars for $3.5\ \si{yr}$ \citep{Thompson2018}. A planet which transits three or more times and produces a transit \snr/ of 7.1 or more is considered detectable. The reliability and completeness of such a survey is replicated and those synthetic planets which would have been vetted as `planetary candidates' by the Kepler Robovetter \citep{Thompson2018}, are kept. Such transiting synthetic planetary systems with four or more planets form the Bern KOBE Multis catalogue. KOBE was developed and introduced in \cite{Mishra2021}. There are $6\,715$ planets around $1283$ stars in this catalogue.

\textit{Bern Compact Multis:} Ongoing transit missions such as CHEOPS and TESS have been successful in characterising compact multi-planetary systems, such as TOI-178 \citep{2021A&A...649A..26L} and TOI-561 \citep{2021MNRAS.501.4148L}. Inspired by these discoveries, we investigated the architecture of compact planetary systems simulated by the Bern Model. Our aim is to understand the architecture and make predictions for such systems based on the core-accretion paradigm \citep{Pollack1996, Alibert2004, Alibert2005}. All planets with periods of $\leq 100\ \si{d}$ and masses of $\geq 0.1\ \mearth$ were retained. Synthetic planetary systems, in this parameter space, with four or more planets form the Bern Compact Multis catalogue, with $2\,412$ planets around $400$ stars included.

\subsection{Observational dataset: A new catalogue}
\label{subsec:observations}
\begin{table*}
\caption{Observed multi-planetary systems: There are 41 planetary systems with 194 planets in this catalogue. Only the first five rows are shown here. The entire table is available online. Online version includes additional identification columns: KIC ID, TIC ID, and GAIA ID. Missing information is indicated by `--'. References for individual systems are given in appendix Sect. \ref{sec:starplanetreferences}.}
\label{table:1}
\tiny
\centering

\begin{tabular}{rclllllll}
	\hline \hline
	\multicolumn{9}{c}{Stellar parameters} \\ \hline 
   Hostname & Multiplicity & $M_\star [M_\odot]$ & $R_\star [R_\odot]$ & $L_\star [L_\odot]$ & $T_\mathrm{eff} [K]$ &              [Fe/H] &         Age [Gyr] & Distance [pc] \\
	\hline
Sun &                   8 &                 $1$ &                 $1$ &                 $1$ &              $5,772$ &                 $0$ &  $04.6 \pm \ 0.1$ &           $0$ \\
Trappist-1 &                   7 &   $0.1 \pm \ 0.002$ &   $0.1 \pm \ 0.001$ &          $5.53e-04$ &    $2,566 \pm \ 026$ &  $+0.04 \pm \ 0.08$ &  $07.6 \pm \ 2.2$ &       $012.0$ \\
TOI-178 &                   6 &    $0.7 \pm \ 0.03$ &    $0.7 \pm \ 0.01$ &   $0.1 \pm \ 01.08$ &    $4,316 \pm \ 070$ &  $-0.23 \pm \ 0.05$ &  $07.1 \pm \ 6.1$ &       $062.7$ \\
HD 10180 &                   6 &    $1.1 \pm \ 0.05$ &    $1.1 \pm \ 0.04$ &   $1.5 \pm \ 00.02$ &    $5,911 \pm \ 019$ &  $+0.08 \pm \ 0.01$ &  $04.3 \pm \ 0.5$ &       $039.0$ \\
HD 219134 &                   6 &    $0.8 \pm \ 0.03$ &   $0.8 \pm \ 0.005$ &   $0.3 \pm \ 00.01$ &    $4,700 \pm \ 020$ &  $+0.11 \pm \ 0.04$ &  $11.0 \pm \ 2.2$ &       $006.5$ \\
	\hline 
\end{tabular}

\vspace*{0.5em}

\begin{tabular}{rccccccc}
	\hline \hline
	\multicolumn{8}{c}{Planetary parameters} \\ \hline 	
   Hostname &                                                  Planet &                                                                    $M_p [M_\oplus]$ &                                                                $R_p [R_\oplus]$ &                                                                      $a_p [AU]$ &                                                                         $e$ &                                                                $i\ [^{\circ}]$ &                                     min. $M_p$ \\
\midrule
Sun &  $ \left( \Centerstack[l]{m,v,e,m,\\j,s,u,n} \right)$ &  $\left( \Centerstack[c]{$00.055 \pm 00.000$,\\$00.815 \pm 00.000$,\\ ...} \right)$ &  $\left( \Centerstack[c]{$0.383 \pm 0.000$,\\$0.949 \pm 0.000$,\\ ...} \right)$ &          $\left( \Centerstack[c]{$0.387 \pm -$,\\$0.723 \pm -$,\\ ...} \right)$ &        $\left( \Centerstack[c]{$0.21 \pm -$,\\$0.01 \pm -$,\\ ...} \right)$ &          $\left( \Centerstack[c]{$7.00 \pm -$,\\$3.39 \pm -$,\\ ...} \right)$ &  $\left( \Centerstack[c]{F,F,F,F,...} \right)$ \\
Trappist-1 &      $\left( \Centerstack[l]{b,c,d,e,\\f,g,h}\right)$ &  $\left( \Centerstack[c]{$01.374 \pm 00.069$,\\$01.308 \pm 00.056$,\\ ...} \right)$ &  $\left( \Centerstack[c]{$1.116 \pm 0.014$,\\$1.097 \pm 0.014$,\\ ...} \right)$ &  $\left( \Centerstack[c]{$0.012 \pm 0.000$,\\$0.016 \pm 0.000$,\\ ...} \right)$ &  $\left( \Centerstack[c]{$0.01 \pm 0.00$,\\$0.01 \pm 0.00$,\\ ...} \right)$ &  $\left( \Centerstack[c]{$89.73 \pm 0.17$,\\$89.78 \pm 0.12$,\\ ...} \right)$ &  $\left( \Centerstack[c]{F,F,F,F,...} \right)$ \\
TOI-178 &          $\left( \Centerstack[l]{b,c,d,\\e,f,g}\right)$ &  $\left( \Centerstack[c]{$01.500 \pm 00.440$,\\$04.770 \pm 00.680$,\\ ...} \right)$ &  $\left( \Centerstack[c]{$1.152 \pm 0.073$,\\$1.669 \pm 0.114$,\\ ...} \right)$ &  $\left( \Centerstack[c]{$0.026 \pm 0.001$,\\$0.037 \pm 0.001$,\\ ...} \right)$ &              $\left( \Centerstack[c]{$- \pm -$,\\$- \pm -$,\\ ...} \right)$ &  $\left( \Centerstack[c]{$88.80 \pm 1.30$,\\$88.40 \pm 1.60$,\\ ...} \right)$ &  $\left( \Centerstack[c]{F,F,F,F,...} \right)$ \\
HD 10180 &          $\left( \Centerstack[l]{c,d,e,\\f,g,h}\right)$ &  $\left( \Centerstack[c]{$13.222 \pm 00.445$,\\$12.014 \pm 00.699$,\\ ...} \right)$ &                  $\left( \Centerstack[c]{$- \pm -$,\\$- \pm -$,\\ ...} \right)$ &  $\left( \Centerstack[c]{$0.064 \pm 0.001$,\\$0.129 \pm 0.002$,\\ ...} \right)$ &  $\left( \Centerstack[c]{$0.07 \pm 0.03$,\\$0.13 \pm 0.05$,\\ ...} \right)$ &                $\left( \Centerstack[c]{$- \pm -$,\\$- \pm -$,\\ ...} \right)$ &  $\left( \Centerstack[c]{T,T,T,T,...} \right)$ \\
HD 219134 &          $\left( \Centerstack[l]{b,c,f,\\d,g,h}\right)$ &  $\left( \Centerstack[c]{$04.620 \pm 00.140$,\\$04.230 \pm 00.200$,\\ ...} \right)$ &  $\left( \Centerstack[c]{$1.544 \pm 0.059$,\\$1.458 \pm 0.048$,\\ ...} \right)$ &          $\left( \Centerstack[c]{$0.039 \pm -$,\\$0.065 \pm -$,\\ ...} \right)$ &     $\left( \Centerstack[c]{$0.00 \pm -$,\\$0.06 \pm 0.04$,\\ ...} \right)$ &  $\left( \Centerstack[c]{$85.19 \pm 0.13$,\\$87.38 \pm 0.10$,\\ ...} \right)$ &  $\left( \Centerstack[c]{F,F,T,T,...} \right)$ \\
\hline 
\end{tabular}
\end{table*}
%
%

%
%

To demonstrate our framework on observed exoplanetary systems, we have curated a new catalogue of known multi-planetary systems\footnote{The catalogue was last updated in April 2021.}. A salient feature of this catalogue (and the philosophy behind this work) is its focus on considering planetary systems as a single unit of a physical system. Unlike \lang{focussing} on individual exoplanets or a single detection technique, our aim is to study the planetary system as a whole. There are two serious challenges to this endeavour. Firstly, the biases present in detection methods tend to prevent a complete, reliable picture of an exoplanetary system from emerging (either via undetected or mischaracterised planets). Secondly, detecting planets on long orbital periods requires long-term, repeated observations, which is considerably challenging. We hope that upcoming missions and future surveys can mitigate these difficulties. 

We included a planetary system in our catalogue if: (a) it has at least four known planets and (b) masses \lang{are} available for at least four planets. For example, Kepler-33, a five planet system, is included because mass measurements are available for four of its planets\footnote{For this study, the distinction between mass and minimum mass is ignored.}. The criterion \lang{of requiring} minimum four planets emerges due to (a) the requirement for enough planets for adequately characterising the architecture and (b) because for systems with lower number of planets, it is perhaps difficult to uniformly assess whether the low multiplicity is an outcome of natural processes or detection biases. \change{To keep the comparison between observations and theory uniform, all catalogues in this series of works only consider planetary systems with four or more planets. The architecture framework can, however, handle two- or three-planet systems as well.} To make this catalogue useful to the wider community and enable future studies, we gathered several key stellar and exoplanetary properties. For host stars, we report the mass, radius, luminosity, effective temperature, metallicity, age, and distance, along with their identification numbers (when available) in the Kepler Input Catalogue (KIC), TESS Input Catalogue (TIC), and GAIA ID. For planets, we report mass or minimum mass, radius, semi-major axis, eccentricity, and inclination. In a conservative approach, errors (reported when possible) are the maximum of the upper and lower error bounds available in the literature. When multiple publications reported planetary parameters, a more recent publication was preferred. When a single publication reported parameters for all planets in a system, then such a consistent set of solution was given preference (e.g. GJ 676 A or Kepler-11). For stellar parameters, if a star was included in KIC, then the values from \cite{Berger2020} are reported. Most other stellar parameters come from the TIC \citep{Stassun2019} or from individual publications.

There are 41 planetary systems that meet our criteria and define our multi-planetary system catalogue (Table \ref{table:1}). With a total of 194 planets in our catalogue, the number of planetary systems with four, five, six, seven, and eight planets is 24, 7, 8, 1, and 1. In this paper, we present the observed planetary systems as they are known today and we do not correct the observations for any detection biases. Instead, to assist in making comparisons with the theory, detection biases will be placed on simulated planetary systems (Sect. \ref{subsec:bernmodel}). Figure \ref{fig:masssma} shows the mass of observed exoplanets as a function of their semi-major axis.

\edit{
While our observed multi-planetary systems catalogue engenders system-level studies, its current form poses several technical difficulties. Foremost, the number of observations is only forty-one. Secondly, multiple detection methods, such as radial velocity or transits (etc.) were employed to observe these planetary systems. Each observation technique suffers from certain limitations and detection biases. This implies that the observed systems in our catalogue do not constitute a homogeneous and complete set of observations. These two limitations of the observations catalogue prohibit us from deducing any statistically strong result. Nevertheless, we used the observed systems for (a) exemplifying system-level approach to real planetary systems and (b) using our framework on observations to explore trends in the architecture of observed systems.}

\edit{
Our results from the observed catalogue may be affected by another source of difficulty. There are two systems in our catalogue that host some planets without known mass measurements (Kepler-33 b and Kepler-80 f and g). Since these two systems have at least four planets with known masses, they have been included in our study. However, this does not impact the results of the present study in a drastic way. All three planets in these systems without mass measurements are either the innermost and/or the outermost planets in their respective systems. Therefore, the missing measurements do not have a strong influence on the characterisable mass architecture. The missing measurement may have a strong effect  if any planet with unknown mass was in between two planets with known masses.}

        \section{Characterizing architecture: A new framework}
        \label{sec:framework}
        
\subsection{Literature review}
\label{subsec:comparemetrics}

\edit{We review some approaches from other studies that have tried to capture planetary system-level properties in this section.}
\cite{Kipping2018} investigated similarity and ordering (of planetary sizes) at the level of an individual system.
Using an entropy based framework on Kepler systems, he concludes that initial conditions are inferable from the present-day architecture. As we go on to  show in this series, our work not only supports this conclusion, but additionally demonstrates the possible links between initial conditions and final architecture. Although the above-mentioned study considers a similar problem to the one we deal with here, our frameworks differ considerably. Built on step-functions and combinatorics, the aforementioned framework does not take into account the magnitude of variation. 


\cite{Alibert2019} proposed a concept of distance between two planetary systems. 
The Alibert distance captures inter-system differences, whereas our framework quantifies intra-system similarities. The Alibert distance is useful to quantify the similarity (or dissimilarity) between two planetary systems and in unsupervised machine-learning algorithms to find clusters in the space of planetary systems. \edit{\cite{Bashi2021}  recently proposed another concept for distance based on a statistical distance.
The `weighted' energy distance is the distance between two planetary systems, with each planet represented on the log-period and log-radius plane, utilising planetary masses (from a mass-radius relationship) as weights.} As with the Alibert distance, the Bashi-Zucker distance  requires two planetary systems and thus it is not suitable for characterising the global architecture for a single planetary system.


\cite{Gilbert2020} proposed seven parameters for quantifying the global structure of planetary systems: dynamical mass (ratio of mass in planets to stellar mass), mass partitioning (normalised mass disequilibrium), mass monotonicity (weighted Spearman correlation coefficient), characteristic spacing (average mutual Hill radii), gap complexity, flatness, and multiplicity ($\nplanet$). Of these measurements, mass partitioning and mass monotonicity have close parallels with our framework.
The input information required to compute mass partitioning, and monotonicity is exactly the same as the input information for our architecture framework, namely, a set of planetary masses. However, we find that the output displays a curious mix of concepts.

Mass partitioning is zero for a system in which all planets have the same mass. When one planet has some mass and all other planets have negligible mass, the mass partitioning for this system is unity. While this parameter captures the two extreme cases, it is difficult to interpret and employ this measure in cases other than these two extremes. Behaving similarly to a correlation coefficient, mass monotonicity has a range of [-1,1]. It is defined as the Spearman correlation coefficient (between mass and distance) multiplied by the mass partitioning (which is weighted by $\nplanet^{-1}$). \edit{Although the work of \cite{Gilbert2020} studies the architecture of planetary systems at the system-level, we seek a framework which can also be used with planetary properties other than mass, such as  radius, bulk density, water mass fraction, eccentricities, and so on.}

\edit{\cite{Millholland2017} and \cite{Wang2017} showed that the peas in a pod pattern reported by \cite{Ciardi2013, Weiss2018} also extends to planetary masses. \cite{Millholland2017}, \change{using planetary masses derived from transit-timing variations}, studied the clustering of planets in the mass-radius plane and found that the sum of distances (in the log mass-size space) between adjacent planets of real systems is much smaller than a bootstrapped randomised population. Based on a set of 29 \change{RV} observed systems, \cite{Wang2017} infer two types of planetary systems. Planetary systems with masses of $\lesssim 30 M_\oplus$  show intra-system mass uniformity, while systems with masses $\gtrsim 100 M_\oplus$  do not follow the peas in a pod pattern -- indicating that} \change{there are only two possibilities for the architecture structure. As we show in this series of works, their hypothesis of only two architecture types is too simple and cannot capture the richness of physics.}


        \subsection{Concept}
        \label{subsec:concept}
        \begin{figure}
        \resizebox{\hsize}{!}{\includegraphics[trim=0.5cm 0.5cm 2cm 0.5cm , clip]{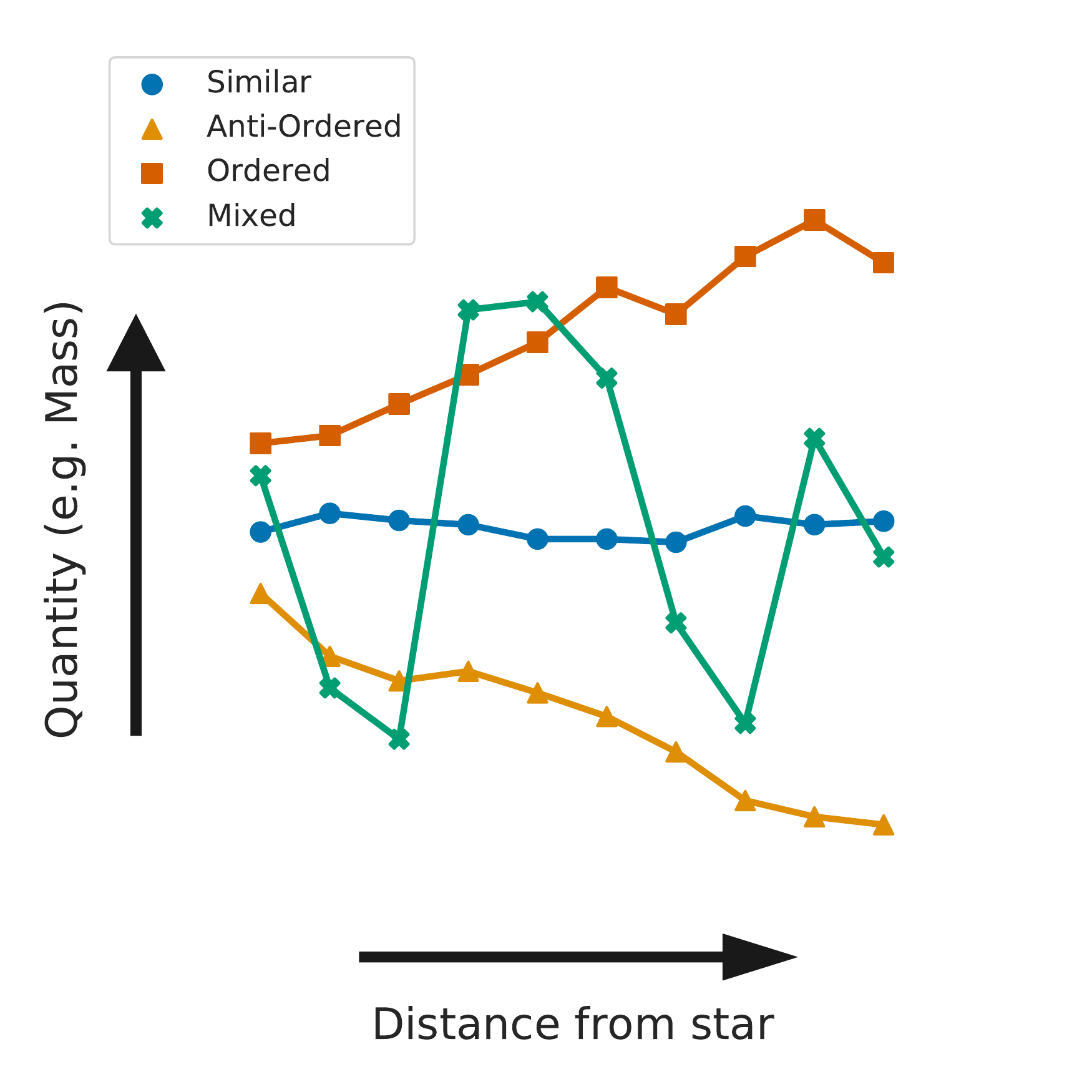}}
        \caption{Classes of architecture.  This schematic diagram shows the four architecture classes: \similar/, \antiordered/, \mixed/, and \ordered/. Depending on how a quantity (e.g. mass or size) varies from one planet to another, the architecture of a system can be identified.}
        \label{fig:schematic_idea}
        \end{figure}

        With our framework, we initially aimed to capture the key aspect about the \pip/ architecture trends. These trends are correlations between adjacent planets or between consecutive pairs of adjacent planets. We want to capture these ideas at the level of a single planetary system through a unified framework. We do this by studying how a quantity, $q_i$, (such as mass, size, or period ratio) varies for all planets within a system. Here, $i$ indexes the planets within a system. For all quantities, we adopt an `inside-out' convention, namely, we start with the innermost planet ($q_{i=1}$) and go to the next adjacent planet ($q_{i=2}$), and so on. By comparing how $q_i$ varies for each planet inside-out, we are actually estimating how $q_i$ varies with distance from the host star.
                
        In comparing a quantity, $q_i$, with distance, four kind of variations emerge. In one scenario, a quantity could show little to no variation. In another case, the value of a quantity may increase with increasing distance or, conversely, the quantity could decrease from one planet to another. Finally, it is also possible for a quantity to not have any clear variations from one planet to another. We identify these four scenarios as the four classes of architectures that can exist at the level of a single planetary system. This idea is depicted in Fig. \ref{fig:schematic_idea}. 
        
        \cite{Mishra2021} suggested that the mass correlations could originate from planet-formation physics and the correlations of size and spacing could be derivative. Therefore, we first apply our framework using planetary masses (except in Sects. \ref{sec:internalcomposition} and \ref{sec:discussion}). As depicted in Fig. \ref{fig:schematic_idea}, when the masses of all planets within a system are similar to each other, we label the architecture of such systems as `similar'. This architecture class corresponds to the \pip/ architecture reported in observations \citep{Weiss2018, Millholland2017}. When the masses of planets tend to increase inside-out, the architecture of such systems is labelled `ordered'. If the planetary mass tends to decrease from the inner planet to the outer, we label the architecture of these systems as `anti-ordered'. Finally, if a large increasing and decreasing variation in the planetary masses is present, we label the architecture of such systems as `mixed'. \edit{The mixed architecture class is also useful in capturing all other architecture patterns which do not fall under the other three architecture classes. \cite{Kipping2018}, for example, has analysed some interesting repeating patterns.}
        By introducing these architecture classes, our framework organises the possibilities for system architecture. 
        
        \lang{One might wonder, at this point, why introduce such a concept and the ensuing mathematical machinery?} While part of this work began as an inspired exploration to categorise our understanding of system architecture, it turns out that there are good physical reasons to pursue this process. As is shown in this and a companion paper, planetary systems that have the same architecture tend to have a host of other properties in common, such as internal structures (core-mass, ice-mass) distributions. Most importantly, systems with a common architecture tend to have same formation pathways, initial conditions, and evolutionary histories. Practically, this means that a quick glance at a system's architecture may reveal a lot more about its formation scenario. 
        
        \change{Our architecture classification framework utilises two quantities --  the \cstext/ and the \cvtext/,  introduced in Sects. \ref{subsec:cs} and \ref{subsec:cv}, respectively.} These two coefficients allow us to quantify the conceptual ideas we have presented above. Together, these coefficients define a new space of possibilities for system architectures.  In Sect. \ref{subsec:classify}, we identify the regions of this architecture space that correspond to the four architecture classes introduced above. As this framework deals with the architecture of multi-planetary system, systems with only one planet are not studied within this framework. 

        \subsection{Coefficient of similarity}
        \label{subsec:cs}
    
    The term `coefficient of similarity' is commonly used in the fields studying statistics of ecology and genetics \citep{simcoeff, 10.1673/031.009.7101}. We borrow the term but develop our own concept and definition.Let $q$ be a planetary quantity such as mass, size, period ratios of adjacent planets, bulk density, eccentricity, and so on\footnote{For quantities which admit zero as a possible value, the \cstext/ may become ill-defined. This is a coordinate singularity and can be dealt with an appropriate treatment (see Eq. \ref{eq:cs_zero} Sect. \ref{subsec:icefrac}).}. The value of this quantity for the $i^\mathrm{th}$ planet in a system is denoted by $q_i$. The coefficient of similarity, $\cs$, measures how $q$ changes from one planet to another, inside-out. For a system with $\nplanet$ planets, it is defined as:
        \begin{equation}
        \label{eq:cs}
        \cs (q) = \frac{1}{\nplanet-1}  \ \sum_{i=1}^{i=\nplanet-1} \  \Bigg(log \ \frac{q_{i+1}}{q_i} \ \Bigg).
        \end{equation}
        There is a clear physical interpretation for $\cs (q)$: the coefficient of similarity measures the average order of magnitude variation in the quantity $q$ from one planet to another. The definition of the \cstext/ allows us to map the architecture of a planetary system on a one dimensional axis. When $\cs(q) \ \approx 0$, then the system's architecture could imply a similarity in $q$. When $\cs(q)$ is positive, then planets within a system are ordered in $q$. Conversely, $\cs(q)$ being negative, implies that the planets are anti-ordered.
        
        
        \change{We have developed a mathematical formalism to study the sensitivity of the \cstext/. In Appendix \ref{sec:limits}, we derive the limiting values of the \cstext/ and present the results here.  For example, when the $q_i$ values for all planets in a system are within $10\%$ of each other, then the maximum possible value of $\cs(q)$ is $0.09$ (see Eq. \ref{eq:cslimit}). For maximum tolerances of $20\%,\ 40\%,\ 60\%, \ \text{and}\  80\%$, the maximum possible value of $\cs(q)$ are $0.18,\ 0.37,\ 0.60,\ \text{and}\ 0.95$ respectively. In Fig. \ref{fig:cscvlimits}, we show the dependence of the $\max \cs(q)$ on $t$.}
        
        The \cstext/ cannot distinguish between two classes of architecture: similar and mixed. Systems which show similarity will have $\cs(q) \approx 0$. However, system with mixed architecture have large increasing and decreasing variations, such that the log of ratios $\tfrac{q_{i+1}}{q_i}$ cancels itself out. Such systems will also have $\cs(q) \approx 0$. We propose the \cvtext/ to distinguish these two architecture classes. \change{The \cstext/ depends on the actual order in which planets exist (inside-out) in a system. As we go on to show, the \cvtext/ does not depend on the ordering of planets in a system.}
        
        \subsection{Coefficient of variation}
        \label{subsec:cv}
        
        The \cvtext/, $\cv$, is a standard descriptive statistic used to measure the magnitude of variation in a set of numbers \citep{Katsnelson1957,Sharma2010, Abdi2010}. 
    It is defined as the ratio of the standard deviation with the mean:
        \begin{equation}
        \label{eq:cv}
        \cv (q) = \frac{\sigma(q)}{\bar{q}}.
        \end{equation}
        
        The \cvtext/ is a positive quantity. When all $q_i$  have the same value then $\cv(q) = 0$. Planetary systems consisting of planets that have a small (or large) variability in their $q_i$ values will have a small (or large) value of the \cvtext/. Now, the distinction between systems showing similarity and mixed architecture is clear. While similar systems will have a low value of the \cvtext/, mixed systems will have a high value of \cvtext/. 
        
        \change{Since this coefficient is a well known statistical measure, there are some derivations for its limit. A classical result from \cite{Katsnelson1957} shows that, for a set of $\nplanet$ numbers, the maximum value of the \cvtext/ is $\sqrt{\nplanet - 1}$. However, this result is only a particular case in our setup. In Appendix \ref{sec:limits}, we develop a mathematical formalism to understand the limits of the \cvtext/ and present the results here. When the $q_i$ values for all planets in a system are within $10\%, \ 30\%, \ 50\%, \ 70\%, \ \text{and} \ 90\%$ of each other, the absolute theoretical upper limit of $\cv (q)$ is $0.10,\ 0.31,\ 0.58,\ 0.98,\ \text{and}\ 2.06$ respectively. Figure \ref{fig:cscvlimits} shows how this upper limit varies with the maximum tolerance, $t$, for a system. }
        
%
%
        
        \subsection{Classifying the architectures of planetary systems}
        \label{subsec:classify}

        \def\figwidth{8cm}
        \begin{figure*}
                \centering
                \includegraphics[width=\figwidth]{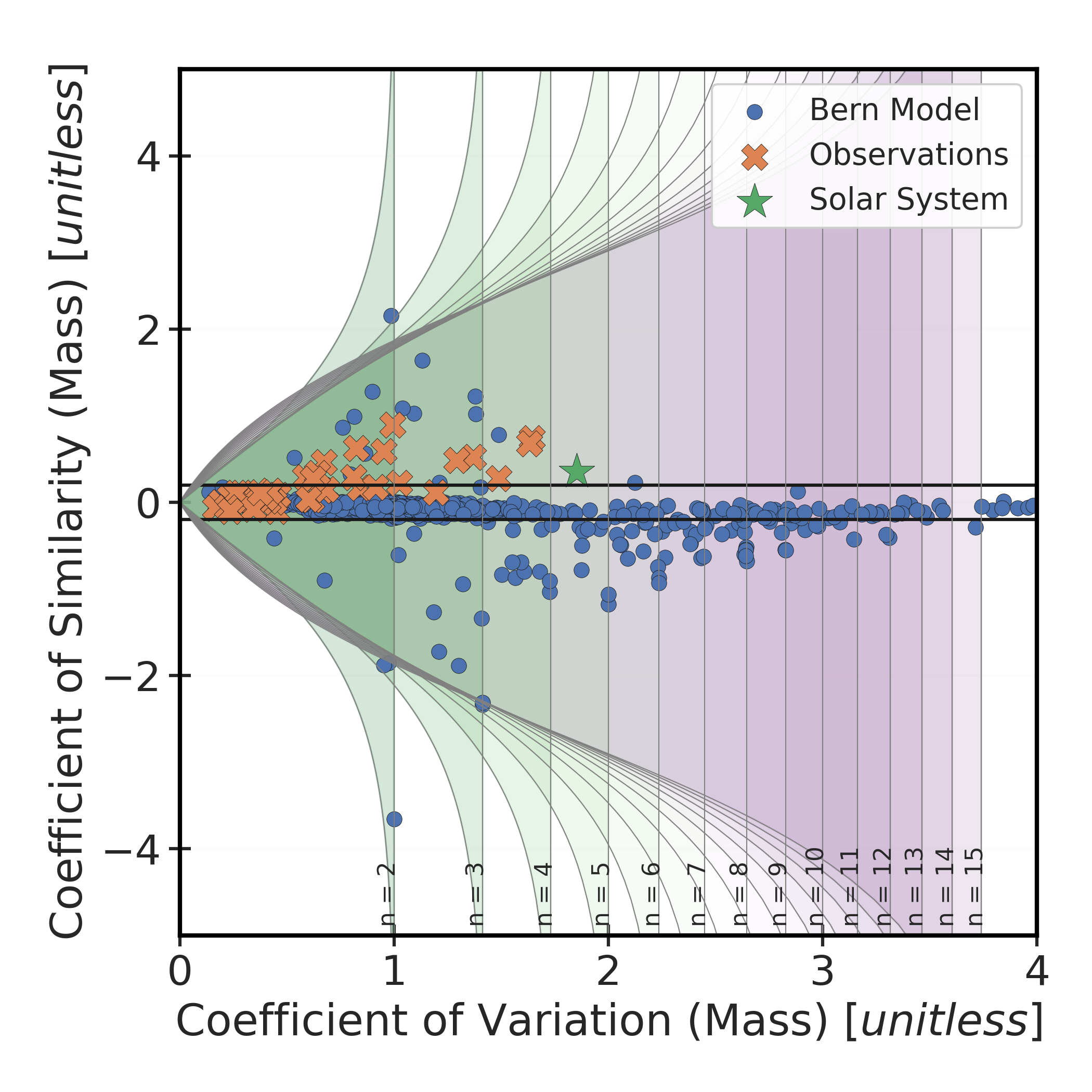}
                \includegraphics[width=\figwidth]{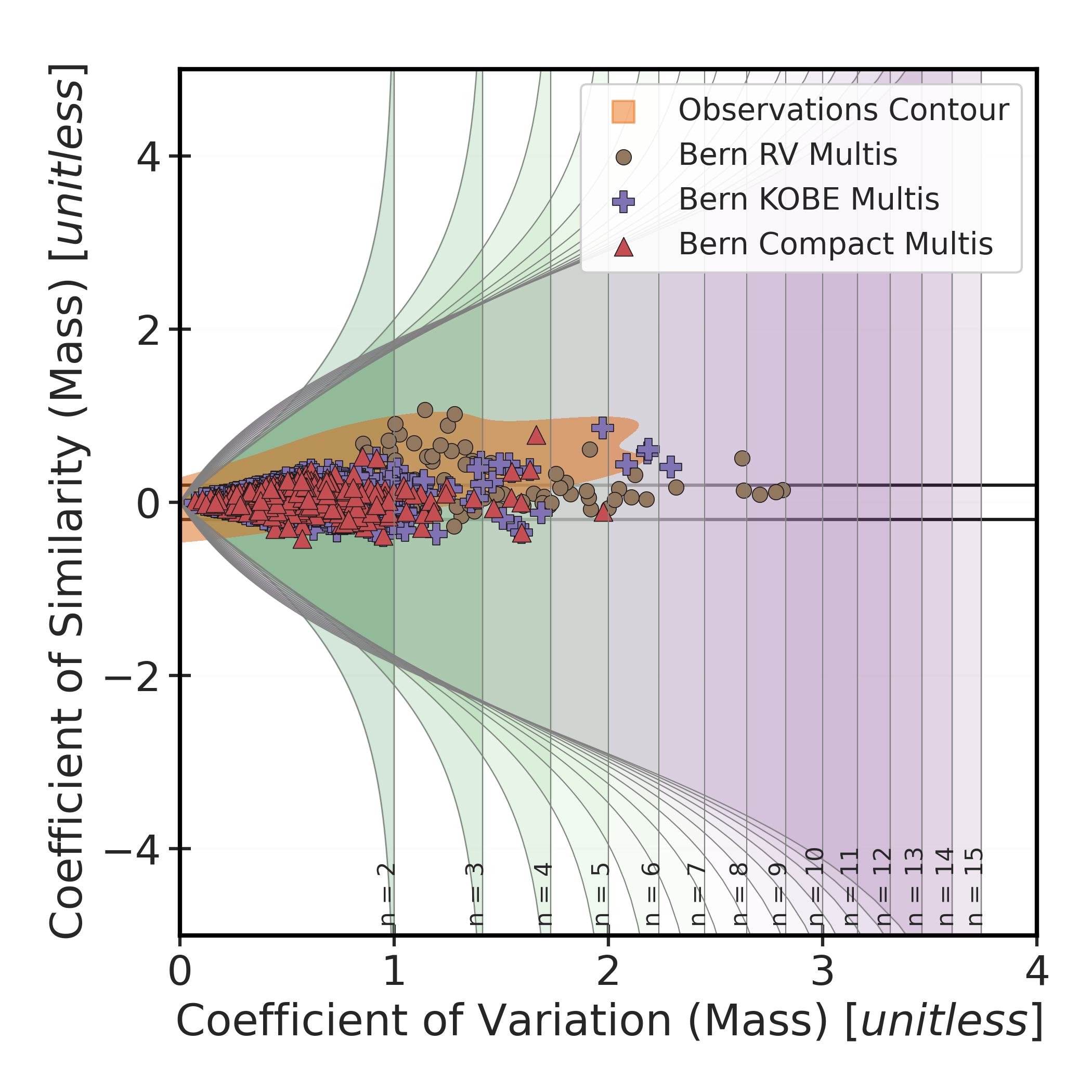}
                \caption{New parameter space:\ Architectures of planetary systems.  Both panels shows the \cstext/ (mass) as a function of the \cvtext/ (mass). The shaded regions show the allowed parameter space for planetary systems. The white gaps (between two shaded regions) mark the mathematically forbidden regions of this architecture space. Different parts of this parameter space are identified with four architecture classes, in accordance with Eq. \ref{eq:classify}. Each point corresponds to an individual planetary system. For visual clarity, the shaded and unshaded regions are drawn only for systems hosting up to fifteen planets. \textit{Left:} Planetary systems from the Bern model and observations. \textit{Right:} Synthetically observed systems depicting the detection biases of radial velocity and transit surveys.}
                \label{fig:csversuscv}
        \end{figure*}
        
        
        We are interested in obtaining a mapping from the scale-invariant coefficients to an architecture class. 
        \edit{In Appendix \ref{sec:classificationboundaries}, we present some considerations that motivate the selection of boundaries between the four classes. The selected boundaries were additionally tested on thousands of mock planetary systems to check their ability to correctly classify the four architecture classes.} We propose the following boundaries for identifying the architecture class based on planetary masses. 
        \begin{equation}
        \label{eq:classify}
                \begin{split}
                        &\textbf{Architecture class}                                            &\textbf{Condition}\\
                        &\text{Anti-ordered}                                                   &\cs(M) < - 0.2\\
                        &\text{Ordered}                                                                       &\cs(M) > + 0.2\\
                        &\text{Similar}                                                                       &|\cs(M)| \le 0.2 \ \text{and} \ \cv(M) \le \frac{\sqrt{\nplanet - 1}}{2}\\
                        &\text{Mixed}  \qquad \qquad                                  &|\cs(M)| \le 0.2 \ \text{and} \ \cv(M) > \frac{\sqrt{\nplanet - 1}}{2}   
                \end{split}
        \end{equation}

    \change{A natural (and welcome) outcome of these criteria is that a two-planet system can never have a mixed class architecture. The boundary between similar and mixed class is half the maximum possible value of the \cvtext/.} For the solar system, $\cs(M) = 0.36$ and $\cv(M) = 1.85$. This framework robustly identifies the architecture of the solar system as ordered\footnote{Even if the masses of each solar system planet were randomly varied within 85\% of their original values, the emerging architecture is still ordered. With 1M trials, varying the masses randomly within 90\% of their original values lead to ordered (for $\approx 99.45 \%$ trials), mixed (for $\approx 0.55 \%$ trials), and similar (for $\approx 0.001 \%$ trials) architectures.}. This classification is in line with the historic understanding of the solar system architecture: small rocky planets on the inside and giant planets on the outside. If, however, Neptune were replaced with an Earth-like planet, the architecture of the solar system would be classified as mixed. Considering only the inner four planets of the solar system, $\cs(M) = 0.10$ and $\cv(M) = 0.85$, would make the architecture of the inner solar system belong to the similar class. The architecture of the outer four giants in the solar system is anti-ordered and we have $\cs(M) = -0.42$ and $\cv(M) = 1.11$. 
        
        Figure \ref{fig:csversuscv} shows the $\cs(M)$ versus $\cv(M)$ space for planetary systems from several catalogues. The Bern model planetary systems occupy all four regions of this architecture space. Observed planetary systems, however, span only a limited region of this parameter space, given\edit{  the low multiplicity of observed planetary systems}. The architecture space spanned by the observed planetary systems (shaded contour) is in agreement with the synthetically observed planetary systems from Bern Compact Multis, Bern KOBE Multis, and Bern RV Multis. 
        
        The architecture for the systems in the synthetically observed catalogue was calculated based only on the planets that were detected (for RV/KOBE) or included (for Bern Compact Multis) in the above-mentioned catalogue. It is theoretically possible for a single Bern model system to exhibit different architectures depending on the planets which are detected or included. The reverse is also true -- the architecture of an observed planetary system may change if new planets are discovered or old controversial candidates are rejected. While the ground truth architecture for observations seems elusive, a comparison with synthetic observations can bring forth patterns which are unexpected. With this in mind, we consider the following example. 
                
        Detection biases, in both radial velocities and transits, generally disfavour the discovery of less-massive and small planets at larger distances. This implies that \antiordered/ architectures are difficult to detect. In fact, we have no known example of a planetary system showing \antiordered/ architecture in our observations catalogue. This is surprising for two major reasons: (a) theory suggests their existence:\ there are several synthetic planetary systems from the Bern Model whose architecture is \antiordered/; (b) theory suggests their discovery:\ all three synthetically observed catalogues contain some (albeit few) \antiordered/ planetary systems. 
    Since the number of systems in our catalogue is too low, we refrain from making any conclusions and, instead, we  await the discovery of \antiordered/ architectures  in the future. However, if such architectures are not found despite considerable efforts, this result will become a strong indicator for shaping our understanding of planet formation. 
        
        Another aspect of this new architecture space is the underlying mathematical structure\footnote{Visualizing this structure is easy (not shown). (a) Construct mock planetary systems with masses, for each mock planet, randomly drawn from a uniform distribution with suitable limits. (b) It is suggested to vary the number of planets in these mock systems randomly. (c) Calculate the $\cs (M)$ and the $\cv (M)$ using equations \ref{eq:cs} and \ref{eq:cv}. (d) Plot $\cs (M)$ versus $\cv (M)$ for this mock population. For large number of systems the plot should be symmetric about $\cs (M) = 0$.}.  In   Fig. \ref{fig:csversuscv},  the shaded areas shown regions where a planetary system, with $\nplanet \in [2,15]$ planets, is allowed. A system with two planets, for example, can only occupy the shaded region labelled `$\nplanet = 2$'. All non-shaded regions (in white -- except the shaded regions for 16 or more planets which is not drawn), on this architecture space, is mathematically forbidden. These are parts of the architecture parameter space that no planetary system, irrespective of its configuration, can occupy. This strong result stems from the mathematical limits that were derived for this work (see Sects. \ref{subsec:cs}, \ref{subsec:cv}, and appendix \ref{sec:limits}).

        For clarity and future convenience, we introduced some terminology to the method. When the architecture framework (i.e. $\cs$ and $\cv$) is applied on planetary \change{bulk} masses, the resulting information tells us the mass architecture of a system, namely, the arrangement and distribution of masses in said system. Similarly, when this framework is applied on radii, it gives us the radius architecture (arrangement and distribution of radii) for the system (Sect. \ref{subsec:radius}). Similarly, we can obtain the bulk-density architecture (Sect. \ref{subsec:bulkdensity}),  core-mass architecture (Sect. \ref{subsec:coremass}), water mass fraction architecture (Sect. \ref{subsec:icefrac}), period-ratio or spacing architecture, eccentricity-architecture, and so on. \textbf{In this series of papers, we identify a system's architecture based on its \change{bulk mass} architecture. Thus, when a system is said to be similar, we are referring to the similarity in terms of the mass architecture.} 
        
        \section{Characteristics of architecture classes}
        \label{sec:architecturetypes}

        \subsection{General comments}

        In earlier studies on the \pip/ architecture, the strength of population-level (i.e. across many planetary systems) trends was quantified using Pearson correlations coefficient \citep{Weiss2018, Zhu2019, Chevance2021, Millholland2021, Mishra2021}. The correlation coefficients were calculated using planetary quantities in the log space (i.e. by first taking the $log_{10}$ of all quantities). This resulted in higher values of the correlation coefficient since quantities have limited range to perambulate in the log space. Consider planetary masses. We calculated the correlation coefficient between the mass of adjacent inner and outer planets in the Bern model population \cite[see Fig. 7 in][]{Mishra2021}. The value of the coefficient is $0.66$ in the log space and $0.16$ in the linear space. This highlights that the planetary masses are more closely clustered in log than in linear space. 
        
        We tested the same correlation for all systems in each architecture class. We expect planetary masses in \mixed/, \ordered/, and \antiordered/ systems should (by definition) have low correlations. On the other hand, \similar/ class architecture should exhibit a strong correlation. Surprisingly, in log space all architecture classes show strong correlations. The coefficient value is $0.67$ for \similar/ class, $0.69$ for \mixed/ class, $0.50$ for \ordered/ class, and $0.58$ for \antiordered/ class architectures. However, in the linear space the coefficient values reflects our expectation: $0.61$ for the \similar/ class, $0.20$ for the \mixed/ class, $0.16$ for \ordered/ class, and $0.05$ for \antiordered/ class. This underscores that strong correlations in the log space may not be indicative of substantive architecture trends. It also shows that our framework is capable of identifying systems in which the '\pip/' architecture is discernible even in the linear space.

        \begin{table}
	\caption{Architecture type of known multi-planetary systems (see Table \ref{table:1} for catalogue and Fig. \ref{fig:archplot} for architecture plot).}
	\label{table:2}
	\tiny
	\centering
	
	\begin{tabular}{rcccc}
\hline \hline
Hostname & Multiplicity & $C_{S}(M)$ & $C_{V}(M)$ & Architecture Class \\
\hline
 Solar System &          $8$ &    $+0.36$ &       1.85 &           Ordered \\
Trappist-1 &          $7$ &    $-0.10$ &       0.45 &           Similar \\
TOI-178 &          $6$ &    $+0.08$ &       0.46 &           Similar \\
HD 10180 &          $6$ &    $+0.14$ &       0.66 &           Similar \\
HD 219134 &          $6$ &    $+0.27$ &       1.49 &           Ordered \\
HD 34445 &          $6$ &    $+0.17$ &       0.84 &           Similar \\
Kepler-11 &          $6$ &    $+0.22$ &       1.03 &           Ordered \\
Kepler-20 &          $6$ &    $+0.00$ &       0.44 &           Similar \\
Kepler-80 &          $6$ &    $-0.00$ &       0.19 &           Similar \\
K2-138 &          $6$ &    $+0.03$ &       0.61 &           Similar \\
55 Cnc &          $5$ &    $+0.52$ &       1.37 &           Ordered \\
GJ 667 C &          $5$ &    $-0.02$ &       0.29 &           Similar \\
HD 158259 &          $5$ &    $+0.11$ &       0.29 &           Similar \\
HD 40307 &          $5$ &    $+0.07$ &       0.33 &           Similar \\
Kepler-102 &          $5$ &    $+0.02$ &       0.41 &           Similar \\
Kepler-33 &          $5$ &    $+0.46$ &       0.67 &           Ordered \\
Kepler-62 &          $5$ &    $+0.15$ &       0.68 &           Similar \\
HD 20781 &          $4$ &    $+0.29$ &       0.59 &           Ordered \\
TOI-561 &          $4$ &    $+0.33$ &       0.64 &           Ordered \\
DMPP-1 &          $4$ &    $+0.29$ &       0.81 &           Ordered \\
GJ 3293 &          $4$ &    $+0.27$ &       0.62 &           Ordered \\
GJ 676 A &          $4$ &    $+0.90$ &       0.99 &           Ordered \\
GJ 876 &          $4$ &    $+0.12$ &       1.20 &             Mixed \\
HD 141399 &          $4$ &    $+0.06$ &       0.40 &           Similar \\
HD 160691 &          $4$ &    $+0.63$ &       0.82 &           Ordered \\
HD 20794 &          $4$ &    $+0.08$ &       0.25 &           Similar \\
HD 215152 &          $4$ &    $+0.07$ &       0.23 &           Similar \\
HR 8799 &          $4$ &    $-0.07$ &       0.17 &           Similar \\
K2-266 &          $4$ &    $+0.03$ &       0.60 &           Similar \\
K2-285 &          $4$ &    $+0.01$ &       0.31 &           Similar \\
Kepler-89 &          $4$ &    $+0.17$ &       0.91 &             Mixed \\
Kepler-106 &          $4$ &    $+0.11$ &       0.26 &           Similar \\
Kepler-107 &          $4$ &    $+0.13$ &       0.42 &           Similar \\
Kepler-223 &          $4$ &    $-0.06$ &       0.22 &           Similar \\
Kepler-411 &          $4$ &    $-0.08$ &       0.34 &           Similar \\
Kepler-48 &          $4$ &    $+0.74$ &       1.64 &           Ordered \\
Kepler-65 &          $4$ &    $+0.68$ &       1.63 &           Ordered \\
Kepler-79 &          $4$ &    $-0.10$ &       0.24 &           Similar \\
WASP-47 &          $4$ &    $+0.59$ &       0.95 &           Ordered \\
tau Cet &          $4$ &    $+0.12$ &       0.37 &           Similar \\
HD 164922 &          $4$ &    $+0.49$ &       1.29 &           Ordered \\
\hline
	\end{tabular}
\end{table}          
        For all 41 observed planetary systems in our catalogue, we report their architecture classes in Table \ref{table:2}. Figure \ref{fig:archplot} shows the architecture of all observed multi-planetary systems in our catalogue. The systems are sorted by their \cstext/ values. The figure also shows the four classes of architecture for a few randomly selected synthetic planetary systems. To understand the characteristics of the different architectures, we study the distribution of planetary masses, radii, and semi-major axes as well as the multiplicity distributions. For planetary systems across all catalogues, this is shown in Fig. \ref{fig:architecturecharacteristics}. We describe the characteristics of different architectures in the following subsections. \edit{The discussion in the next subsection involves results derived from both observed and synthetic planetary systems. In addition, we present a gallery of mass-distance diagrams showing the four architecture classes in Appendix \ref{sec:gallery}.}

    \begin{figure}
        \resizebox{\hsize}{!}{\includegraphics[trim= 0.5cm 0.5cm 0.5cm 0.5cm,clip]{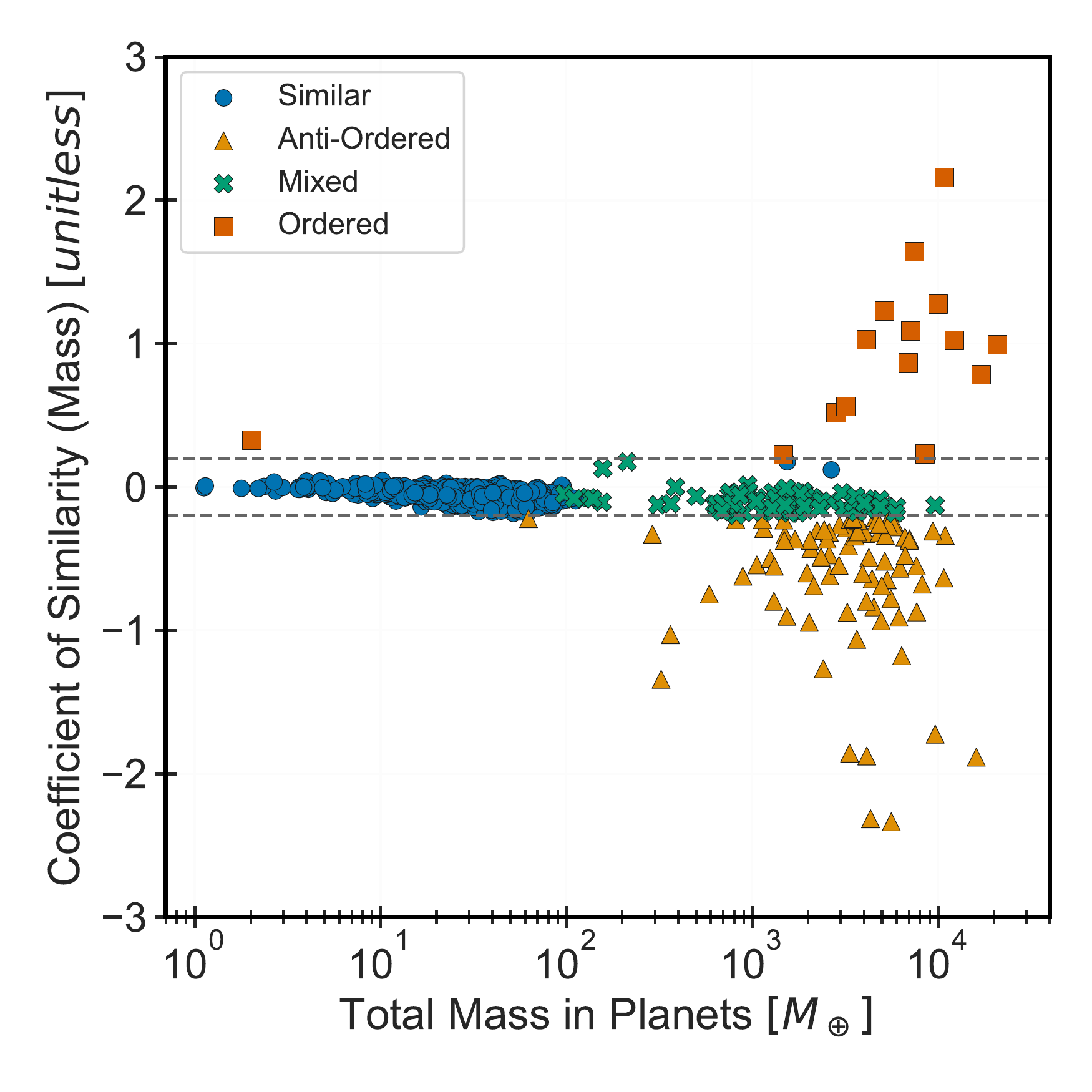}}
        \caption{Four classes of system architecture. The diagram shows the \cstext/ for a system as a function of the sum of mass of each planet in a system. Dashed horizontal lines correspond to $\cs = \pm 0.2$. This diagram emphasises the four classes of planetary system architecture, namely: \antiordered/, \similar/, \mixed/, and \ordered/. It also shows that the \cstext/ can not distinguish between \similar/ and \mixed/ architectures.}
        \label{fig:csvsmass}
        \end{figure}
        \change{Figure. \ref{fig:csvsmass} shows the \cstext/ of masses as a function of the total planetary mass in a system for all synthetic planetary systems from the Bern model. This figure shows several key aspects. Firstly, it illustrates the four architecture classes as separate clouds of scattered points strengthening the proposed four classes of planetary system architecture. Secondly, it shows that the architecture framework is scale-invariant, that is, the system architecture is sensitive only to the relative distribution of a quantity -- and not its absolute value. For example, while most similar system have $\lessapprox 100 \mearth$ mass in their planets (suggesting a lack of giant planets), there are some \similar/ systems with mass values of $\approx 2000 \mearth$  for their planets and host giant planets. Likewise, most \ordered/ systems host giant planets and have $\gtrapprox 2000 \mearth$ mass in their planets, there is an \ordered/ systems without any giant planets. Also, it illustrates that the \cstext/ partitions planetary systems into three groups: \antiordered/, \similar/ and \mixed/ in one group, and \ordered/. This demonstrates that the \cvtext/ is necessary to distinguish between the \similar/ and \mixed/ systems. Finally, the diagram shows that the architecture class of a system has strong links with the total mass of planets in the system. This hints that there must be general patterns in the formation pathways of systems of the same architecture. This topic is discussed in \papertwo/, from this series.}
        
        \subsection{Frequency of architecture}
        \label{subsec:frequency}
        
        \begin{figure}
        \resizebox{\hsize}{!}{\includegraphics[trim= 0.5cm 0.5cm 0.5cm 0.5cm,clip]{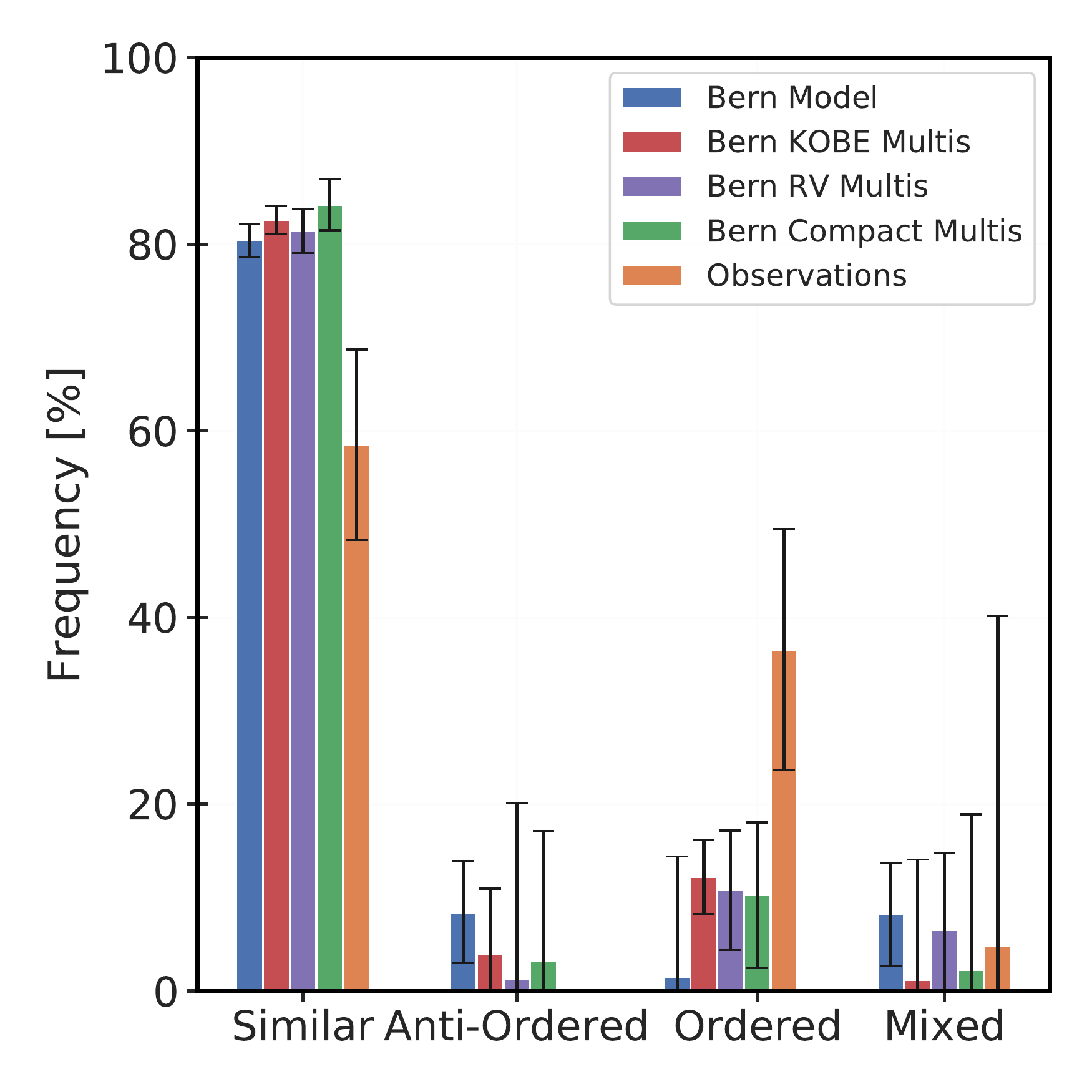}}
        \caption{Frequency diagram for the architecture classes. Currently, there are no known examples of observed planetary systems with \antiordered/ architecture. The length of error bars visualises the total number of systems in each bin as: ${100}/{\sqrt{\mathrm{bin\ counts}}}$.}
        \label{fig:archfreq}
        \end{figure}    
        
        The frequency of each architecture class across all catalogues is shown in Fig. \ref{fig:archfreq}. Similar systems are the most common architecture classes emerging from simulations, with a frequency of $\approx80.2\%$. About $\approx8\%$ of synthetic systems show \mixed/ and \antiordered/ architectures. Ordered architecture is a rare outcome in simulations ($\approx 1.5\%$). In observations, \similar/ class is the most common architecture ($\approx59\%$). Fifteen observed exoplanetary systems (out of forty-one) are part of the \ordered/ architecture class ($\approx37\%$). About $\approx5\%$ of observed planetary systems show \mixed/ architecture. There are no known examples of observed system with \antiordered/ architecture. 
        
        
        Comparing the frequency of architecture classes for observed systems with synthetically observed systems brings out some peculiar features. Firstly, theoretical catalogues seem to suggest that observations should find more \similar/ systems and fewer \ordered/ systems. The frequency of \similar/ (\ordered/) systems in our observed catalogue is significantly lower (higher). Secondly, while the frequency of \mixed/ systems seems to be in agreement with synthetic observations, this agreement is not statistically significant.  

        These discrepancies probably arise from the incompleteness prevalent in our observations catalogue. Transit surveys are conducted in a manner which allows the completeness and reliability of these survey to be estimated. The completeness of RV surveys, on the other hand, is very difficult to estimate. Further, the observation techniques used to find the exoplanets in our observations catalogue are heterogeneous, consisting of RV, transits, transit-timing variations, and direct imaging; this complicates the estimation of completeness. The PLATO mission is an upcoming space mission that is equipped to allow for statistical estimates of cosmic occurrence rates of planetary system architecture in our galaxy \citep{Rauer2014}. If more exoplanetary systems are uniformly detected and characterised, then it would be possible to estimate the occurrence rate of the different classes of system architecture. While such a result would constitute an important knowledge about our Universe, it could also become an excellent way of constraining our knowledge of initial conditions for planetary formation and the physical processes which shape the system architecture. The frequency of architecture class in simulations is a direct consequence of the initial conditions and the physical processes modelled in the Bern model.

        \def\figwidth{8.1cm}
        \def\figheight{20cm}
        \begin{figure*}
                \centering
                \includegraphics[width=8.6cm, height=\figheight]{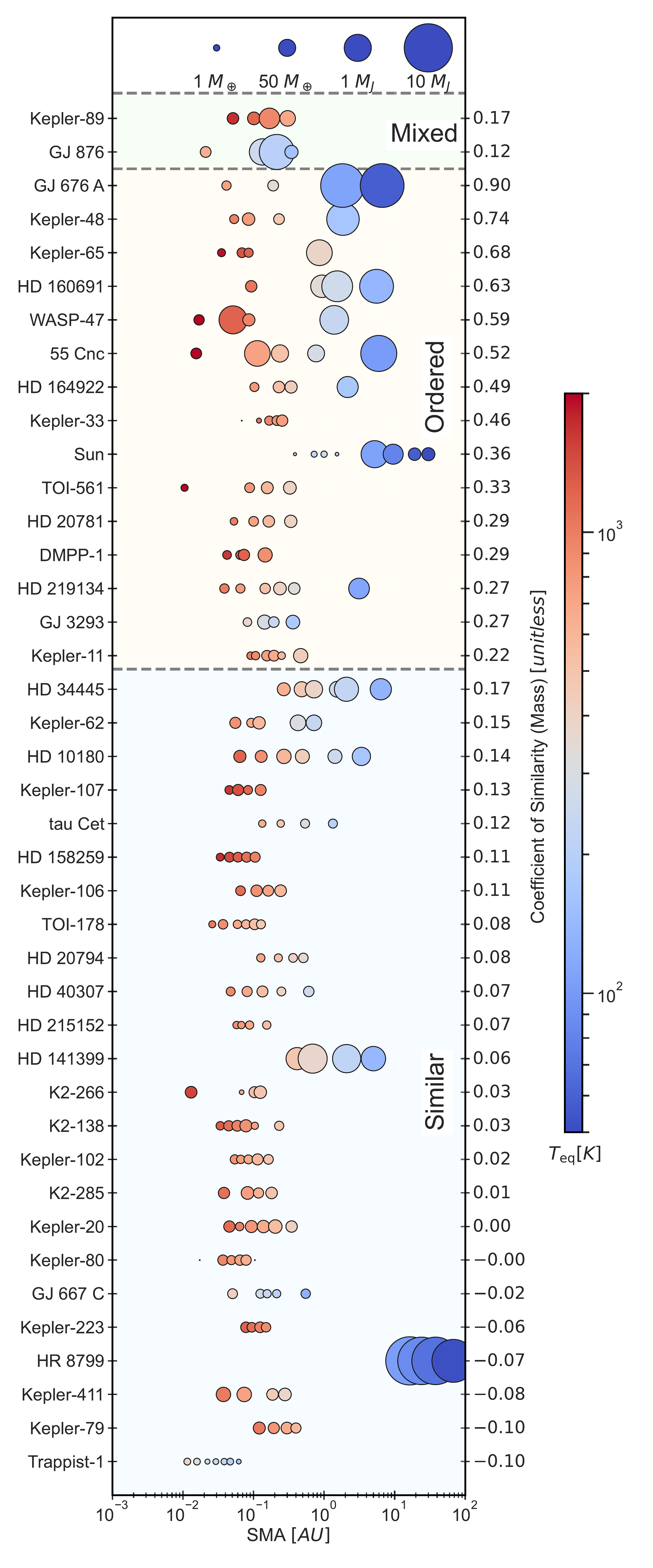}
                \includegraphics[width=\figwidth, height=\figheight]{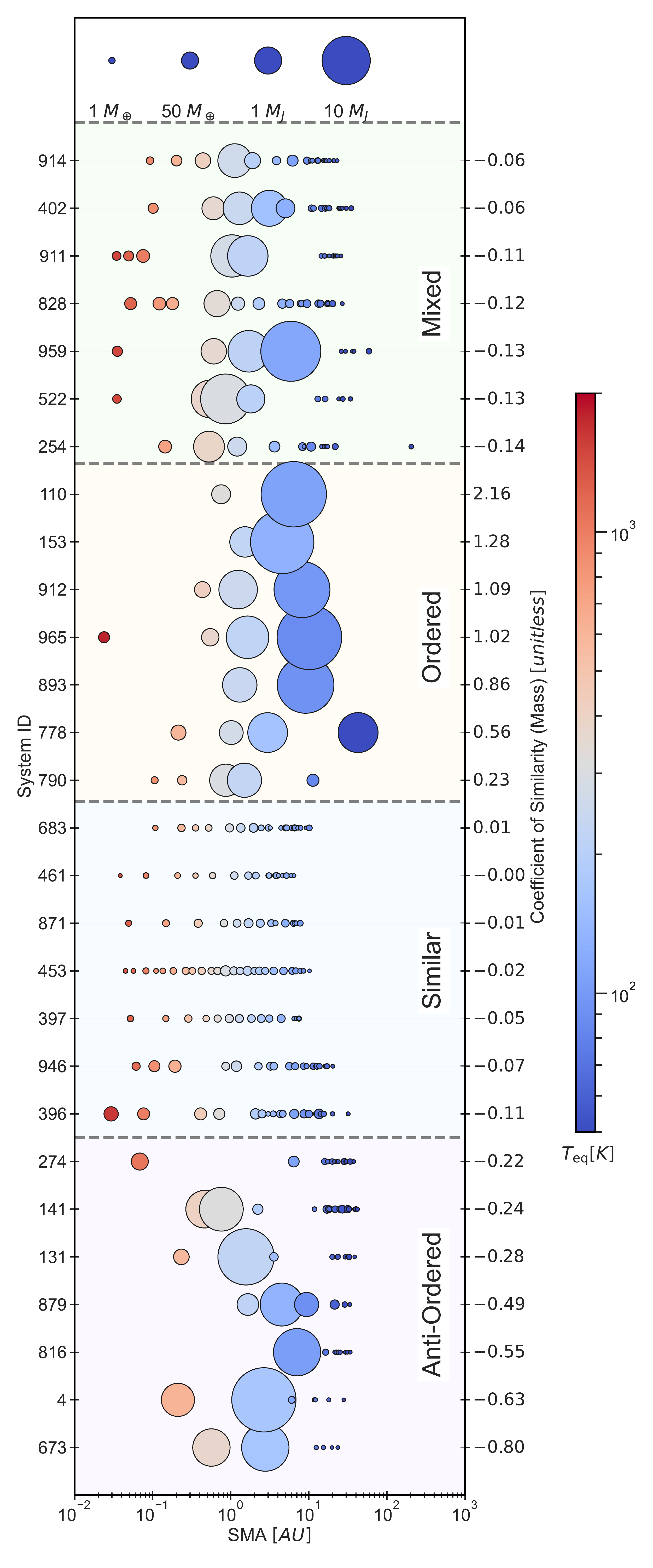}
                \caption{Architecture plot showing the architecture of observed (left) and randomly selected synthetic planetary systems (right). Each row is for one planetary system and the circles in that row represent planets. The area of the circle encodes planetary mass, and the colour shows the equilibrium temperature. The \cstext/ for each system is shown on the right y-axis. The x-axis shows the semi-major axis, \edit{which is different for the two panels.}}
                \label{fig:archplot}
        \end{figure*}
        
        \def\figwidth{4.5cm}
        \begin{figure*}
                \centering
                \includegraphics[width=\figwidth, trim=0.6cm 0.8cm 0.6cm 0.6cm,clip]{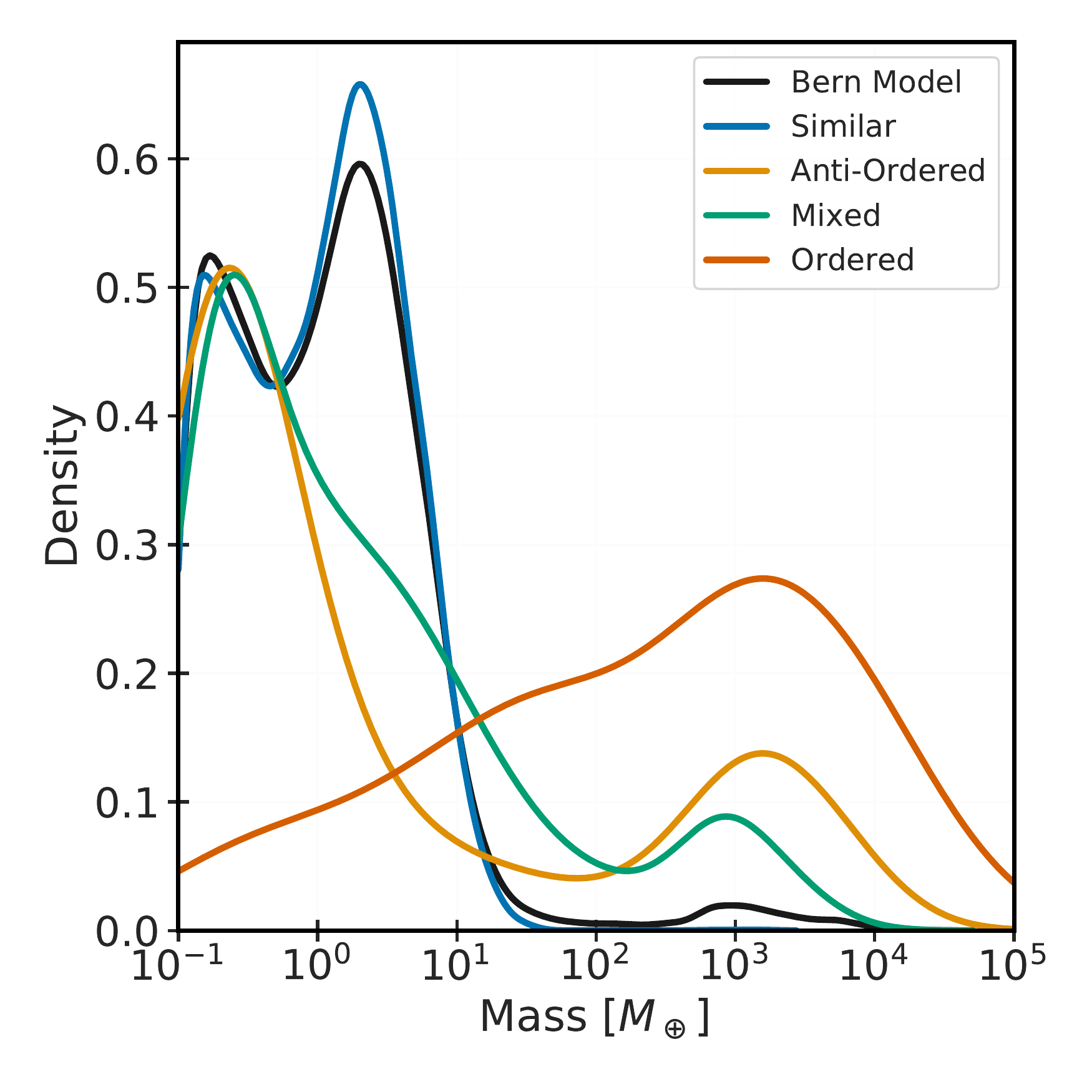} 
                \includegraphics[width=\figwidth, trim=0.6cm 0.8cm 0.6cm 0.6cm,clip]{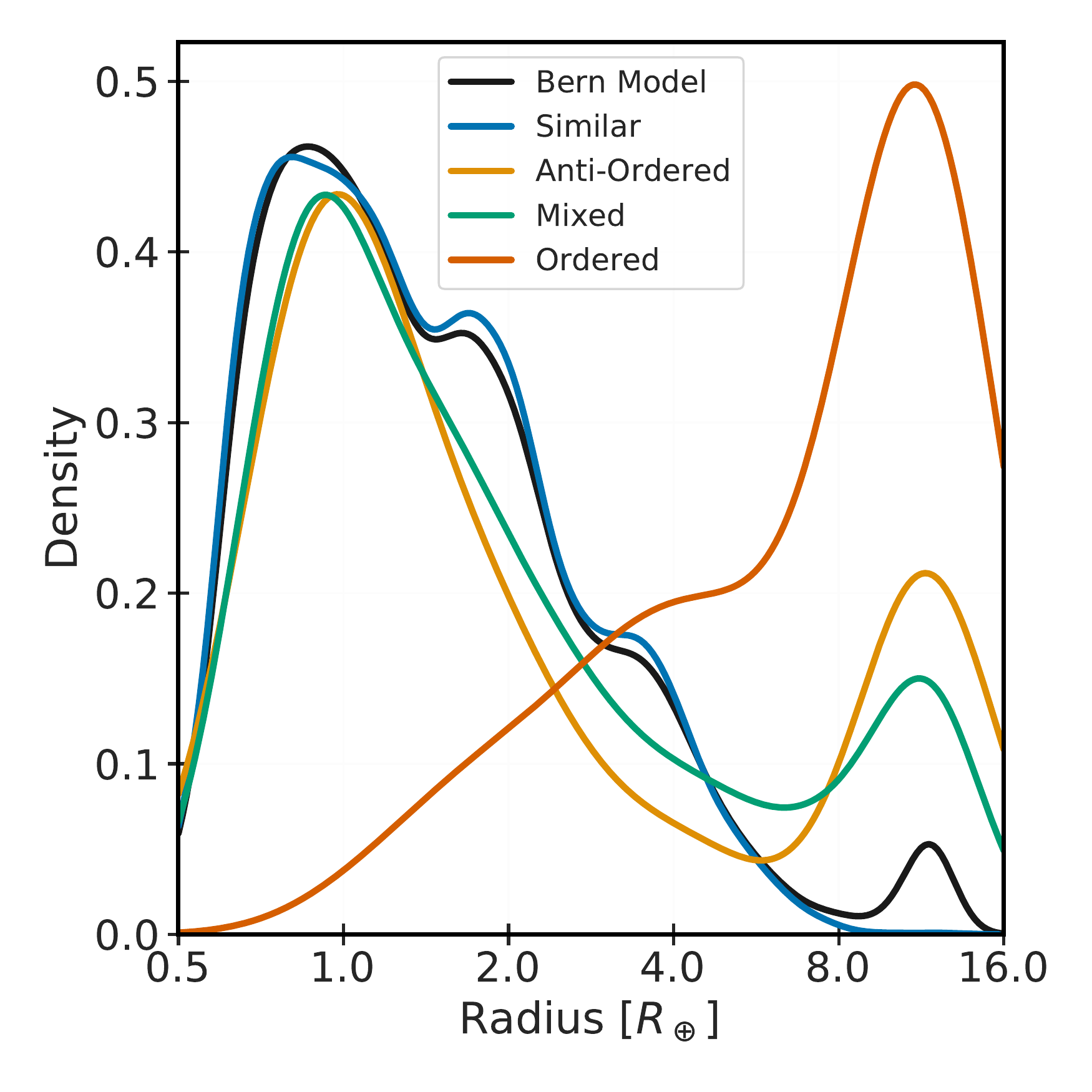}           
                \includegraphics[width=\figwidth, trim=0.6cm 0.8cm 0.6cm 0.6cm,clip]{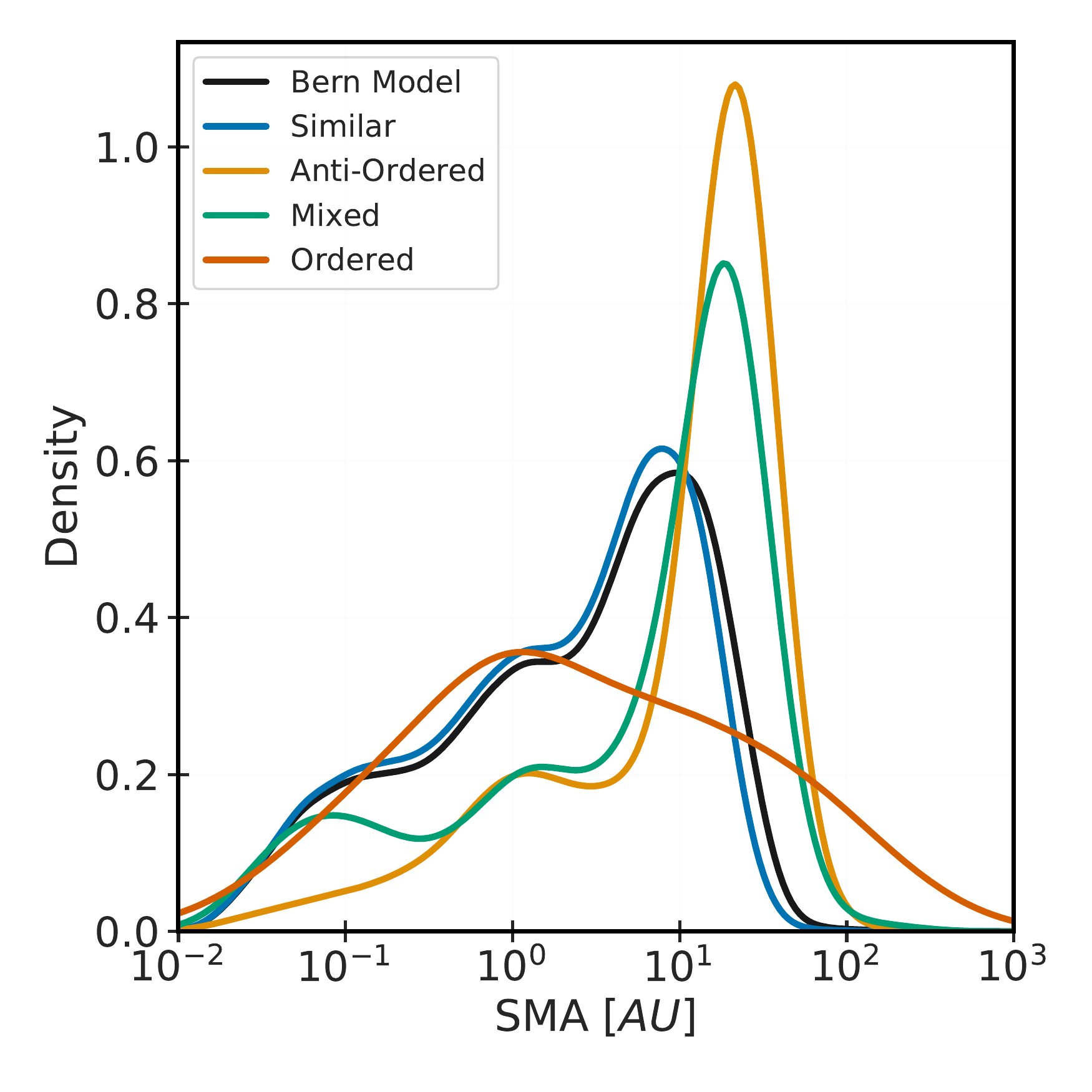}
                \includegraphics[width=\figwidth, trim=0.6cm 0.8cm 0.6cm 0.6cm,clip]{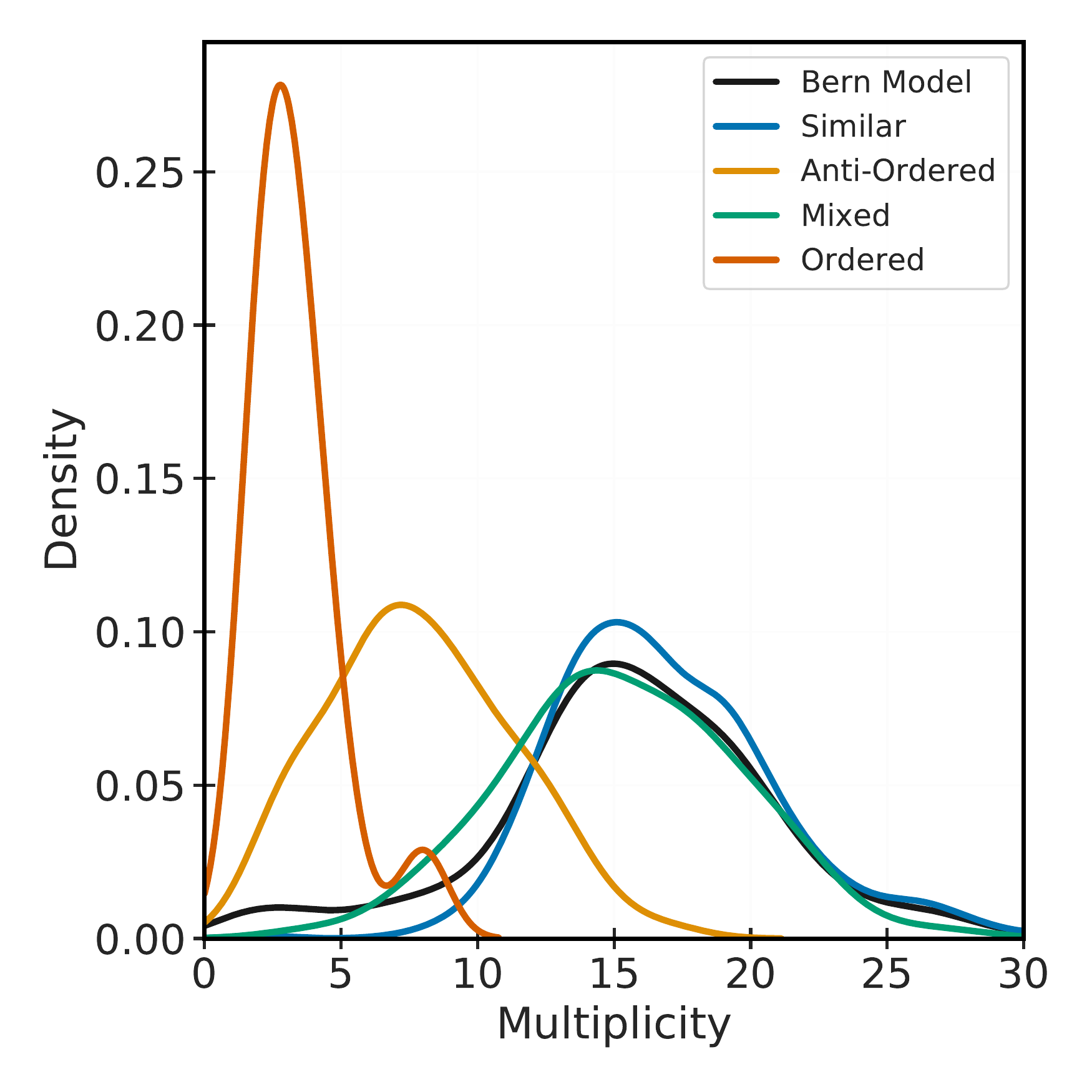}
                
                \includegraphics[width=\figwidth, trim=0.6cm 0.8cm 0.6cm 0.6cm,clip]{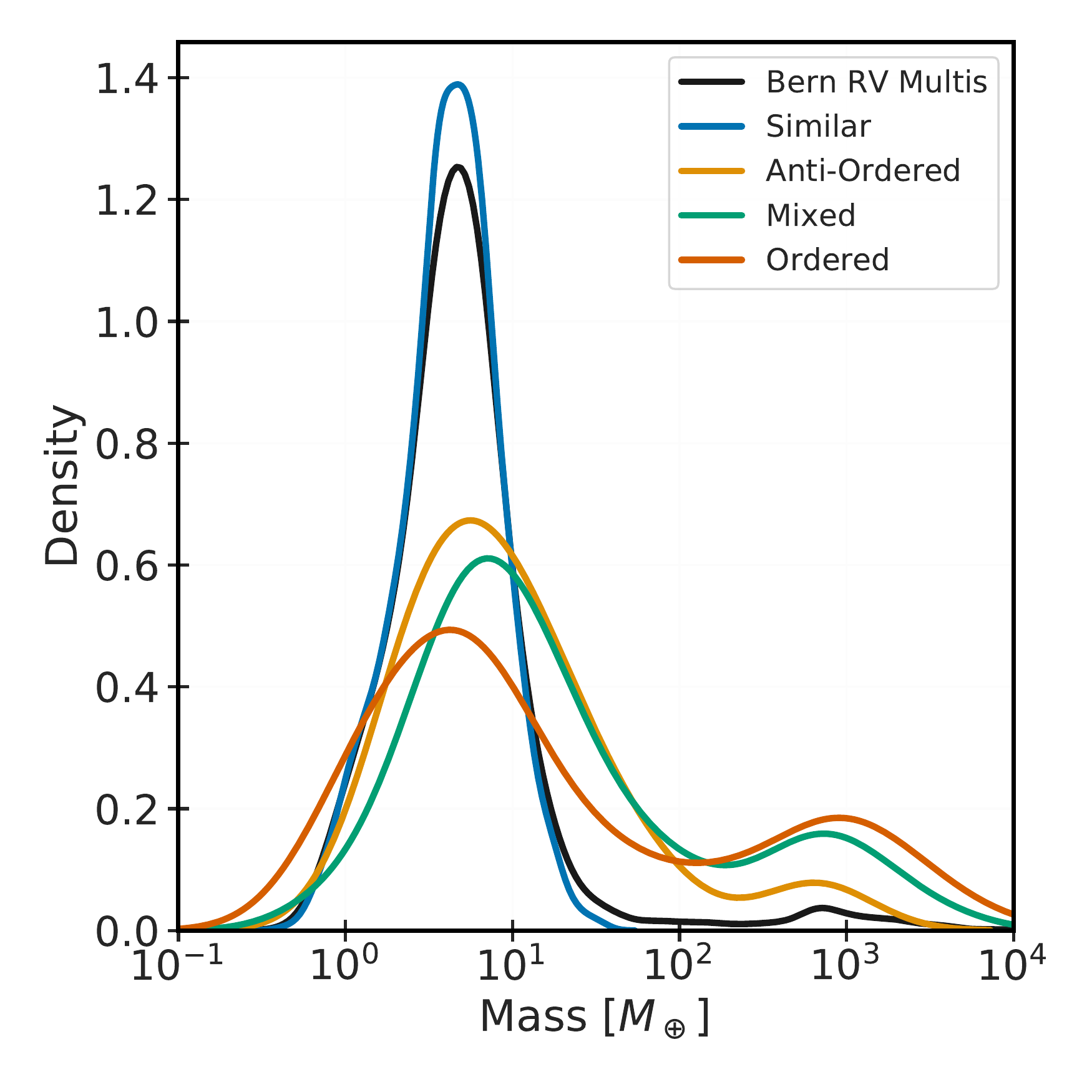}
                \includegraphics[width=\figwidth, trim=0.6cm 0.8cm 0.6cm 0.6cm,clip]{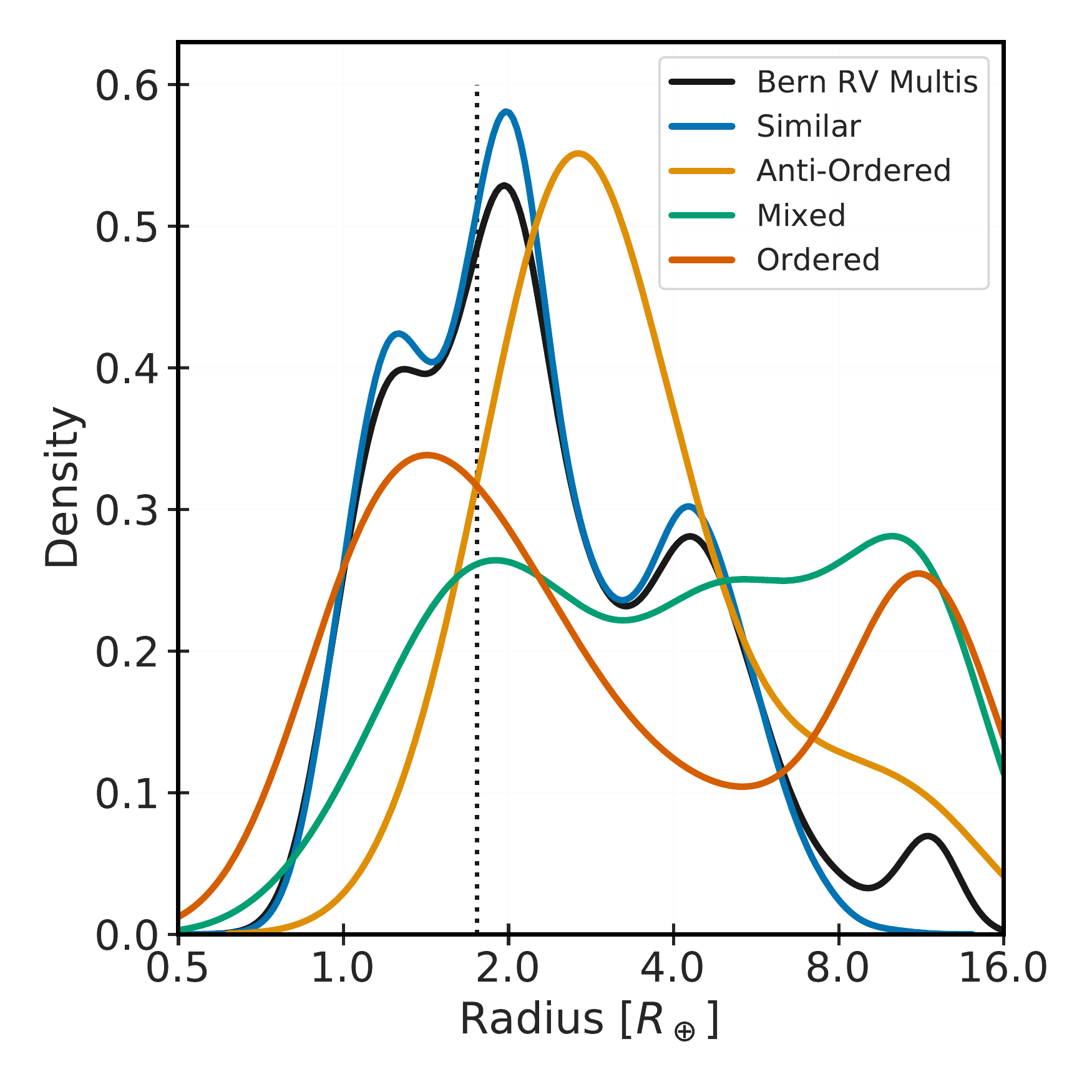}                
                \includegraphics[width=\figwidth, trim=0.6cm 0.8cm 0.6cm 0.6cm,clip]{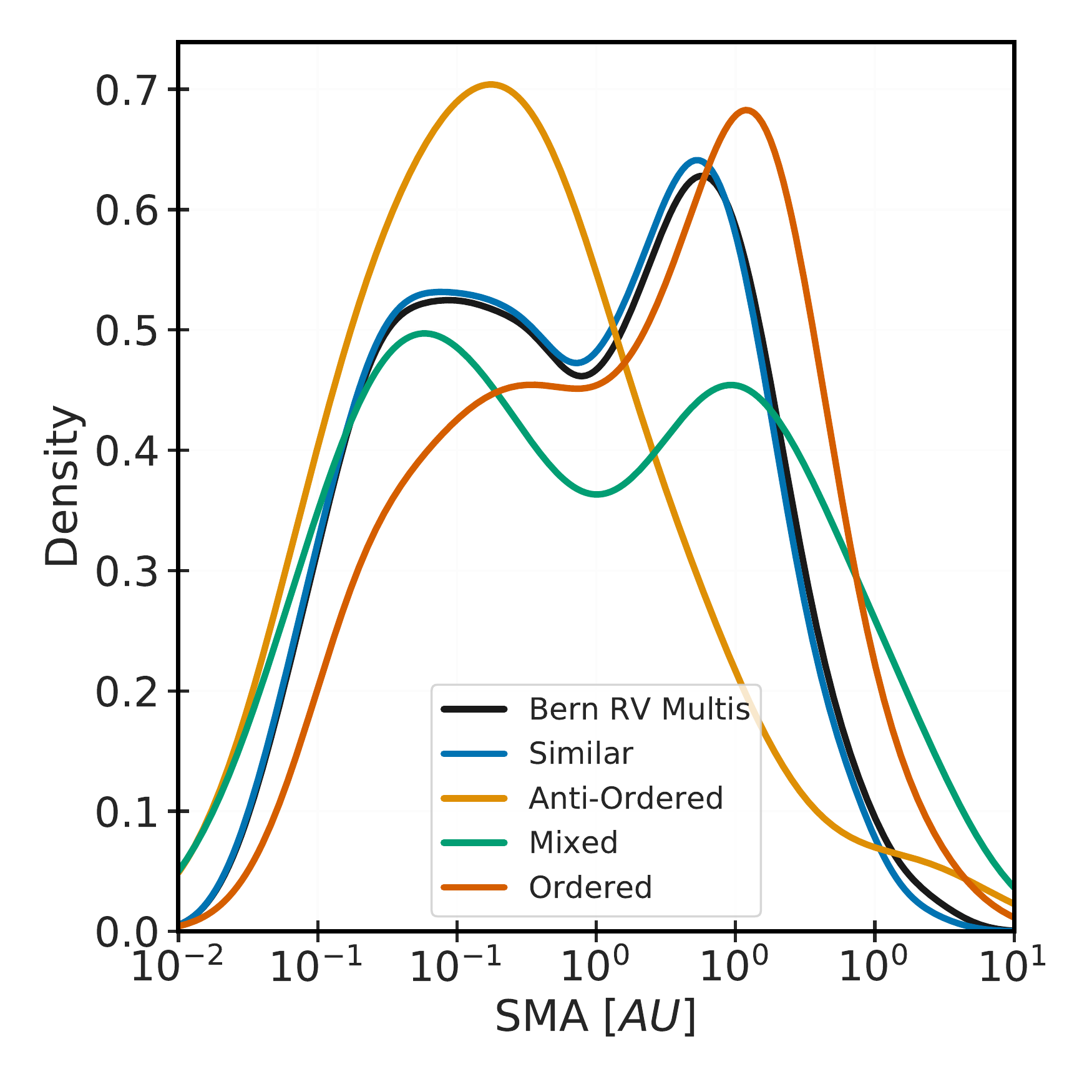}           
                \includegraphics[width=\figwidth, trim=0.6cm 0.8cm 0.6cm 0.6cm,clip]{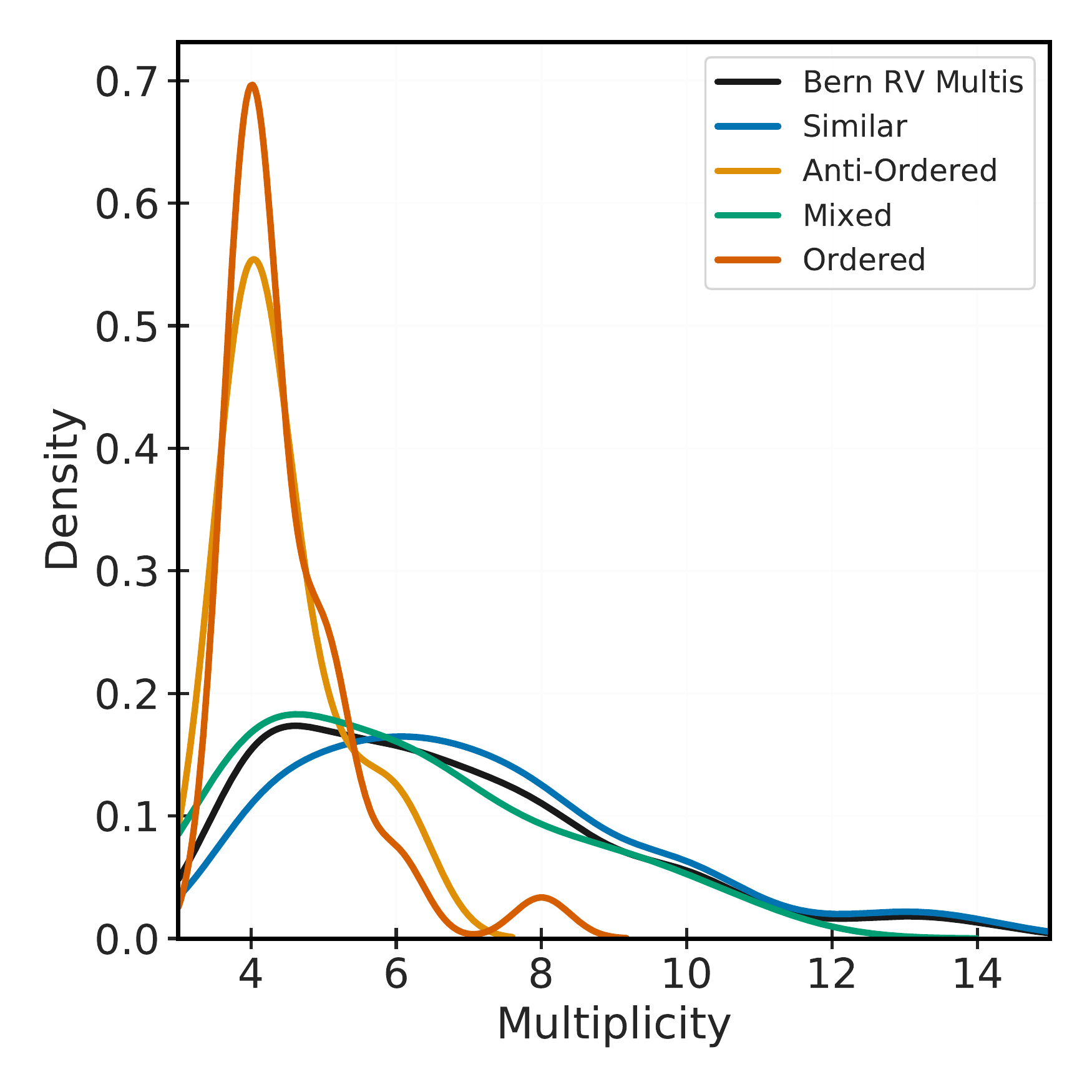}
                
                \includegraphics[width=\figwidth, trim=0.6cm 0.8cm 0.6cm 0.6cm,clip]{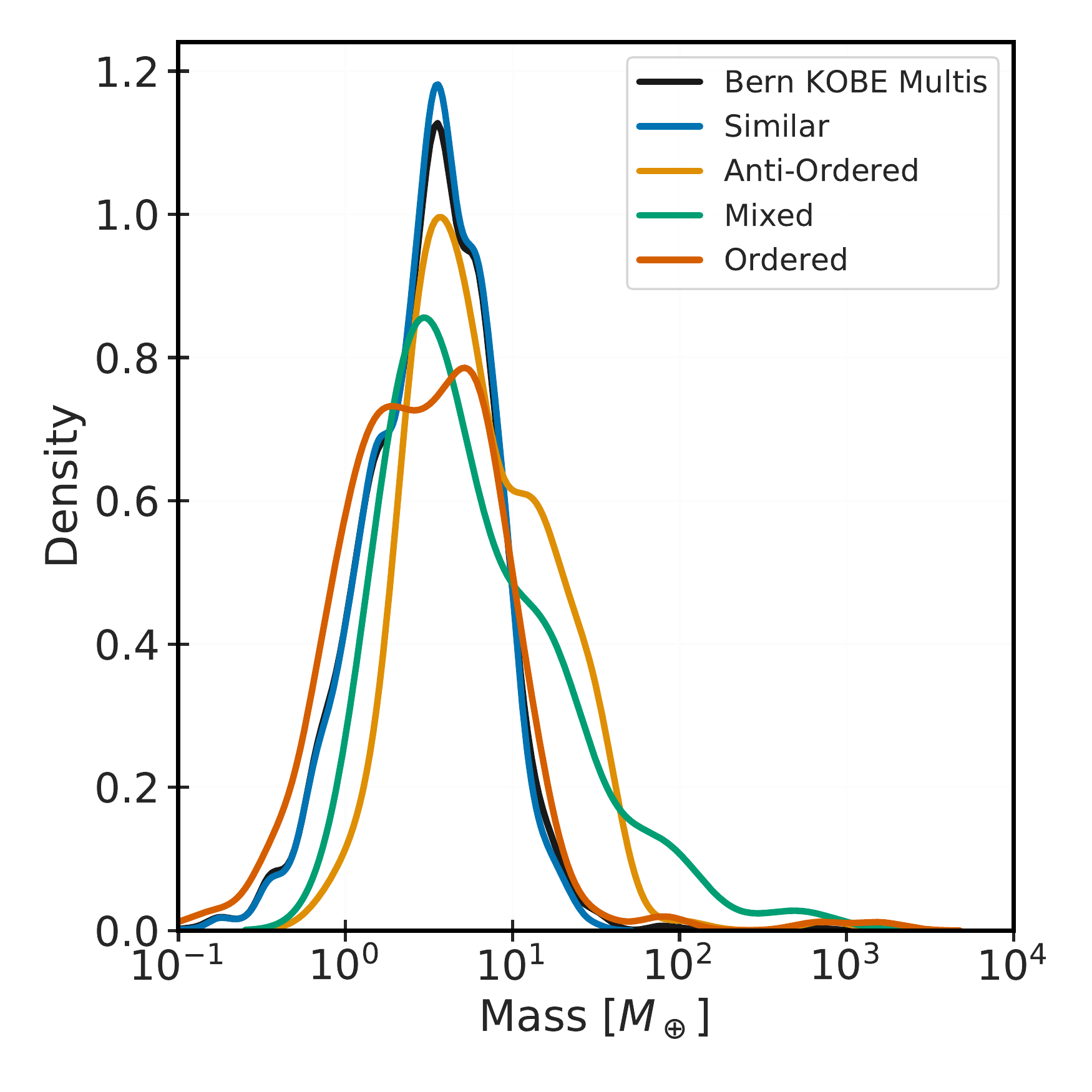}
                \includegraphics[width=\figwidth, trim=0.6cm 0.8cm 0.6cm 0.6cm,clip]{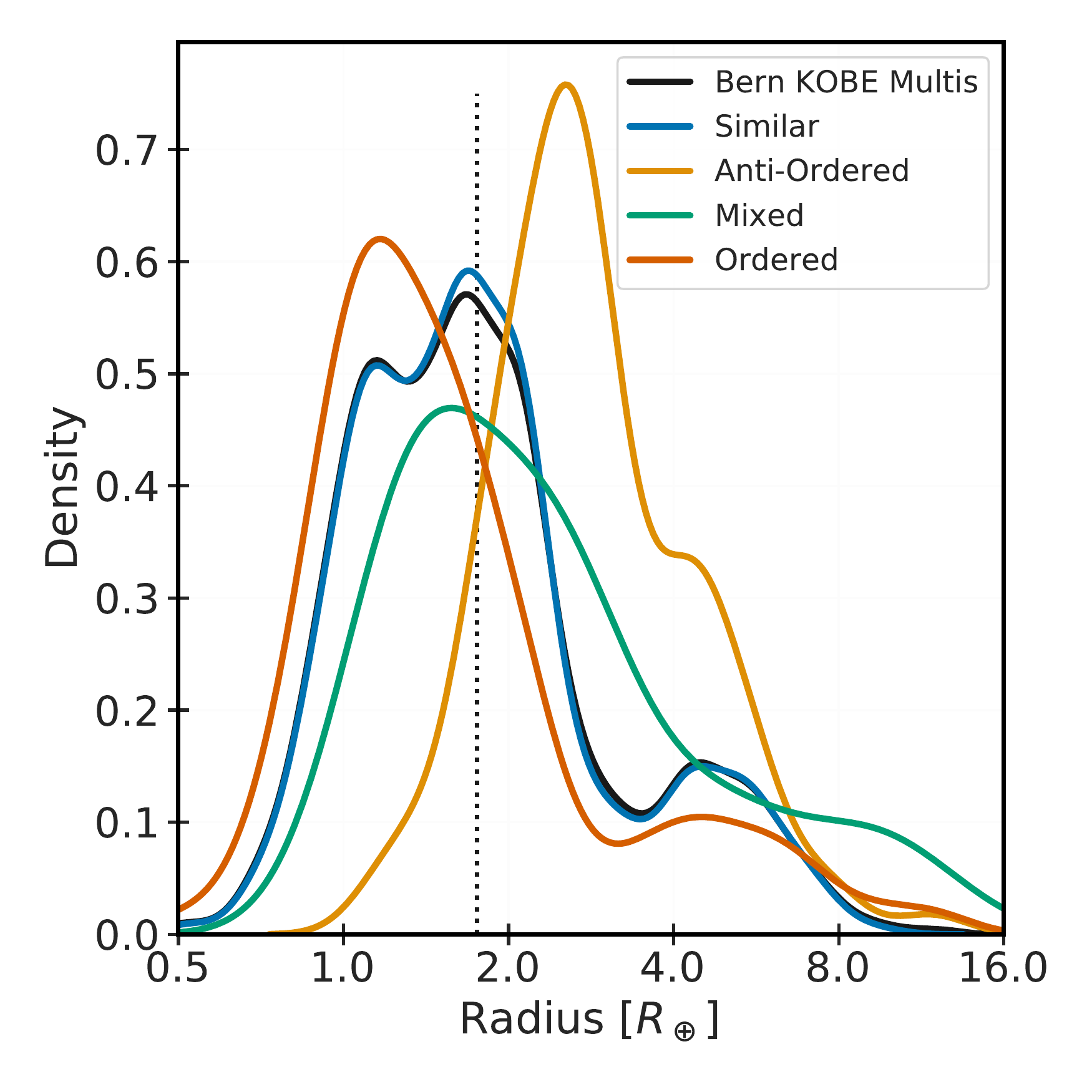}              
                \includegraphics[width=\figwidth, trim=0.6cm 0.8cm 0.6cm 0.6cm,clip]{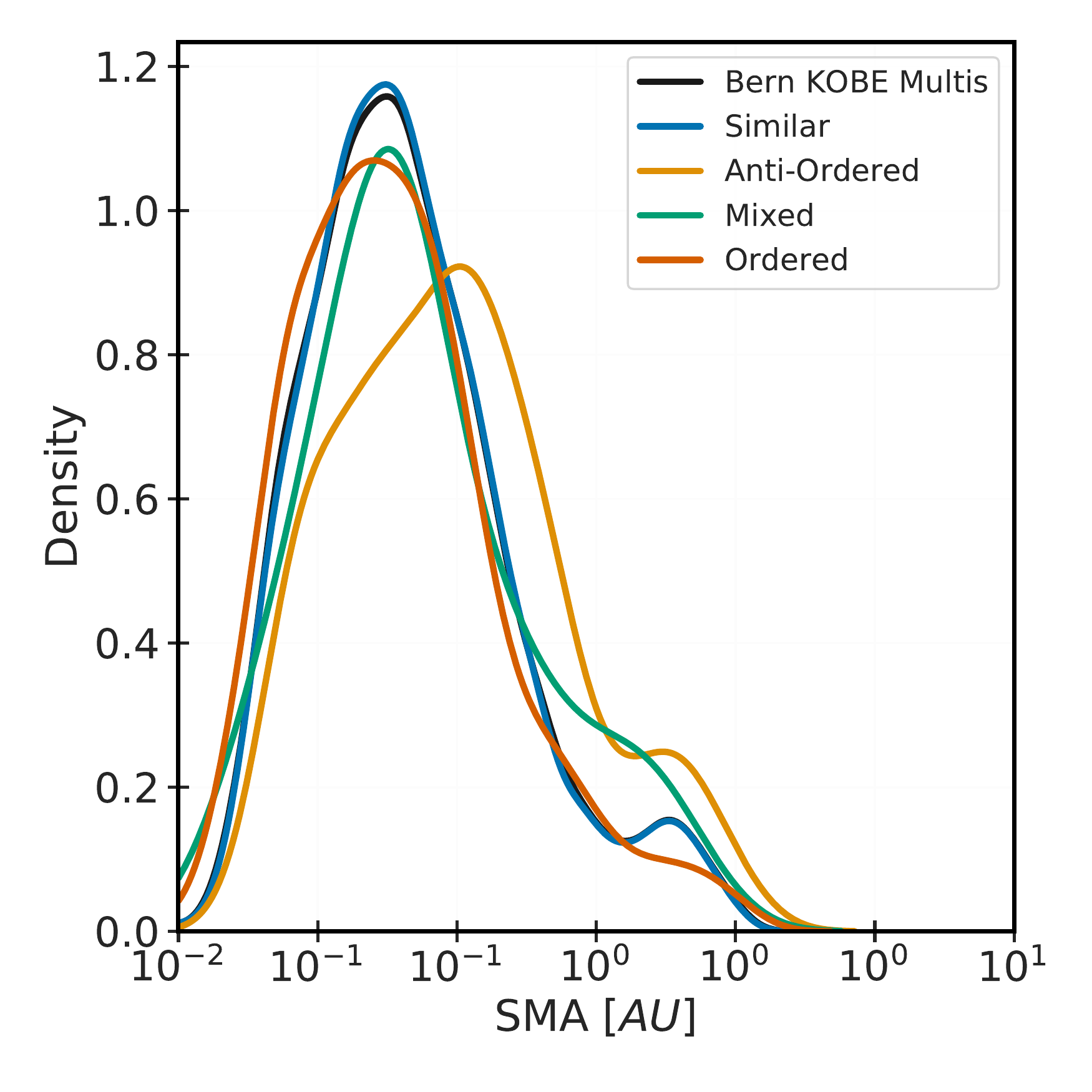}         
                \includegraphics[width=\figwidth, trim=0.6cm 0.8cm 0.6cm 0.6cm,clip]{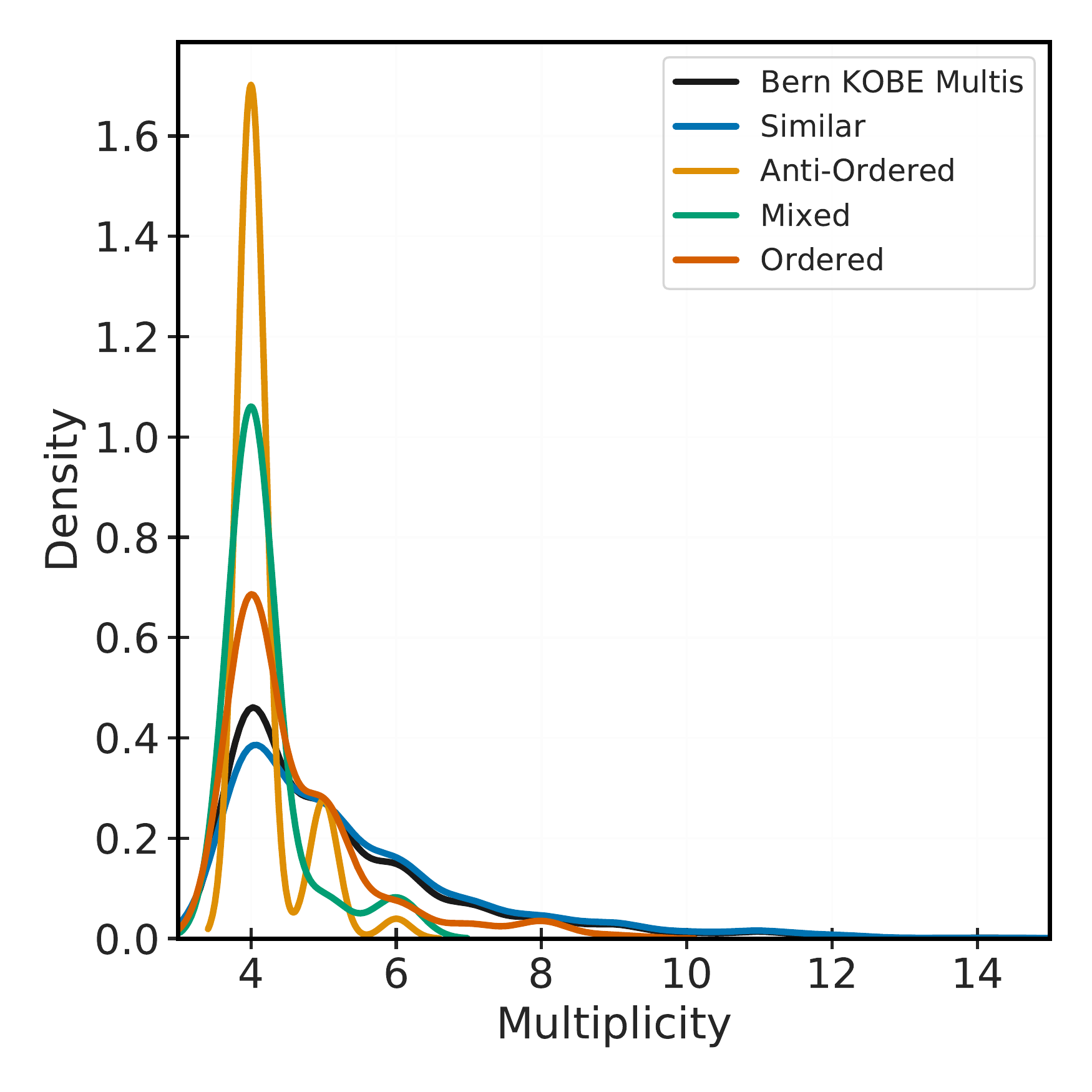}
                
                \includegraphics[width=\figwidth, trim=0.6cm 0.8cm 0.6cm 0.6cm,clip]{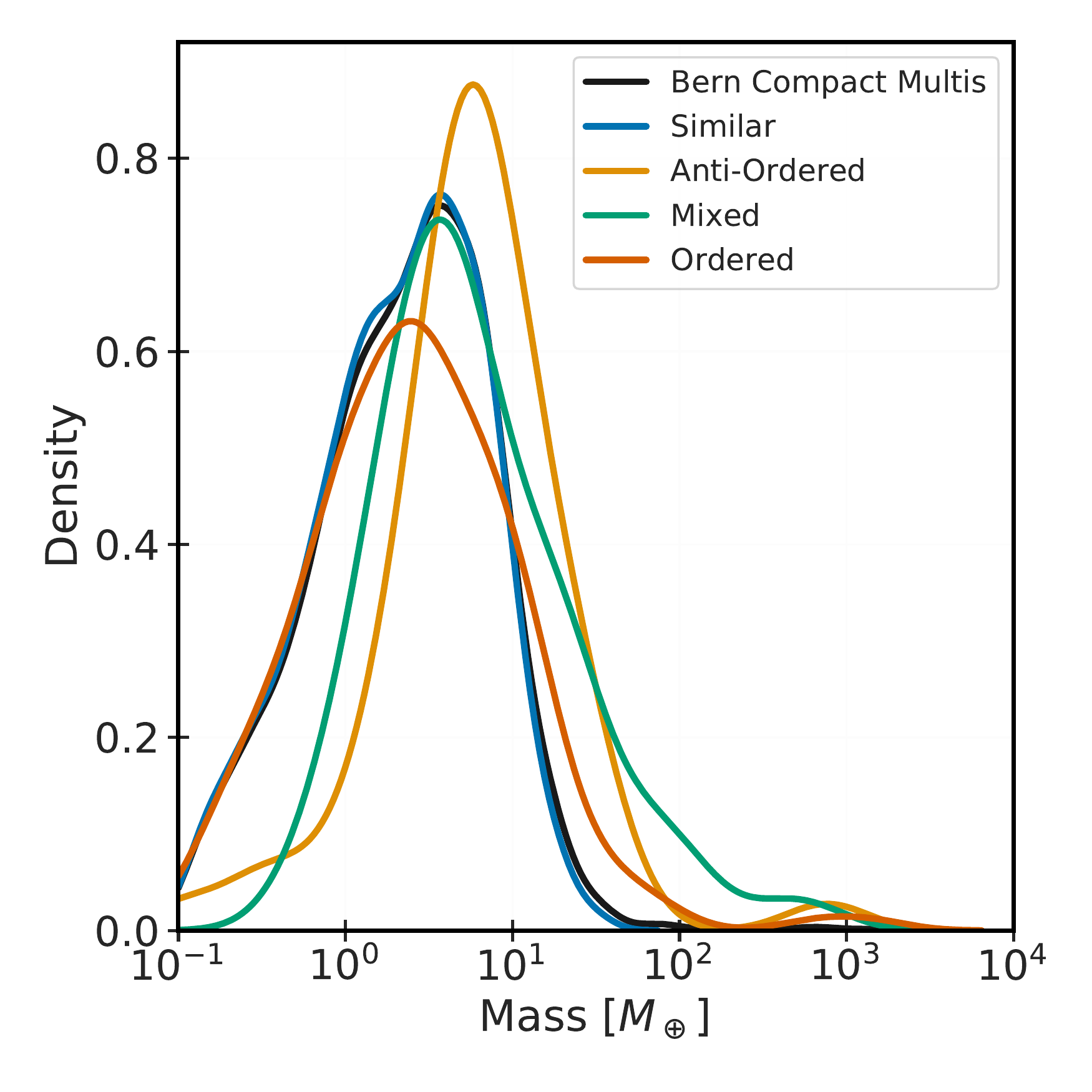}
                \includegraphics[width=\figwidth, trim=0.6cm 0.8cm 0.6cm 0.6cm,clip]{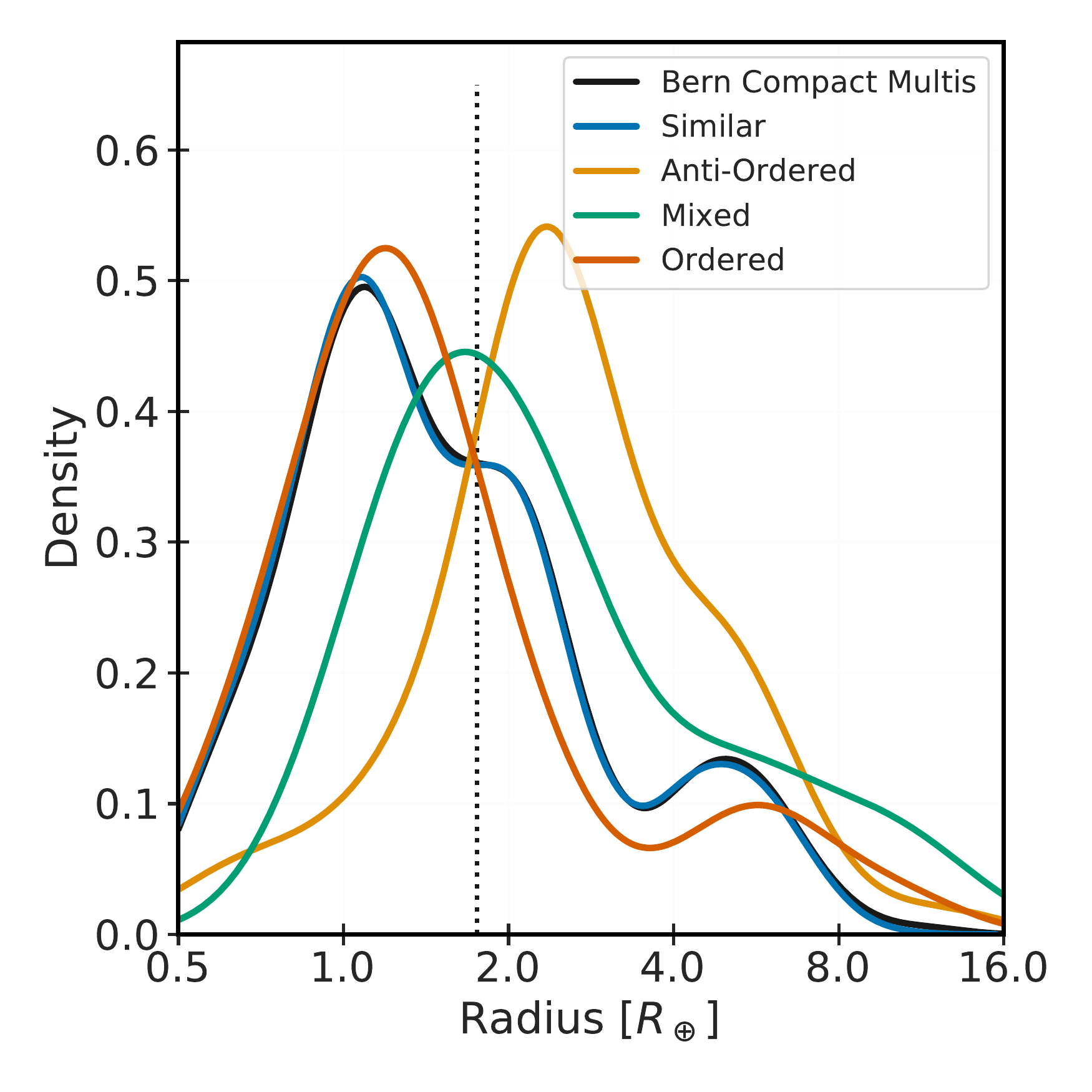}
                \includegraphics[width=\figwidth, trim=0.6cm 0.8cm 0.6cm 0.6cm,clip]{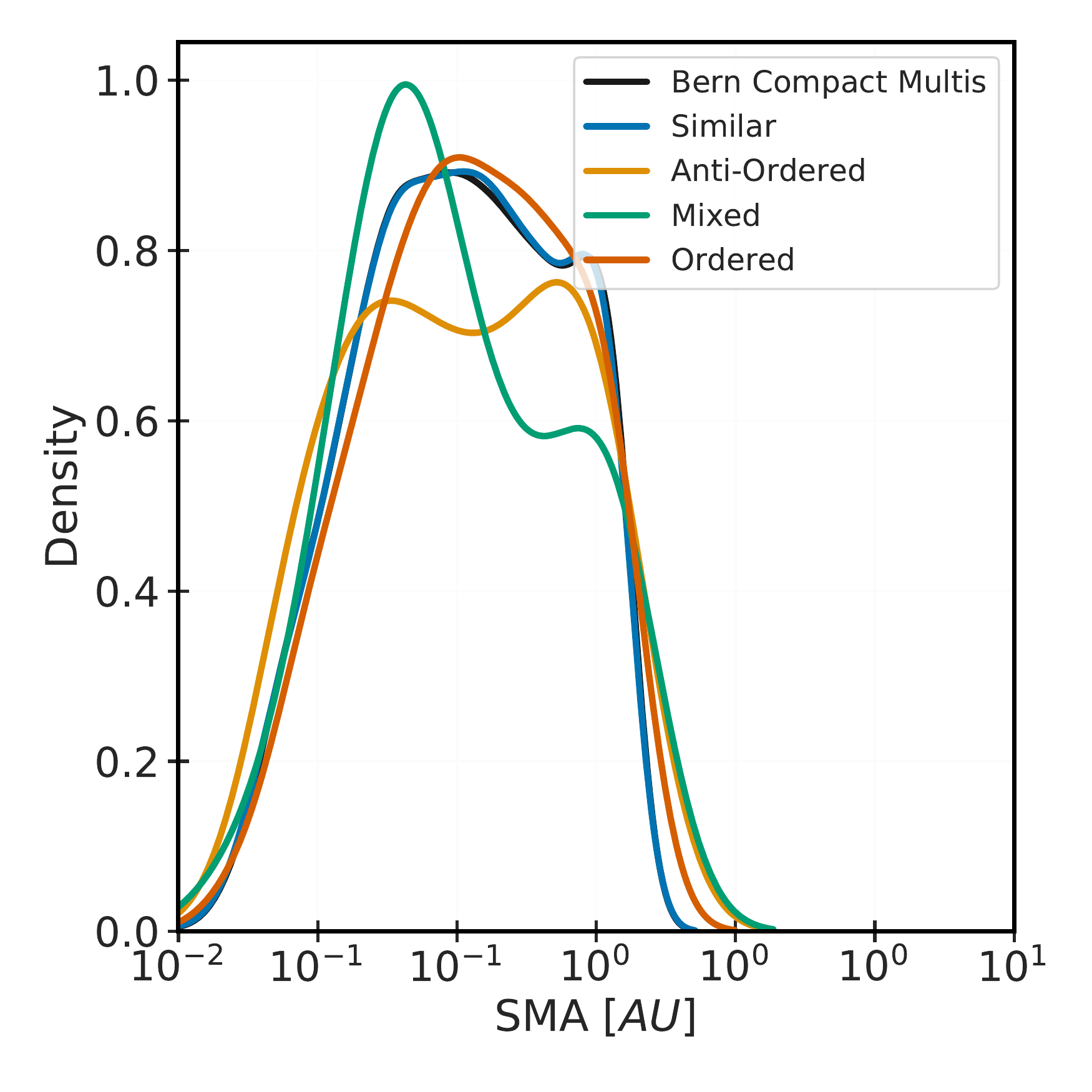}              
                \includegraphics[width=\figwidth, trim=0.6cm 0.8cm 0.6cm 0.6cm,clip]{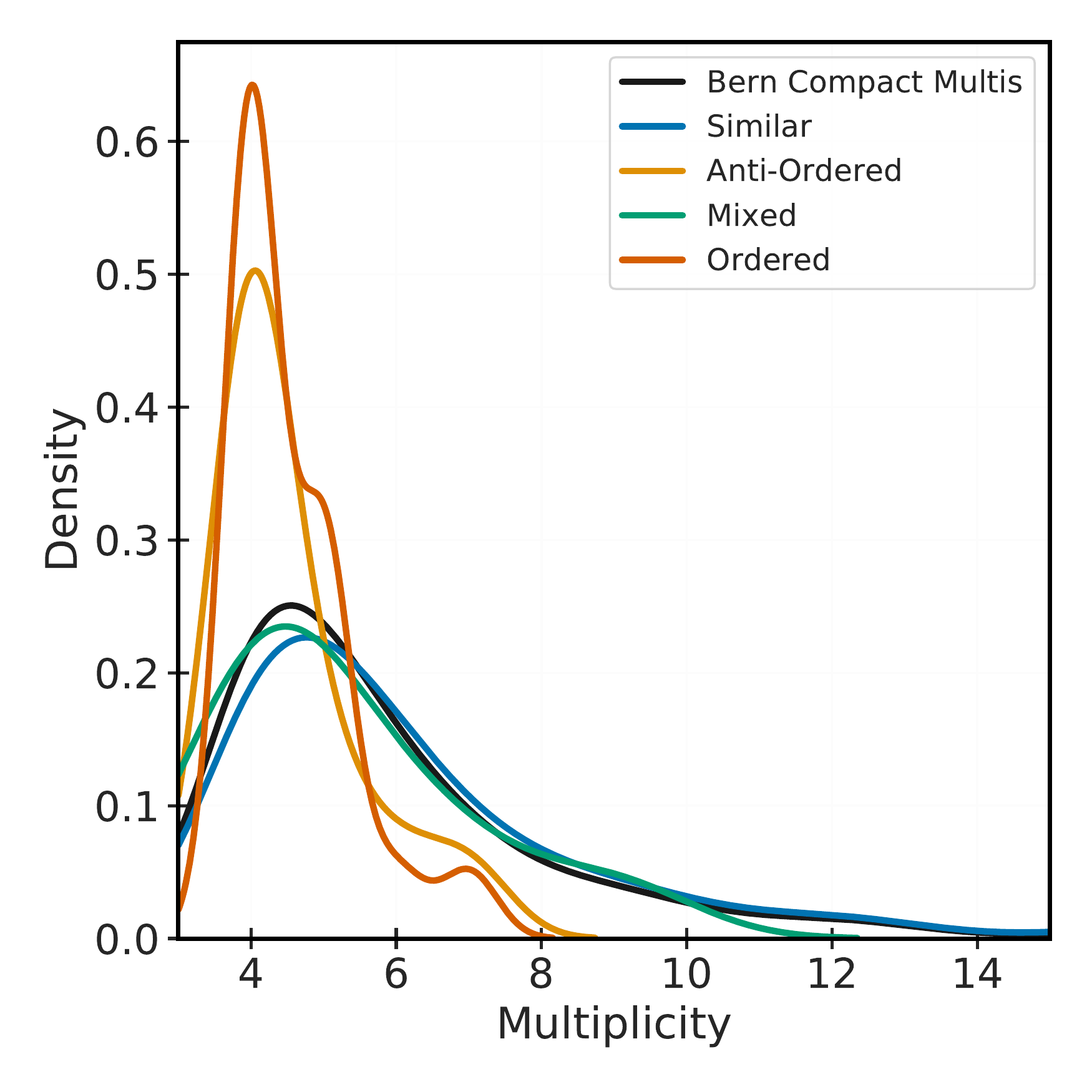}
                                
                \includegraphics[width=\figwidth, trim=0.6cm 0.8cm 0.6cm 0.6cm,clip]{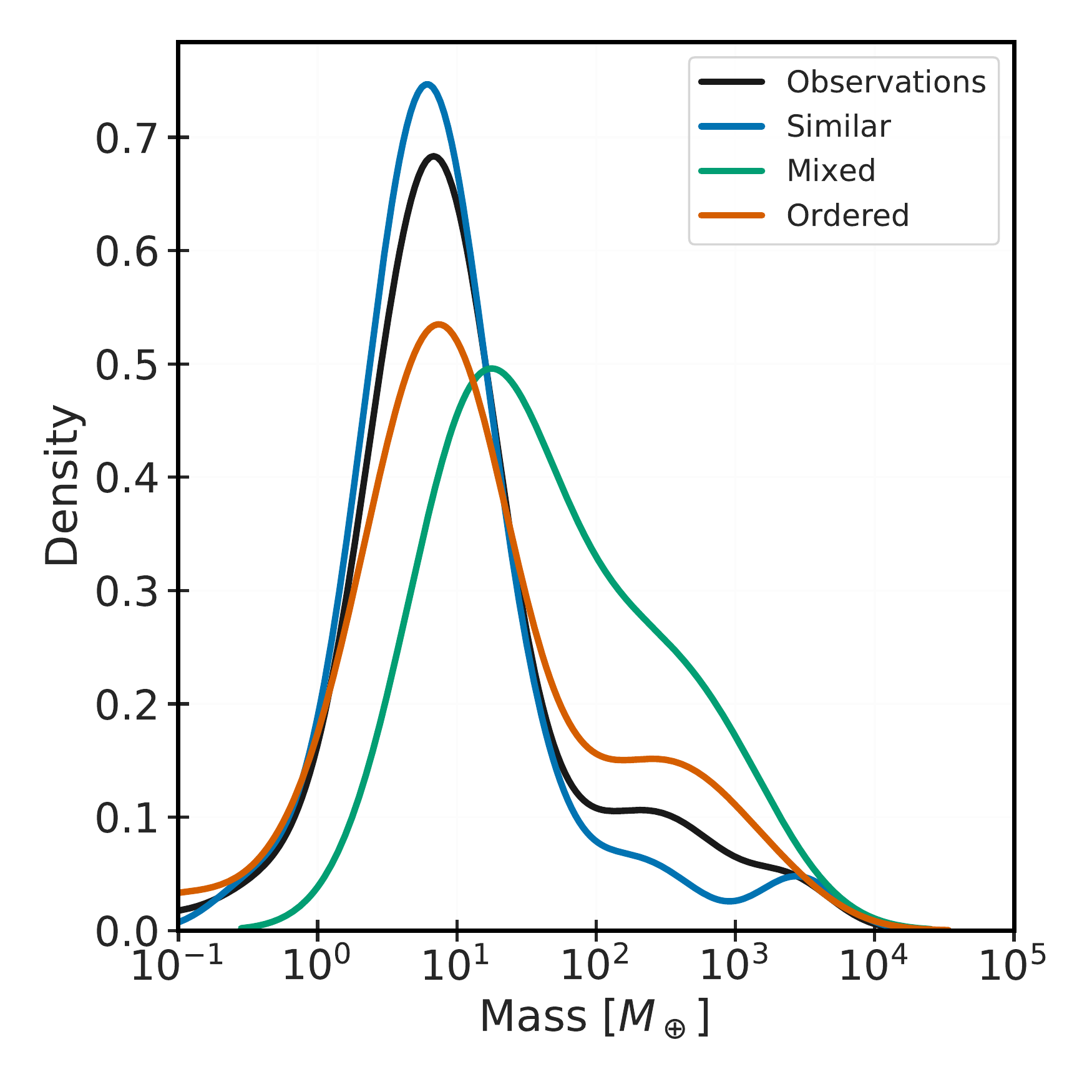}
                \includegraphics[width=\figwidth, trim=0.6cm 0.8cm 0.6cm 0.6cm,clip]{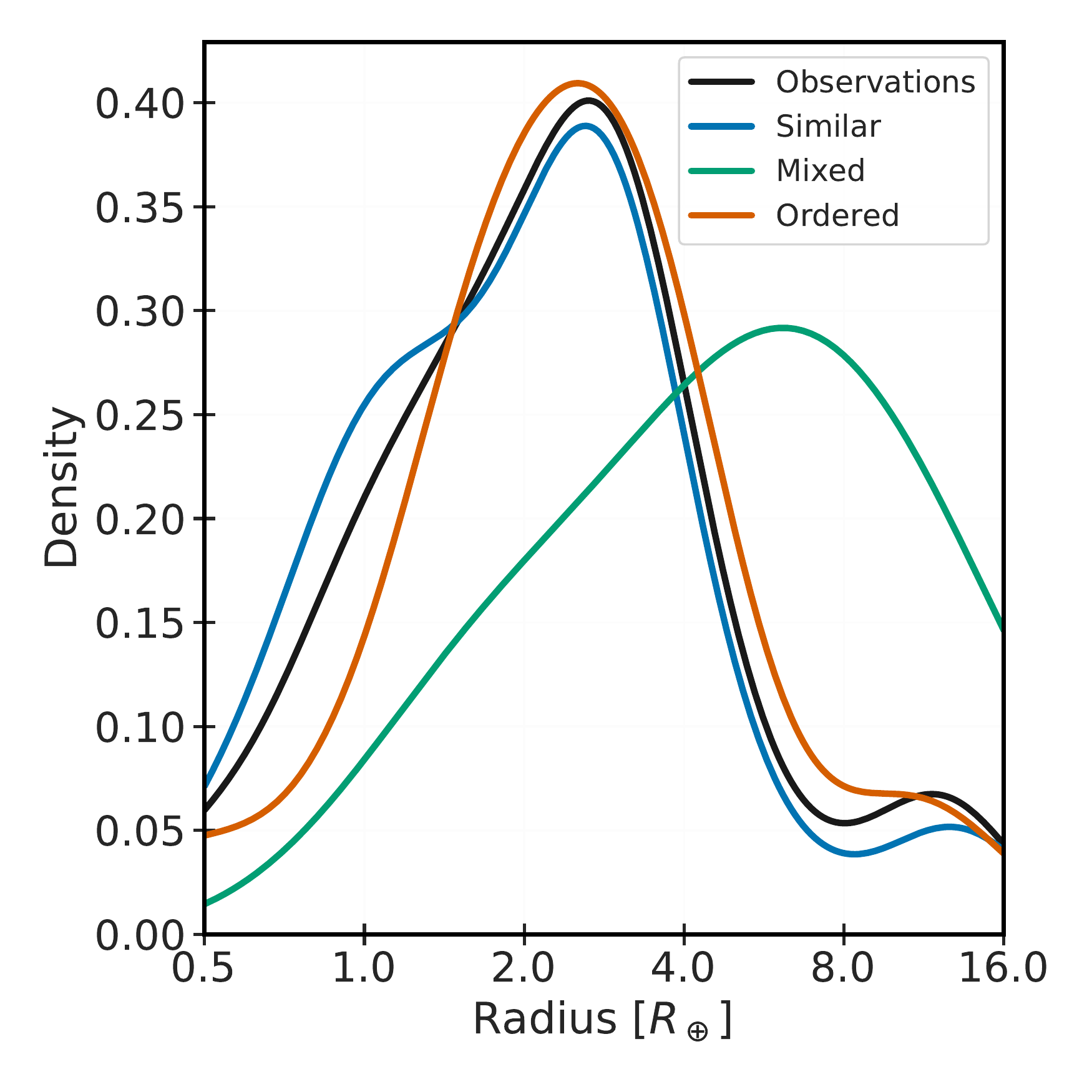}            
                \includegraphics[width=\figwidth, trim=0.6cm 0.8cm 0.6cm 0.6cm,clip]{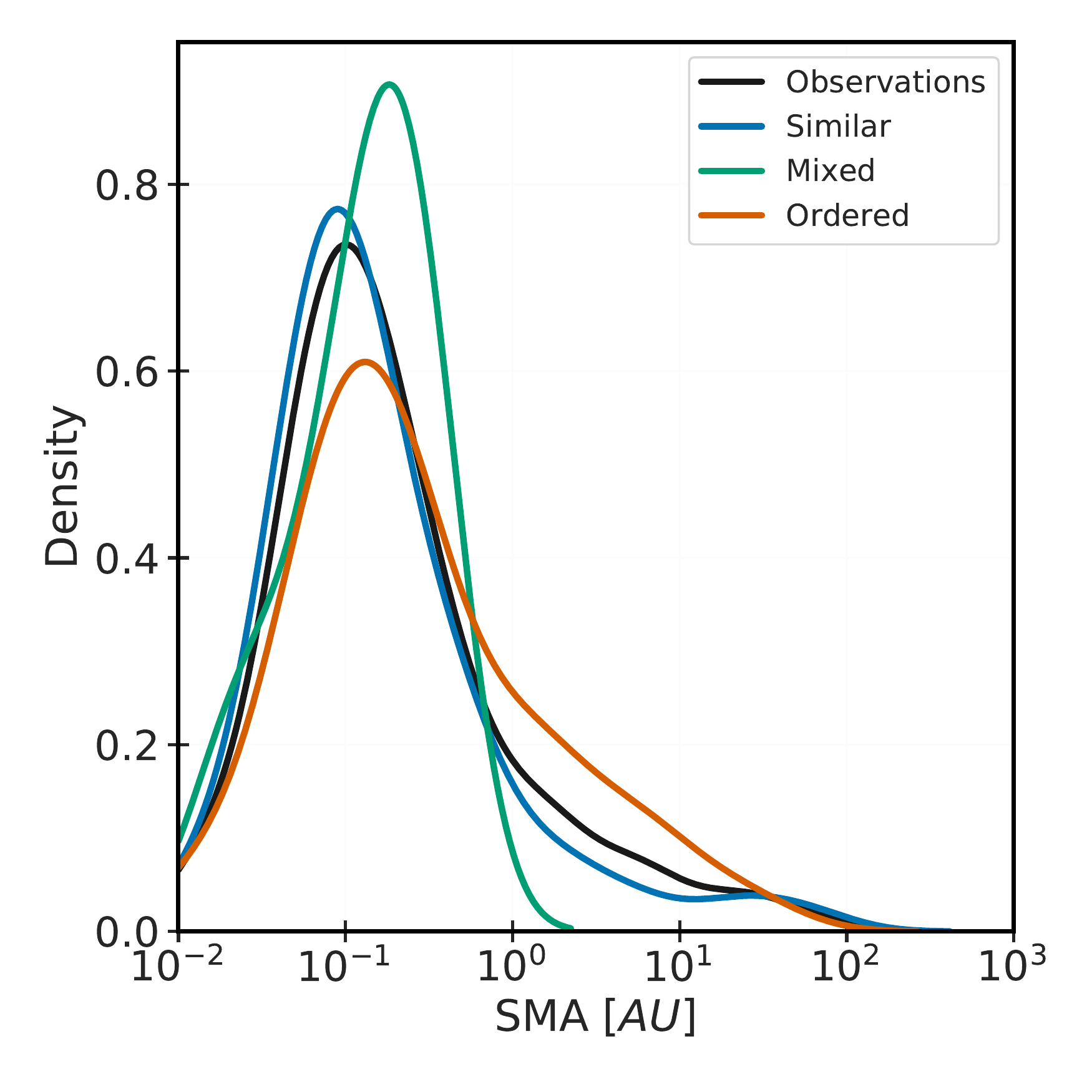}               
                \includegraphics[width=\figwidth, trim=0.6cm 0.8cm 0.6cm 0.6cm,clip]{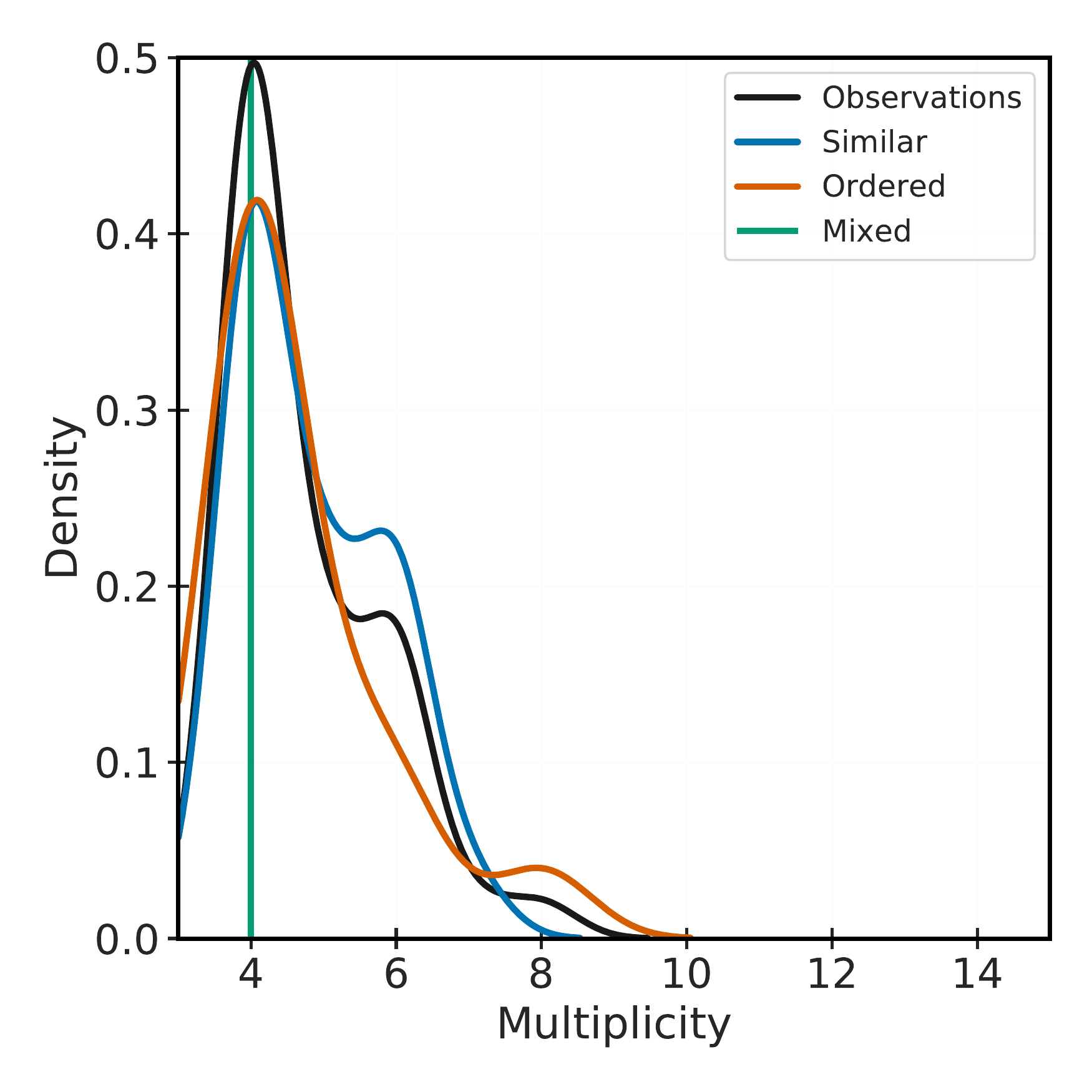}
                \caption{Characteristics of the architecture classes. \edit{These plots show the distribution of various quantities (columns) as function of different catalogues (rows). Left to right: Distributions of mass, radius, distance, and multiplicity in the following catalogues (top to bottom): Bern model, Bern RV Multis, Bern KOBE Multis, Bern Compact Multis, and observations. All catalogues are described in Sect. \ref{sec:catalogues}. Some notable features from these plots are discussed in Sect. \ref{sec:architecturetypes}}. All individual distributions are normalised such that the area under each curve sums to unity.  The dotted vertical line in the radius distributions marks $1.75 \rearth$ -- approximately, the location of the well-known gap in the radius distribution \citep{Fulton2017}. Since there are only two \mixed/ systems with the same multiplicity ($\nplanet = 4$) in our observations catalogue, a vertical line replaces the density kernel.
                \edit{The \change{Gaussian} density kernels in all other cases were estimated using Scott's rule \citep{2015mdet.book.....S}.}}
                \label{fig:architecturecharacteristics}
            \end{figure*}
        
        \subsection{Architecture class: \similar/}
        
        Planetary systems have a \similar/ architecture when all planets in the system  have masses that are approximately similar to each other. These planetary systems are the archetypical examples of the peas in a pod trend. There are several well-known planetary systems exhibiting \similar/ architecture, such as Trappist-1 \citep{2021PSJ.....2....1A}, TOI-178 \citep{2021A&A...649A..26L},  Kepler-20 \citep{2016AJ....152..160B}, and so on. This architecture is the most common outcome of planetary formation and is also the most frequent architecture class in our observed catalogue.

Similar systems in the Bern model are composed of several low-mass planets. They tend to have limited diversity in planetary masses when compared with the observed systems. The mass distribution, for \similar/ systems in the Bern model, shows that there are many low-mass ($< 1 \mearth$) planets in these systems. \edit{This peak is missing in observations as well as synthetic observations as low mass exoplanets are difficult to observe.} This could, however, be remedied in future as current radial velocity spectrographs reach the $\approx 20$ cm/s precision necessary for discovering exoplanets in the super-Earths and Earths mass range \citep{Lillo-Box2021, Netto2021}. The radius distribution of \similar/ systems implies that these systems are prominently composed of rocky planets, super-Earths and sub-Neptunes\footnote{Throughout this paper, we use planetary classes (e.g. rocky, super-Earths, etc.) from the radius based classification of \cite{Kopparapu2018}}. 
        
The Bern RV Multis show a bimodal planetary distance distribution for \similar/ systems (as well as for \mixed/ and \ordered/). The approximate location of the gap is $0.28\ \si{au}$ or $55\ \si{d}$ (for a solar mass star). This bi-modality is not visible in our observed catalogue. Planets in \similar/ and \mixed/ systems in the Bern Model also show a dip around this location. In the Bern Model, inwardly migrating giant planets ($\gtrsim 100\ \mearth$) tend to stop around $0.4\ \si{au}$ or $100\ \si{d}$. Inside this region, low-mass planets are populous. We attribute this bi-modality to these two populations of planets. This bi-modality probably arises because planets switch their orbital migration from type I to type II depending on their masses \citep{Emsenhuber2021A}. This bi-modality cannot be seen in Bern Compact Multis because we only include planets with periods less than $100 \si{d}$. For Bern KOBE Multis, the completeness of the Kepler mission for large distant planets is poor \cite[see Fig. C.2 in][]{Mishra2021}. However, a dip at this location in Bern KOBE Multis is visible. It would be interesting to see if such a bi-modality is also present in the Kepler catalogue. \edit{We tested the significance of this bi-modality with Hartigan's dip test \citep{hartigan1985}. The dip test \change{is suggestive of} the bi-modality for the Bern RV Multis and Bern KOBE Multis (p-value < 0.05) and insignificant for the other catalogues.} 
        
        \edit{A system's architecture is sensitive only to the relative distribution of a quantity (such as mass) amongst its planets and not the absolute distribution. HR 8799 offers an example \citep{2008Sci...322.1348M} as a relatively young system with four directly imaged giant planets. Our framework identifies the architecture of this systems as \similar/. Most observed \similar/ systems are composed of low-mass planets ($\lesssim 100 \mearth$), making HR 8799 a unique exception. This shows that the architecture framework is sensitive only to the relative variations in the mass. Additionally, there are only two systems (out of 1000) in our simulated catalogue where a \similar/ architecture arises from only giant planets. Even then, these two synthetic systems have only two giant planets much closer to the star than the HR 8799 planets. The Bern Model does not produce many HR 8799-like systems. This suggests that a system with \similar/ architecture made up of only giant planets is probably rare. One possibility could be that systems (e.g. HR 8799) with such architecture are probably difficult to form via core accretion pathway \citep{2018haex.bookE..36K}. Such systems may require additional formation mechanisms such as protoplanetary disk instabilities \citep{Schib2021, 2010Icar..207..509B, 2010ApJ...710.1375K}.}

        \subsection{Architecture class: \mixed/}
        Planetary systems where the planetary masses (inside-out) show broad increasing and decreasing variations have \mixed/ architecture.  
        GJ 876 and Kepler-89 host planetary systems with a \mixed/ class architecture. GJ 876 is an M dwarf low luminous ($\approx 0.01~ L_\odot$) star hosting four planets with masses between $8 - 888 \mearth$. The outer three planets are in a Laplace mean-motion resonance \citep{Millholland2018}. Kepler-89, on the other hand, is an early F, highly luminous ($\approx 3.5~ L_\odot$) star. It hosts a compact four planet system with masses between $10 - 100 \mearth$. Despite the starkly different stellar properties, the architecture of these two systems is analogous: $\cs(M) = 0.12~\text{and}~0.17$, $\cv(M) = 1.2~\text{and}~0.9$, respectively. While the \cstext/ is low for both systems, the \cvtext/ is larger than $\sqrt{3}/2$, which helps us identify the architecture of these systems as \mixed/ class. Indeed, Fig. \ref{fig:archplot} indicates that this identification is correct.
        
        \edit{The frequency of this} architecture class in the Bern model is $\approx 8.2\%$. The Bern model's synthetic \mixed/ architecture planetary systems (Fig. \ref{fig:archplot} right) tend to have numerous Earth-mass planets outside $10~\si{au}$. This parameter space (mass-distance plane, Fig. \ref{fig:masssma}), however, remains inaccessible to most exoplanet detection techniques. These systems are \change{also} composed of super-Earths, sub-Neptunes, Neptunes, and Jovian planets. The bi-modality in distance distribution (discussed before) is prominent for these architectures in Bern RV Multis. \change{We found a Harigan's dip statistic of 0.03 and p-value of $\sim0.2$ \citep{hartigan1985}.}
        
        \subsection{Architecture class: \antiordered/}
        Planetary systems where the planetary mass shows an overall decrease with distance have an \antiordered/ architecture. There are no observed examples of this architecture class in our catalogue. \edit{The frequency of this} architecture class in the Bern model is $\approx 8.4\%$. About $\approx 4\%$ of systems in Bern KOBE Multis, $\approx 3.2\%$ of systems in Bern Compact Multis, and $\approx 1.2\%$ of systems in Bern RV Multis have this architecture. This shows that it is an observationally challenging system architecture to detect. \change{However, even if $1\%$ of observed exoplanetary systems are Anti-Ordered we should already have  found about 30-40 such systems. More work is necessary to identify the handful of these systems from the already observed systems.} Many currently known single hot Jupiter systems may host additional small, distant, and as yet undetected planets -- revealing these potentially \antiordered/ systems.

        Anti-ordered systems in the Bern Model are mostly composed of low mass planets $\lesssim 5 \mearth$ and giants $\gtrsim 100 \mearth$. In the Bern Model, the radius distribution of this architecture class peaks for Rocky and Super-Earths planets. It decreases for sub-Neptunes and Neptunes and then increases again for Jovian planets. Many of the low-mass planets that make up this architecture class are outside $10 \si{au}$, making their detection very challenging. The multiplicity distribution shows that these systems tend to have fewer planets than \similar/ or \mixed/ architecture. This is an indication that the formation pathway of these architectures differs considerably from the other two types of architecture. Planets from \antiordered/ architectures show \edit{a weak} distance bi-modality feature (discussed earlier in this work). This is understandable since these architectures consist of massive planets in the inner parts and less massive planets in the outer parts of the system. The distance bi-modality seems to arise from low mass planets (migrating via type I) inside 0.28au or 55 days and giant planets (migrating via type II) outside 0.28au or 55 days. This adds further strength in attributing the distance bi-modality to planetary migration.
                
        \subsection{Architecture class: \ordered/}
        
        Planetary systems where the planetary masses shows an overall increase with distance have an \ordered/ architecture. The increasing mass may be monotonic (e.g. TOI-561, HD 20781, DMPP-1,HD 160691, HD 164922) or non-monotonic (e.g. the Solar System, Kepler-11, 55 Cnc, Kepler-48, Kepler-65). Ordered architecture is \edit{a rare} outcome for the Bern model. \edit{Observations are generally biased against discovering small and less massive planets which are farther away from their host star. Such biases, however, make \ordered/ systems} the second most common architecture class. Fifteen systems in our catalogue exhibit this architecture. Unsurprisingly, the most notable known example of this architecture class is the Solar System.
        
        The mass and radius distributions of \ordered/ architecture in the Bern Model shows considerable difference from other architecture. The mass distribution peaks around $1000 \mearth$. Most of the Bern model's \ordered/ systems tend to have at least one giant planet. These systems are also composed of sub-Neptunes, Neptunes, and Jovian planets. 
                
\section{Internal composition across architecture classes}
\label{sec:internalcomposition}

        So far we have seen the new architecture framework (Sect. \ref{sec:framework}) and some characteristics of the four classes of architecture (Sect. \ref{sec:architecturetypes}). In this section, we study the connection between the \change{bulk mass} architecture classes and the internal composition of the planets. \change{This section demonstrates that the same architecture framework can be used to study the multi-faceted nature of planetary system architecture -- from bulk mass architecture to density architecture.} We study \change{several different aspects of the planetary internal composition: (a) radius architecture (Sect. \ref{subsec:radius}); (b) bulk density architecture (Sect. \ref{subsec:bulkdensity}); (c) Core/Envelope mass architecture (Sect. \ref{subsec:coremass}); and (d) fraction of volatiles and water ice in core architecture (Sect. \ref{subsec:icefrac})}. We explore these connections for planetary systems in the simulated (Bern model) and synthetically observed catalogues (Bern RV Multis, Bern KOBE Multis, Bern Compact Multis).  \edit{All results in this section are derived from synthetic planetary systems only.}

        \subsection{Radius architecture}
    \label{subsec:radius}
        \def\figwidth{7cm}
    
    \begin{figure}
    \centering
    \resizebox{\hsize}{!}{\includegraphics[trim=0.3cm 0.3cm 0.3cm 0.3cm,clip,height=\figwidth]{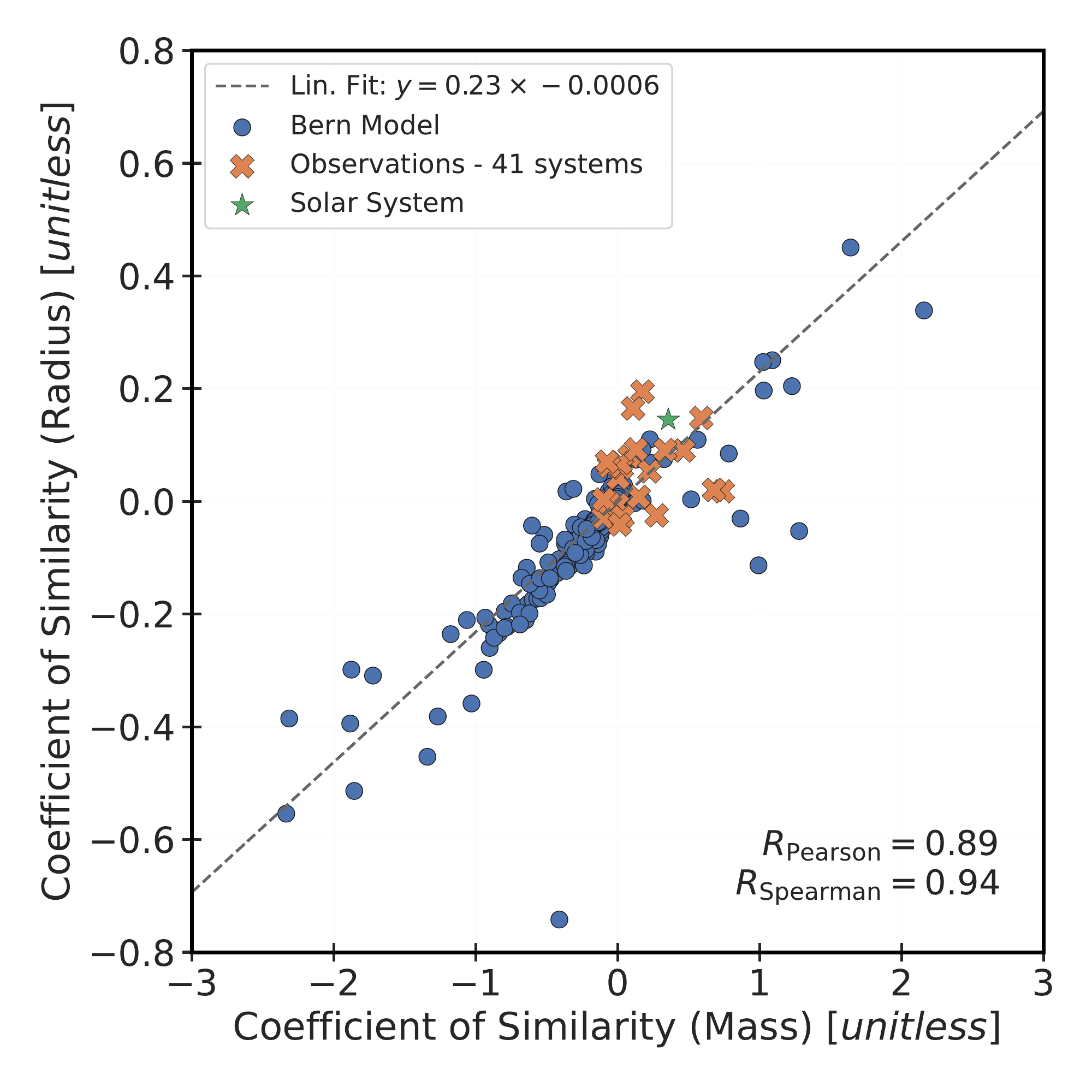}}
    \resizebox{\hsize}{!}{\includegraphics[trim=0.3cm 0.3cm 0.3cm 0.3cm,clip,height=\figwidth]{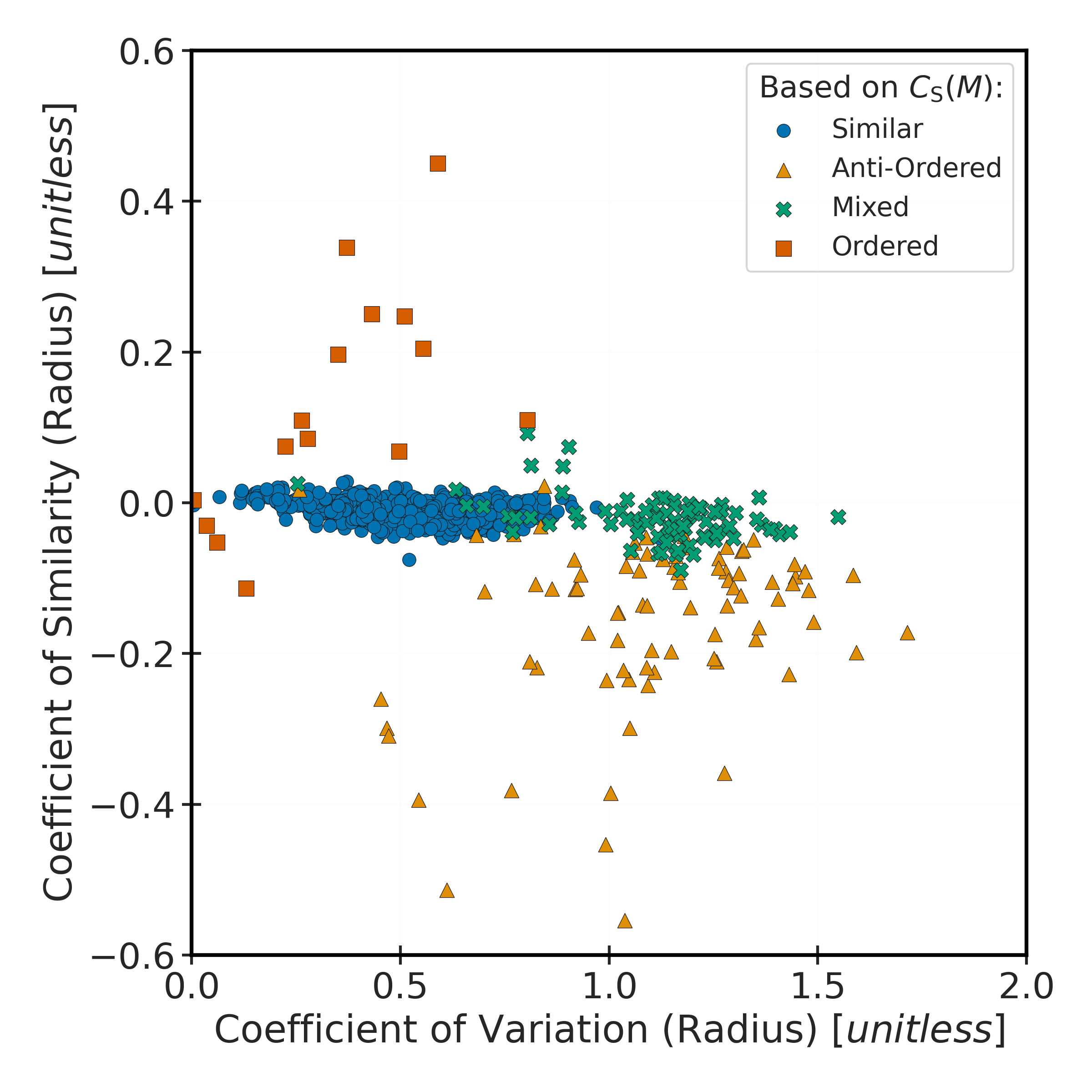}}
    \caption{Radii architecture. Top: The diagram shows the \cstext/ of radii as a function of the \cstext/ of masses, for synthetic and observed planetary systems. The dashed line shows the corresponding linear fit. Bottom: Radius architecture of synthetic planetary systems contrasted with the mass architecture. In the bottom panel, the marker colour and shape indicates the bulk mass architecture of a system and its position on the diagram suggests its radii architecture.}
        \label{fig:unify}
    \end{figure}

\cite{Weiss2018} showed that the size of adjacent exoplanets were similar -- coining the phrase `peas in a pod' to describe this architecture. \cite{Millholland2017, Wang2017} extended these ideas to planetary masses, showing that the masses of adjacent planets are also correlated. In \cite{Mishra2021}, we suggested that the peas in a pod trends in terms of size effectively emerge from the mass trends. Here, we attempt to set our assumption on firmer ground. 
        
Figure \ref{fig:unify} (top) shows the \cstext/ for radii as a function of the \cstext/ of masses, for systems with two or more planets. This allows us to compare the system-level radius architecture with the system-level mass architecture. We easily see that most systems seem to follow a linear relationship. The Pearson correlation coefficient is 0.89, indicating a strong positive correlation between the mass and radius architecture. The coefficient value increases to 0.96, when systems with only three or more planets are considered. Since the mass-radius relation is not a bijective function (i.e. one-to-one correspondence), there are some systems that show a strong deviation from the linear relation. 
        
Figure \ref{fig:unify} (bottom) shows the radii architecture for the synthetic planetary systems\footnote{A future study could investigate the boundaries for robust architecture identification, as in Eq. \ref{eq:classify}, but based on radius instead of mass. Such a classification is readily applicable since radius measurements tend to be uniformly available and are better agreed upon amongst several observers. Data-driven approaches such as machine learning could be useful in such an endeavour.}. This shows that most systems that are \ordered/ (or \antiordered/) in mass are also \ordered/ (or \antiordered/) in terms of radius. The figure also shows that systems which are \similar/ or \mixed/ in mass architecture have $\cs(R) \approx 0$. Systems with mass similarity have lower $\cv(R)$ compared to systems with mass mixture, suggesting that for most systems, the radius architecture closely follows the mass architecture. \change{At the planetary level the radius of a planet is correlated with its mass via the planet's chemical composition \citep{2014ApJ...792....1L}. Our architecture framework shows that such relationships also exist at the system level.} A few mass-\ordered/ systems show similarities in radius. These few systems have the following common features: two mass-\ordered/ giant planets with similar sizes (masses $\sim$ several $M_J$'s, and radius $\approx 1 R_J$). This illustrates that while mass architecture and radius architecture are related, they are not always identical. 
        
        We conclude that the peas in a pod radius correlations generally arise from the underlying mass architecture. We consider the mass architecture primal because planets, foremost, accrete mass from the protoplanetary disk and, consequently, are characterised by a size that is in accordance with their internal structure.                 
                
        \subsection{Density architecture}
        \label{subsec:bulkdensity}
                \def\figwidth{5.5cm}
                \begin{figure*}
                        \centering
                        \includegraphics[trim=0.2cm 0.3cm 0.3cm 0.2cm,clip,height=\figwidth]{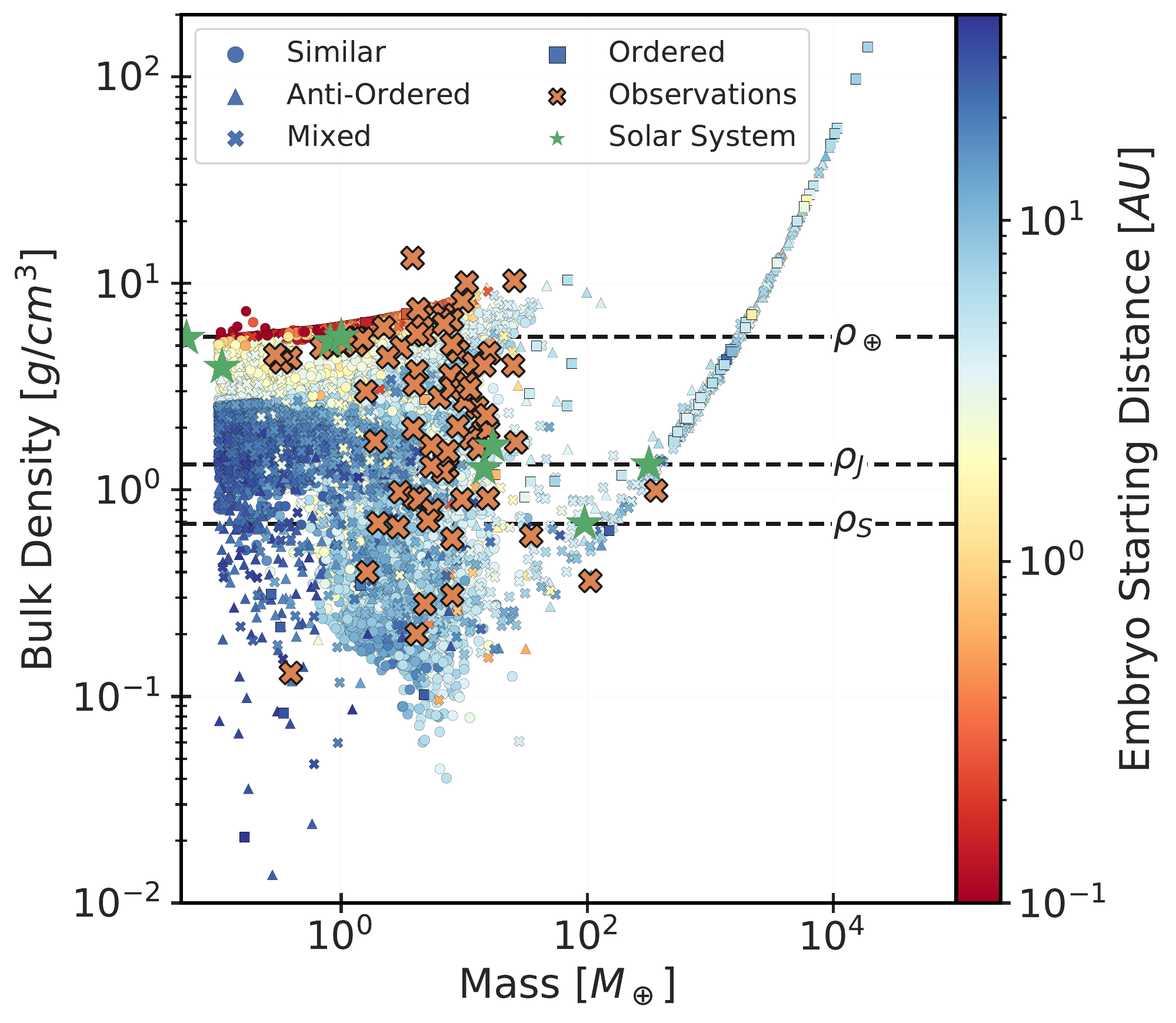}
                        \includegraphics[trim=0.5cm 0.5cm 0.5cm 0.5cm,clip,height=\figwidth]{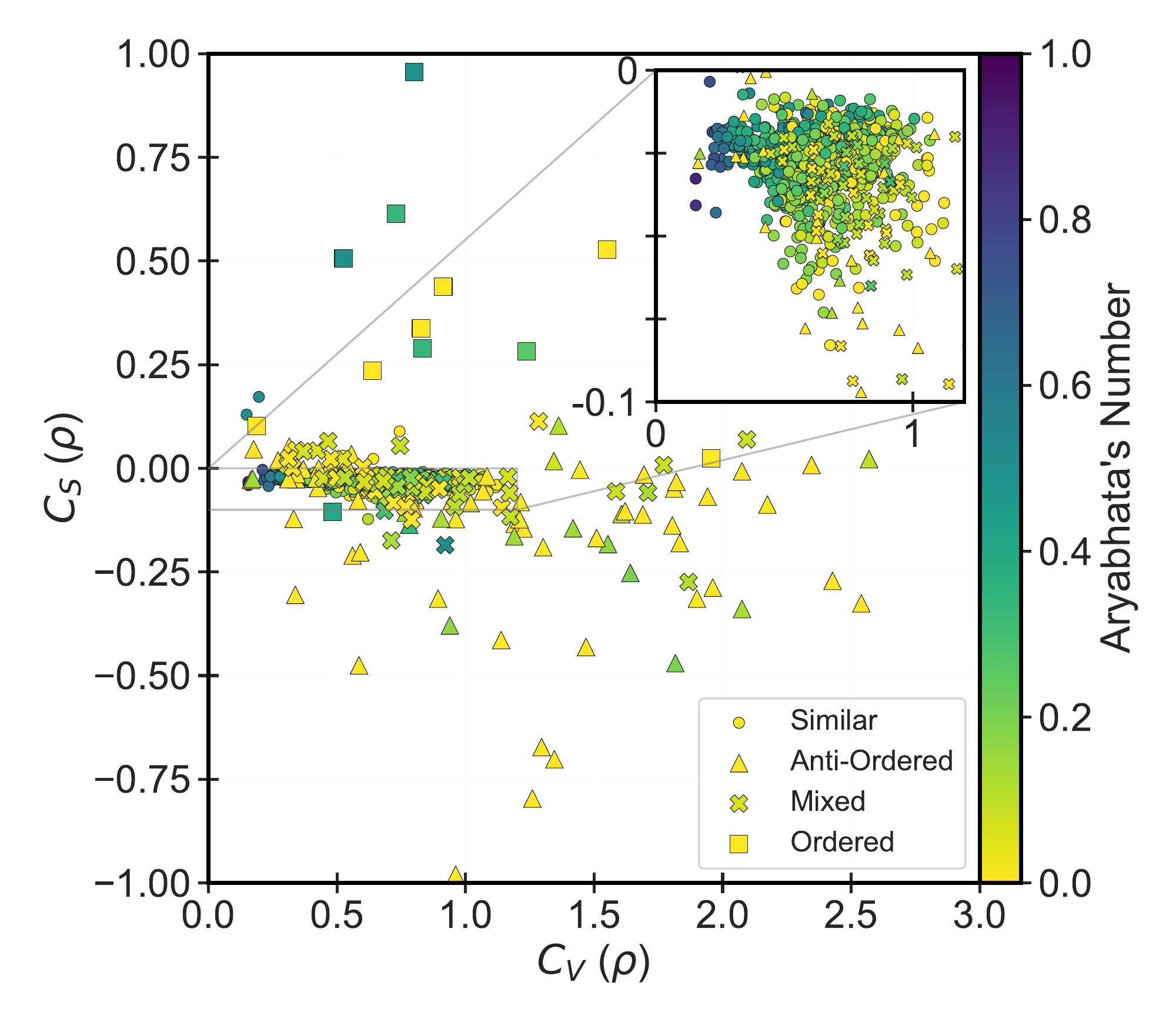}
                        \includegraphics[trim=0.5cm 0.5cm 0.5cm 0.5cm,clip,width=\figwidth]{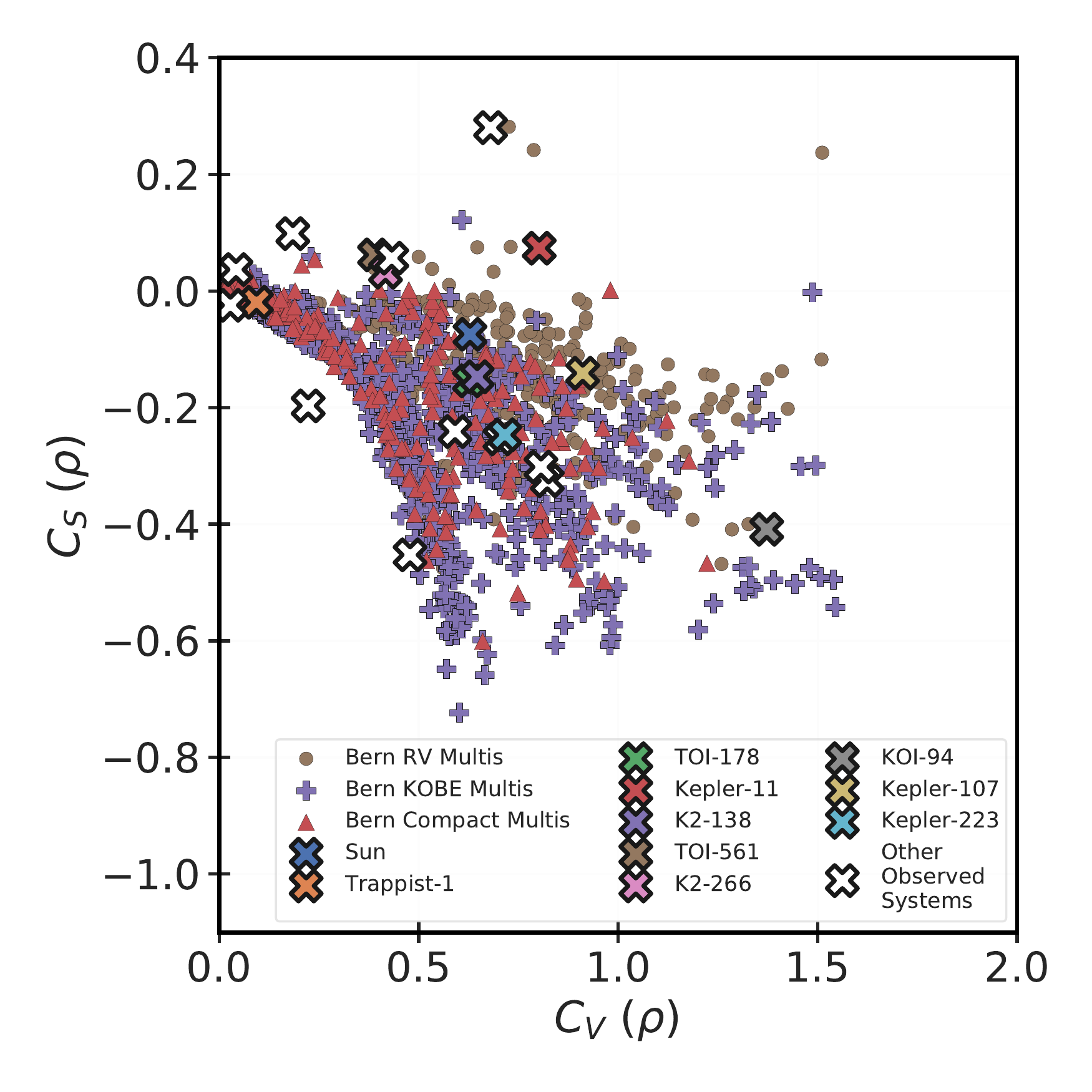}
                        \caption{Density architecture. Left: Bulk density of simulated and few observed planets as a function of their mass and starting locations (for synthetic planets). The marker indicates the mass architecture of the system to which a synthetic planet belongs to. Middle: Density architecture, of synthetic planetary systems, as seen through the \cstext/ versus the \cvtext/ plot. The marker shape and colour indicates their host system mass architecture and the system's Aryabhata's number (see \papertwo/), respectively. Right: Density architecture of planetary systems from the simulated observed catalogue and few observed planetary systems.}
                        \label{fig:density}
                \end{figure*}   
        
        Bulk density (or simply density) is a directly measurable quantity which is sensitive to the internal structure of a planet. This makes density an important characteristic for understanding planetary structure. The density of a planet depends on many parameters and many physical processes. For example, a planet's mass may depend on its accretion history, starting location, amount of material in disk, competition with other planets, and so on. Giant impacts may also affect a planet's density, as explained in \cite{Bonomo2019}. In this section, we study the arrangement and distribution of planetary density around their host star, namely, the density architecture of a system. 
        
        Figure \ref{fig:density} (left) shows the density of a planet, simulated via the Bern model, as a function of its mass and starting location. The figure also shows the density of solar system planets and few observed exoplanets (from our catalogue). The plot can be roughly divided into two halves: (a) planets with a mass of $<100 \ \mearth$ and (b) planets with a mass of $> 100 \ \mearth$. In our simulations, most planets which started inside the ice line tend to have terrestrial Earth-like densities. These planets are $0.5-3 \rearth$ and $\lessapprox 10 \mearth$. Planets starting around or outside the ice line generally accrete more volatile rich material and H/He gas. These planets have lower densities due to their larger sizes. Planet which started outside the ice line (3-10 \si{au}) show a broad diversity in their densities. As they accrete more gases, their density decreases further. These planets are roughly $2-10 \ \rearth$ and are characterised by masses that vary by four orders of magnitude. Planets more massive than 100 $\mearth$ seem to lie on a single curve. Since the size of these planets remains the same ($\approx 1 R_{J} \, \text{or} \,  11 \rearth$,), their densities increases linearly with their masses. Planets that started in the outer regions (30-40 \si{au}) cluster on the density-mass plane. These planets have low densities ($<2 \si{g/cm^{3}}$) and low masses ($\lessapprox 1 \ \mearth$). 
        
        The density architecture for simulated systems in the Bern Model is shown in Fig. \ref{fig:density} (middle). An important relation between mass architecture and density architecture is seen. Some systems which are \ordered/ (or \antiordered/) in mass are also \ordered/ (or \antiordered/) in density, that is, these systems have large positive (or negative) $\cs(\rho)$. In other words, simulations suggest that planetary systems can also be \ordered/ or \antiordered/ in density. A system is \ordered/ in density when the inner planets have small densities and the outer planets have larger densities -- and vice-versa for density \antiordered/ systems. Systems with mass architectures of \similar/ and \mixed/ are strongly clustered around $\cs(\rho) \approx 0$ and $\cv(\rho) < 1$. The inset shows that \similar/ mass systems tend to have small $\cv(\rho), $  while  \mixed/ mass systems have larger $\cv(\rho)$. This implies that some systems that are \similar/ (or \mixed/) in mass show some similarity (or mixture) in density. A system with a \textit{similar} density architecture will host planets that have approximately similar densities. However, the region $\cs(\rho) \approx \cv(\rho) \approx 0$ is empty, indicating the absence of planetary systems where the density of planets (inside out) is approximately the same. While there are exceptions, overall, for many systems, the density architecture seems to follow their mass architecture. 
        
        This approximate link between the mass and density architecture stems from massive planets (> 100 $\mearth$) whose densities increase with their mass (see \ref{fig:density} (left)). Systems which do not host any massive planet are mostly similar in their mass architecture and have $\cs(\rho) \approx 0$. The inset shows that the Aryabhata's number increases as a system approaches the $\cs(\rho) \approx \cv(\rho) \approx 0$ region (see \papertwo/ for the definition of Aryabhata's number). If a system has more surviving planets that started from inside the ice line, then the densities of these planets will be more similar to each other. This means that the density architecture of a system shows some dependence on the starting location of a planet. 
        
        We also investigated if the relation between the mass and density architectures is observable. Figure \ref{fig:density} (right) shows the density architecture for systems from our synthetically observed catalogues. Also shown is the density architecture of some observed exoplanetary systems for which the mass and radius measurements were available. The density architecture of synthetically observed catalogues shows a trend which is quite unlike Fig. \ref{fig:density} (middle). There is an unexpectedly good agreement between the synthetically observed systems and the observed planetary systems. We attribute the peculiar shape of this plot to the difficulty of detecting distant planets. Transit and RV observations favour the detection of planets within $\sim 1$ au. Many close-in planets tend to have Earth-like densities, while planets farther out have lower densities (due to either their volatile rich or gaseous composition). Overall, this would lead to an observed density architecture where inner planets have higher densities and outer planets have lower densities. A situation such as this will be characterised by negative $\cs(\rho)$, which is readily seen from Fig. \ref{fig:density} (right).
        
        In summary, many synthetic systems show a relationship between their mass architectures and their density architectures. Bern model systems that are \ordered/ or \antiordered/ in their mass also tend to be \ordered/ or \antiordered/ in their densities. The dispersion of planetary bulk densities in \textit{similar} class systems is lower than \mixed/ class systems. This relation seems to emerge from massive planets whose densities increases linearly with their masses (since they cannot grow their sizes any more). These relations can be considered as a prediction from this work. As future observations probe the outer parts of an exoplanetary system, we may anticipate the discovery of several systems whose mass and density architectures are closely linked.

        \subsection{Core and envelope mass architecture}
        \label{subsec:coremass}

        \def\figwidth{6cm}
        \begin{figure*}
        \centering
        \includegraphics[trim=0.5cm 0.5cm 0.5cm 0.5cm,clip,width=\figwidth]{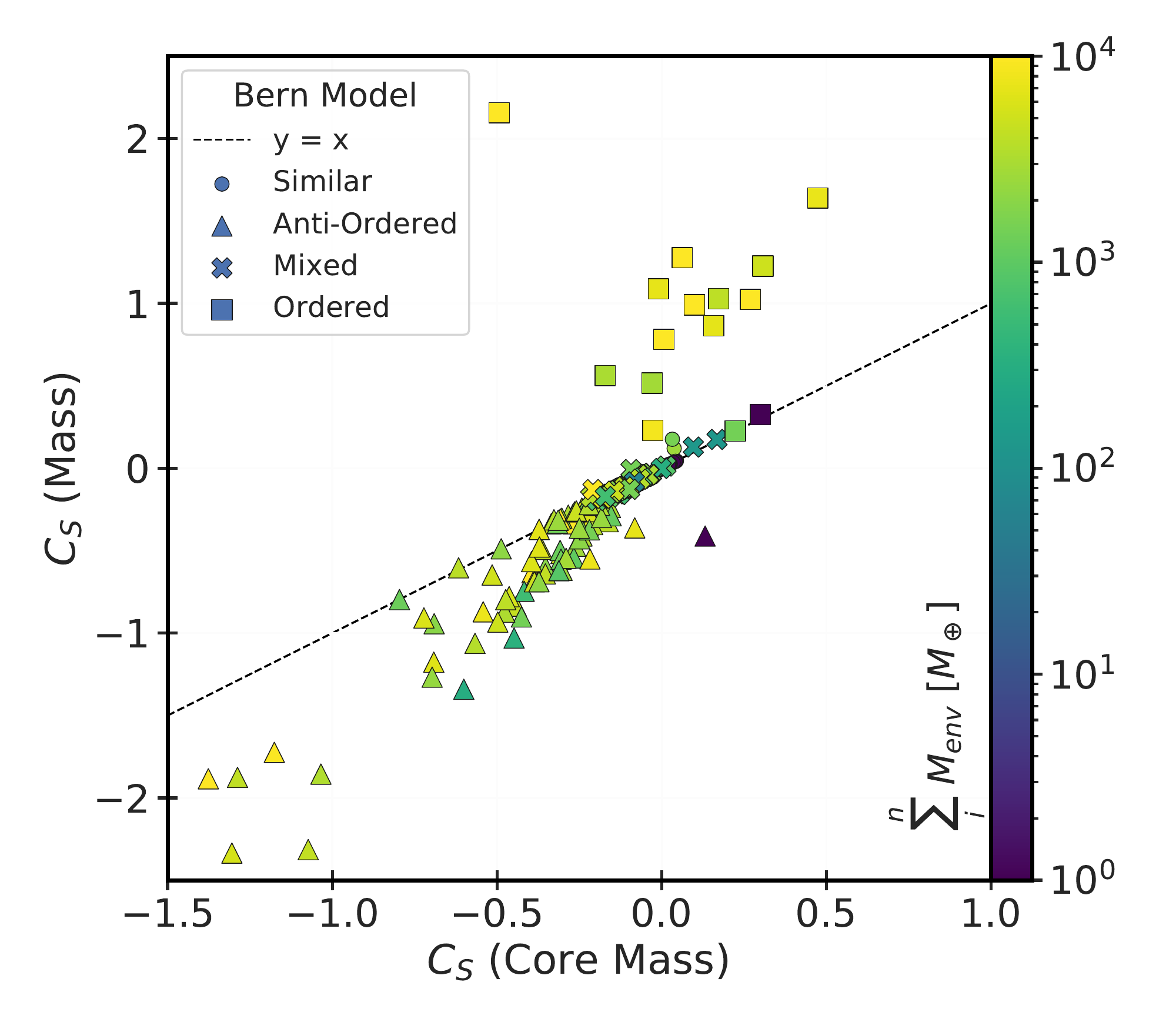}
        \includegraphics[trim=0.5cm 0.5cm 0.5cm 0.5cm,clip,width=\figwidth]{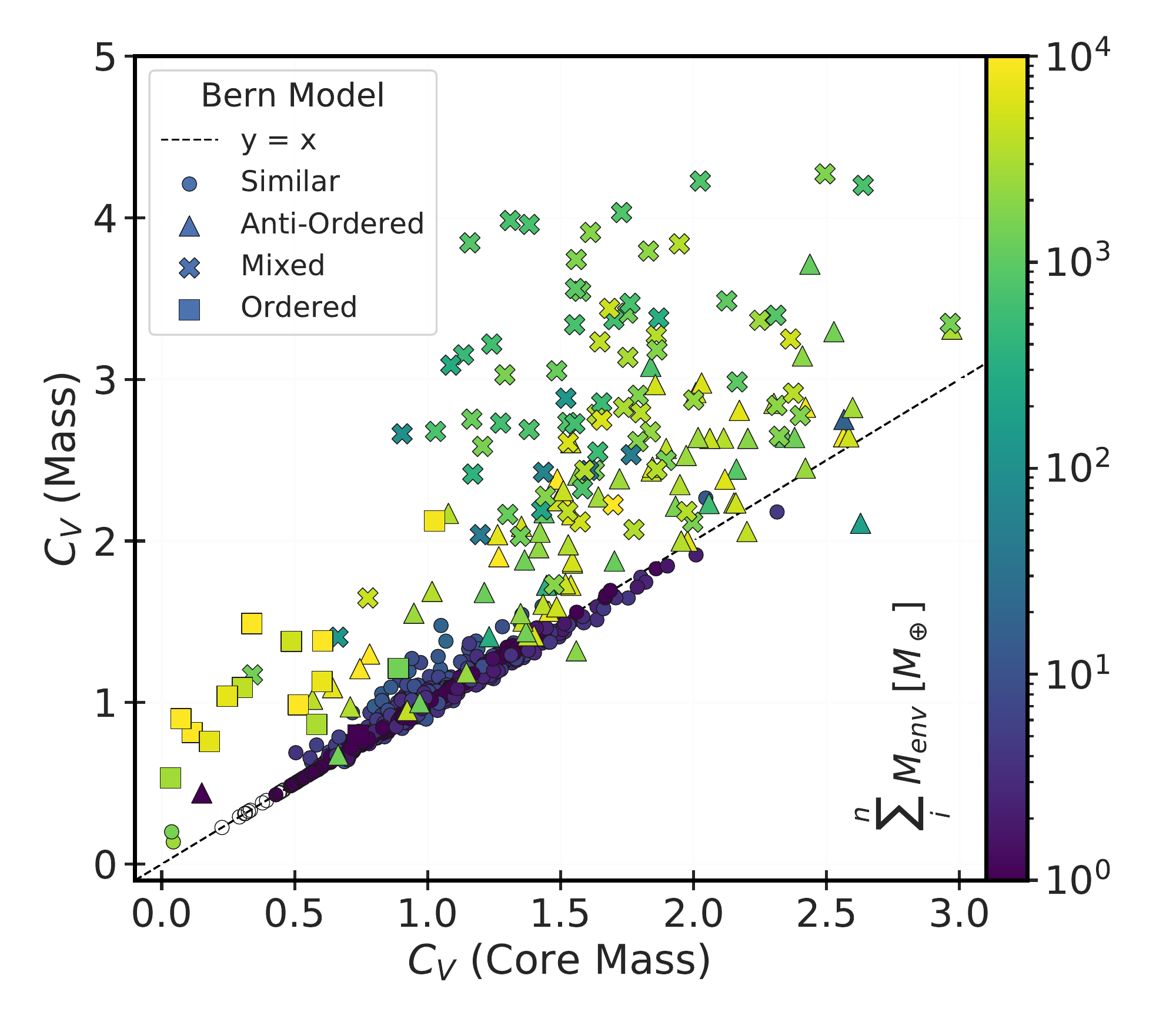}
        \includegraphics[trim=0.5cm 0.5cm 0.5cm 0.5cm,clip,width=\figwidth]{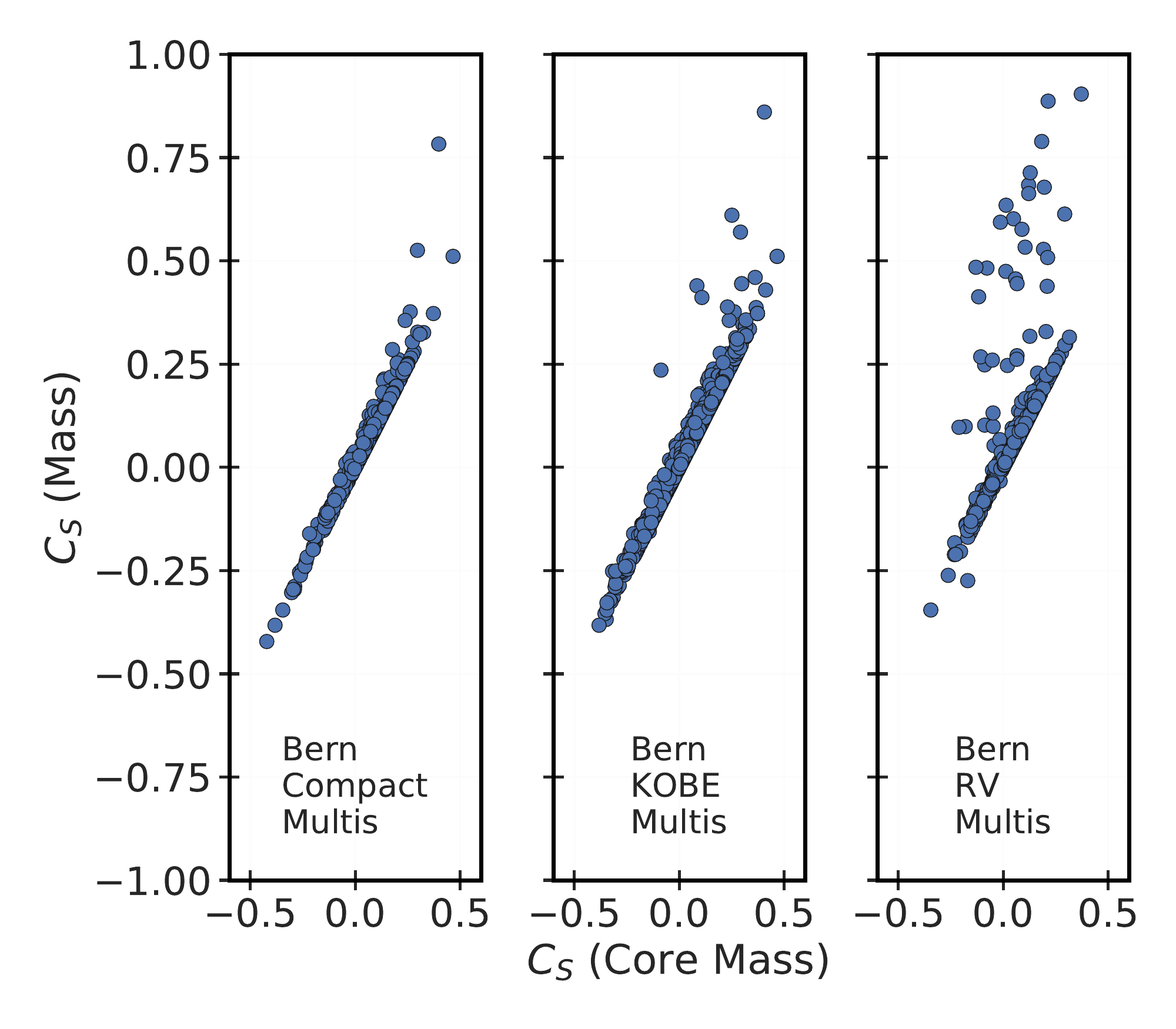}
        \caption{Mass architecture as a function of core-mass architecture. \edit{Panels compare the mass architecture with the core-mass architecture via the \cstext/ (left) and \cvtext/ (middle)}. \change{In the left panel, the points corresponding to \similar/ systems are very tightly clustered on the $y = x$ line and are not visible due to over-plotting of points from other architectures. This signifies the core-mass architecture is very strongly correlated with the mass architecture for \similar/ systems.} The sum of mass in the envelope of each planet in a system is indicated in colour. The right panel plots the \cstext/ for masses and core masses for systems in the synthetically observed catalogues.}
        \label{fig:massarch_coremassarch}
        \end{figure*}

        In this section, we show that (a) most simulated planetary systems inherit their architecture from the underlying core mass architecture; (b) the accretion of gases tends to accentuate the underlying core mass architecture, and (c) the observed mass architecture of a planetary system is a gateway to studying the core mass architecture of the system, since the two are strongly correlated. Exceptions to the first two statements tend to arise for those systems undergoing strong, multi-body dynamical effects such as planet-planet scattering. 

        The fraction of mass which is partitioned into a planet's core and its envelope is governed by planetary formation physics. The end result is dictated by an interplay of several concurrent processes \cite[see][for discussion]{Emsenhuber2021A,Mordasini2012(models)}. In the core-accretion scenario, giant planets are formed when planetary cores can undergo run-away gas accretion \citep{Pollack1996, Alibert2004, Alibert2005}. Proto-planets that have failed to trigger runaway gas accretion comprise a diverse group of planets: Earths, Super-Earths, mini-Neptunes, and Neptunes. 
        
        The bifurcation of a planet's mass into its core and its envelope can probe the formation history. For example, in our simulations, most giant planets ($\gtrapprox 1\ \mj$) have about $1\%$ of their mass in their cores and the rest is in their gassy envelope. On the other hand, low mass planets ($\lessapprox 10\ \mearth$) hardly accrete any gaseous envelope. However, the mass in a planet's core and envelope is not an observable. Even for the solar system planets, internal structure models guide our knowledge of core and envelope masses \cite[see][for a review on Uranus and Neptune]{Helled2020}.
        
        
        As giant planets dominated by their H/He envelopes are rare, we expect a strong correlation between the mass architecture (i.e. the arrangement and distribution of planetary masses) and the core-mass architecture (i.e. the arrangement and distribution of core-masses) to exist also at the system level. In Fig. \ref{fig:massarch_coremassarch}, we show the \cstext/ and the \cvtext/ of planetary mass as a function of the \cstext/ and the \cvtext/ of core mass. The colour indicates the total mass of envelope accreted by all planets in a system. 
        
        Comparing the \cstext/ for planetary masses and core masses (Fig. \ref{fig:massarch_coremassarch},  left panel), we observe that a large fraction of systems ($> 90\%$) follow the \textit{y=x} line. This implies that for most planetary systems, the arrangement and distribution of core masses is imprinted on the mass architecture of the system. Systems which show large deviations from the \textit{y=x} line have generally accreted a large amount of gaseous envelope. This suggests that the formation of one or more giant planet is partly responsible for the deviations. We also observe another important feature. Planetary systems that are \ordered/ in mass are also often \ordered/ in their core-masses. Conversely, mass \antiordered/ systems tend to be \antiordered/ in their core masses as well. In addition, \ordered/ systems are either on or above the \textit{y=x} line, whereas \antiordered/ systems are either on or below this line. This suggests that the accretion of gases generally accentuates the underlying core mass architecture.
        
        Considering the \cvtext/ for masses and core masses (Fig. \ref{fig:massarch_coremassarch}, middle), we see that most of the planetary systems lie either on or above the \textit{y=x} line. The $\cv$ value measures the amount of variation in a set of numbers. This suggests that the variation in total masses, for most systems, is either similar or larger than the variation in the core masses. This is understandable, since the amount of gas accreted by a planet shows a strong correlation with the mass of the planet's core. However, there are a handful of systems where the variation in total mass is less than the variation in core masses. Systems that are \similar/ in the mass architecture are strongly clustered over the \textit{y=x} line. This stems from the low amount of gas ($0 - 20 \mearth$) accreted by planets in these systems. Figure \ref{fig:massarch_coremassarch} (middle) shows that \mixed/ class systems, as opposed to \similar/ systems, form a separate cluster. Physically, this difference is arising from the larger amount of gas ($50 - 5\,000 \mearth$) accreted by planets in these systems. 
        
        Here, the question arises as to whether the strong correlation between mass architecture and core-mass architecture is observable. In Fig. \ref{fig:massarch_coremassarch} (right), we show $\cs (M)$ as function of $\cs (\mcore)$ for the three synthetically observed catalogues. All three catalogues probe the inner regions of a planetary system. The figure shows that the correlation between mass architecture and core mass architecture is strong in all three catalogues. This suggests that the observed mass architecture of a planetary system can be used to study the underlying core-mass architecture of the system. This is potentially useful to distinguish among competing models of planet formation.

        \def\figwidth{9cm}
        \begin{figure*}
                \centering
                \includegraphics[width=\figwidth]{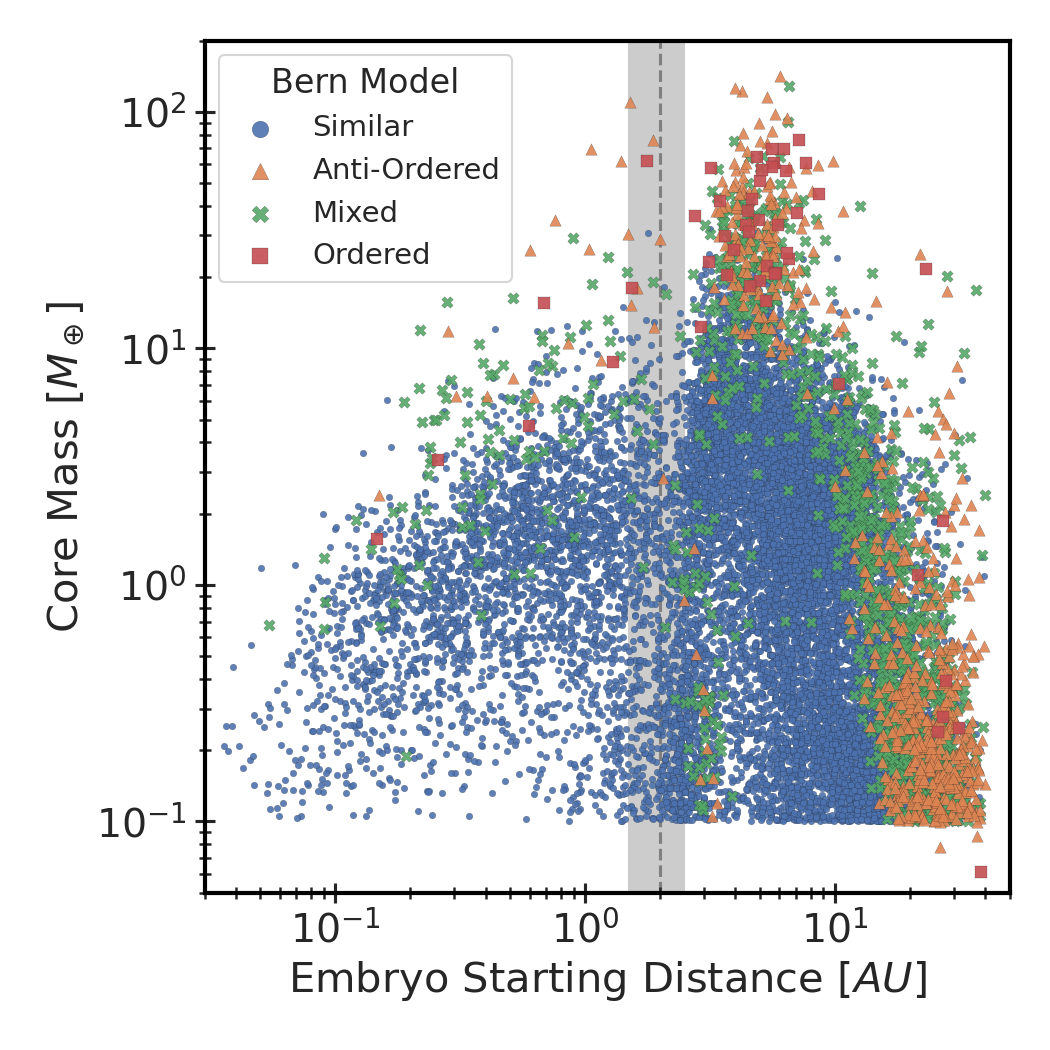}
                \includegraphics[width=\figwidth]{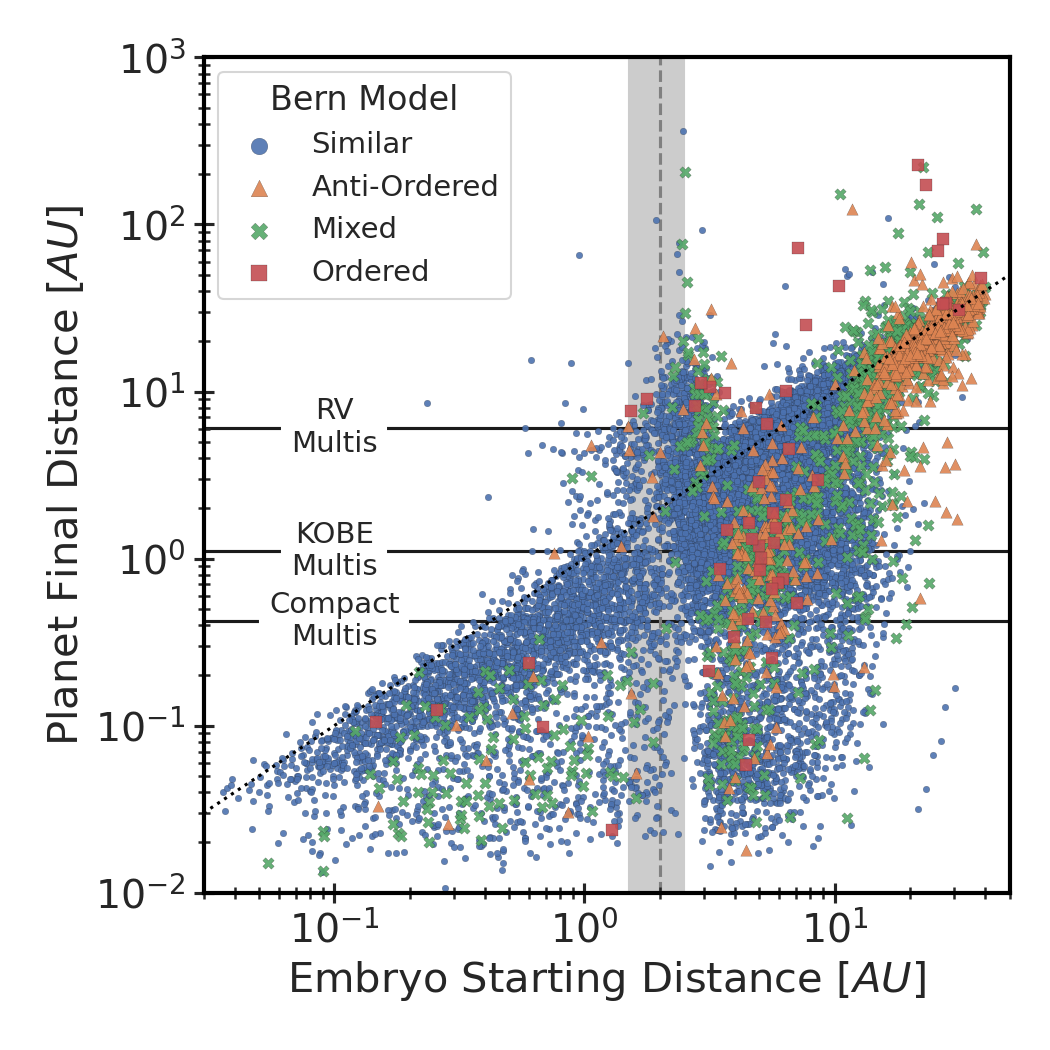}
                \caption{Role of starting location. Plot shows the planetary core mass (left) and final distance (right) versus the starting distance. The marker style indicates the architecture of the system to which the planet belongs. The vertical grey shaded region indicates the evolving locations of the ice line \citep{Burn2019}. The dotted line in the right panel shows the \textit{y=x} line.}
                \label{fig:startingdistance}
        \end{figure*}
        
        \subsubsection{Role of embryo starting location}
        
        We have seen that the core mass architecture of a system strongly governs the overall architecture of the system. The arrangement of planets in a system also reflects the final distances of these planets. It is, therefore, instructive to understand some key aspects which shape these two important properties. The core mass and the final distance of a planet are strongly influenced by, among other effects, the distance at which an embryo starts in our simulations. Figure \ref{fig:startingdistance} shows the core mass (left) and the final distance (right)  as a function of the starting distance. In the Bern model, lunar mass ($0.01 \mearth$) protoplanetary embryos are initialised with a random starting location between the inner edge of the disk and 40 \textit{au}. We also recall that failed embryos (objects with a total masses $< 0.1 \mearth$) are removed from our analysis. 
        
        \cite{Emsenhuber2021A, Burn2021} analysed the nature of planetary migration using migration maps. Both studies show the existence of so-called convergence zones. Within these zones, planets can migrate outwards. However, outside this zone inward migration is prevalent. The existence of such convergence zones suggests that there ought to be regions of planet over-densities; this are essentially regions where planets are radially `stuck.' These studies attribute the presence of these zones to dust opacity transitions and disc structures, finding that these zones evolve with the disc. For a $0.01 \msun$ disc, around a solar mass star at 1Myr, these zones are: (a) for low-intermediate mass planets ($\lessapprox 1 \mearth$) extending from disk inner edge to about $1  \si{au}$ and (b) for intermediate mass planets ($1 - 10 \mearth$) around 2-3 au.       
        
        Figure \ref{fig:startingdistance} (left) shows that even for embryos that start at the same initial distance, the mass accreted by a planetary core can differ by two to three orders of magnitude. These differences arise from (a) varying solid disc masses; (b) competition for accretion in the feeding zone \citep{Alibert2013}; (c) dynamical state of solids in the disc resulting from planetesimal-planetesimal, planetesimal-protoplanet, planetesimal-gas disc interactions, and so on. Nevertheless, the starting distance seems to play a significant role in this scenario. The ice line seems to divide the parameter space into two regions: fewer planets inside the ice line have low mass cores ($\lessapprox 1 \mearth$), while many planets outside the ice line have low-mass cores. 
        
        Inside the ice line, most planets have cores of $1 - 10 \mearth$. Planets that start very close to the star ($\lessapprox 0.1 au$) are unable to accrete a lot of material owing to their small Hill spheres. This explains their small cores masses. Inside the ice line, planets belonging to systems of \mixed/, \antiordered/, and \ordered/ architecture tend to have more massive cores than planets belonging to \similar/ systems. Around the ice line, planets show a large variety of core masses ranging from $0.1 \mearth$ to $100 \mearth$. Outside the ice line we see the same trend as before: planets that are in \similar/ systems, for the same starting location, usually have less massive cores than planets which belong to systems of other architectures. 
                
        The final distance of a planet depends on several factors such as: (a) migration type (type I or type II), (b) planet's mass, (c) local disc properties, and (d) multi-body effects such as \textit{N}-body scattering. The joint distribution of a planet's final and starting locations shows an intriguing trend. Generally, for many planets, the final distance strongly correlates with their starting location. Orbital migration allows planets to move (mostly) inwards -- positioning many planets below the \textit{y=x} line. \textit{N}-body effects (such as planet-planet scattering or outward migration) may scatter some planets further away from their host star. These planets are located above the \textit{y=x} line. Curiously, many planets which end up farther away than their starting location were initialised around the ice line and are mostly low massive ($\lessapprox 20 \mearth$). We attribute this over-density to the outward migration convergence zone around the ice line discussed above.  
        
        \edit{Another important finding is that planets inside the ice line in \similar/ systems probably formed in situ. Fig. \ref{fig:startingdistance} (right) shows that most planets, inside the ice line, which did not migrate inwards are part of \similar/ architecture systems. Conversely, most of the planets which have migrated inwards seem to belong to systems that have \mixed/, \antiordered/, and \ordered/ architectures.} Outside the ice line, many planets have migrated inwards. Most planets starting around 20 au (or more) accrete little mass in their cores and show little radial displacement \citep{2012ApJ...751..158H,2013MNRAS.431.3444C}.  The properties of these embryos may draw some influence from our modelling choice as well. The \textit{N}-body integrator in this model is used for 20 Myr. Longer integration times may allow some embryos to have more massive cores via giant impacts.
        
                        
        \subsection{Core water-ice mass fraction architecture}
        \label{subsec:icefrac}

        \def\figwidth{5.8cm}    
        \begin{figure*}
        \centering
        \includegraphics[trim=0.5cm 0.5cm 0.5cm 0.5cm,clip,width=\figwidth]{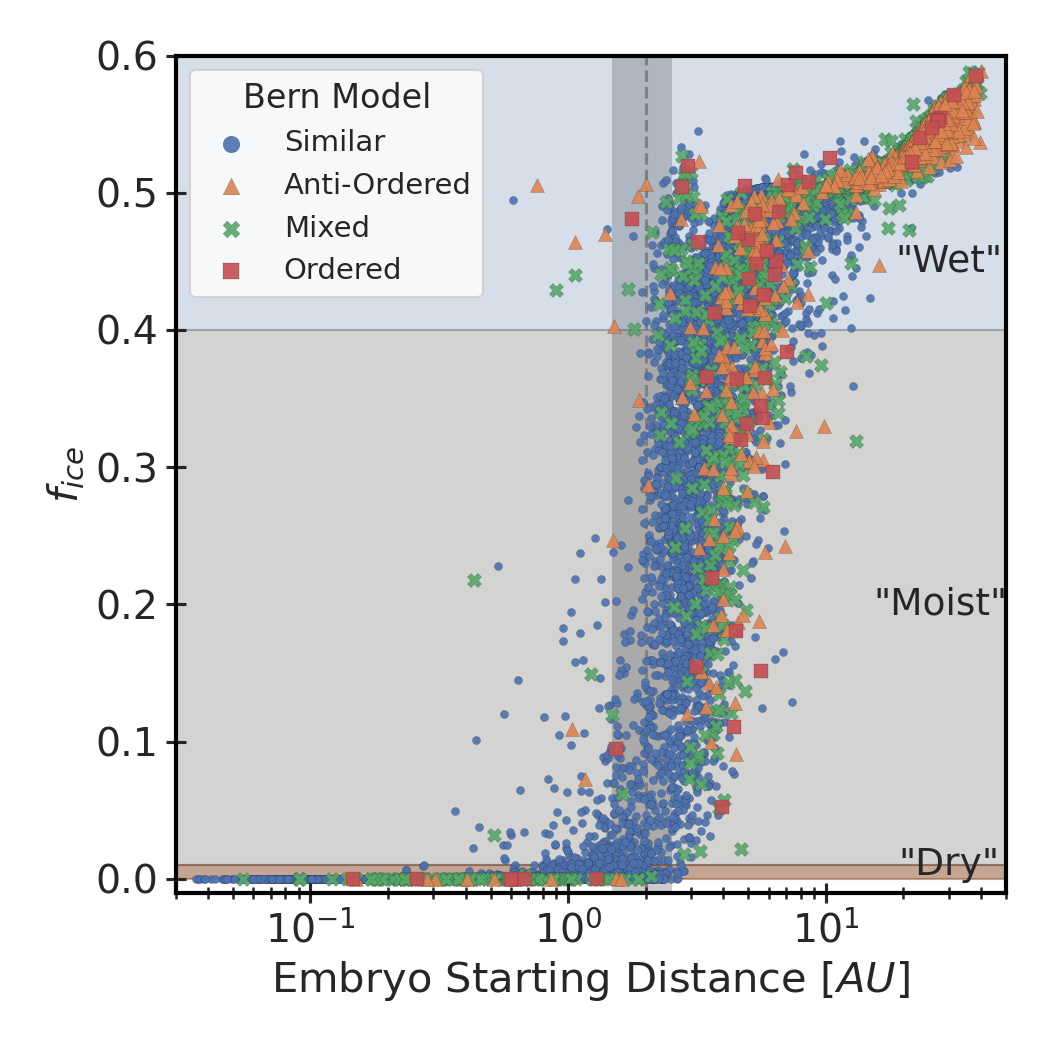}
        \includegraphics[trim=0.5cm 0.5cm 0.5cm 0.45cm,clip,width=6.5cm]{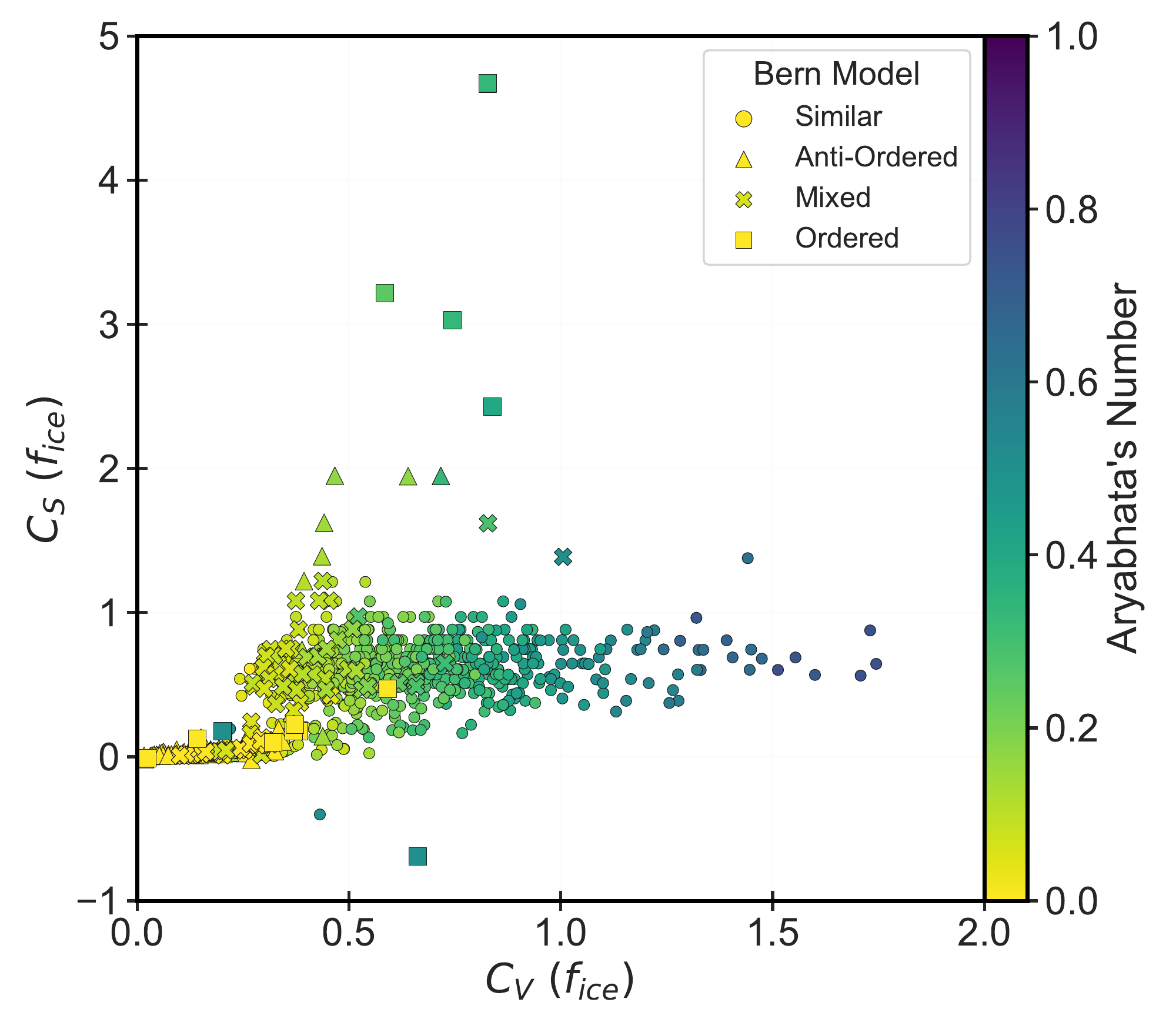}       
        \includegraphics[trim=0.5cm 0.5cm 0.5cm 0.5cm,clip,width=\figwidth]{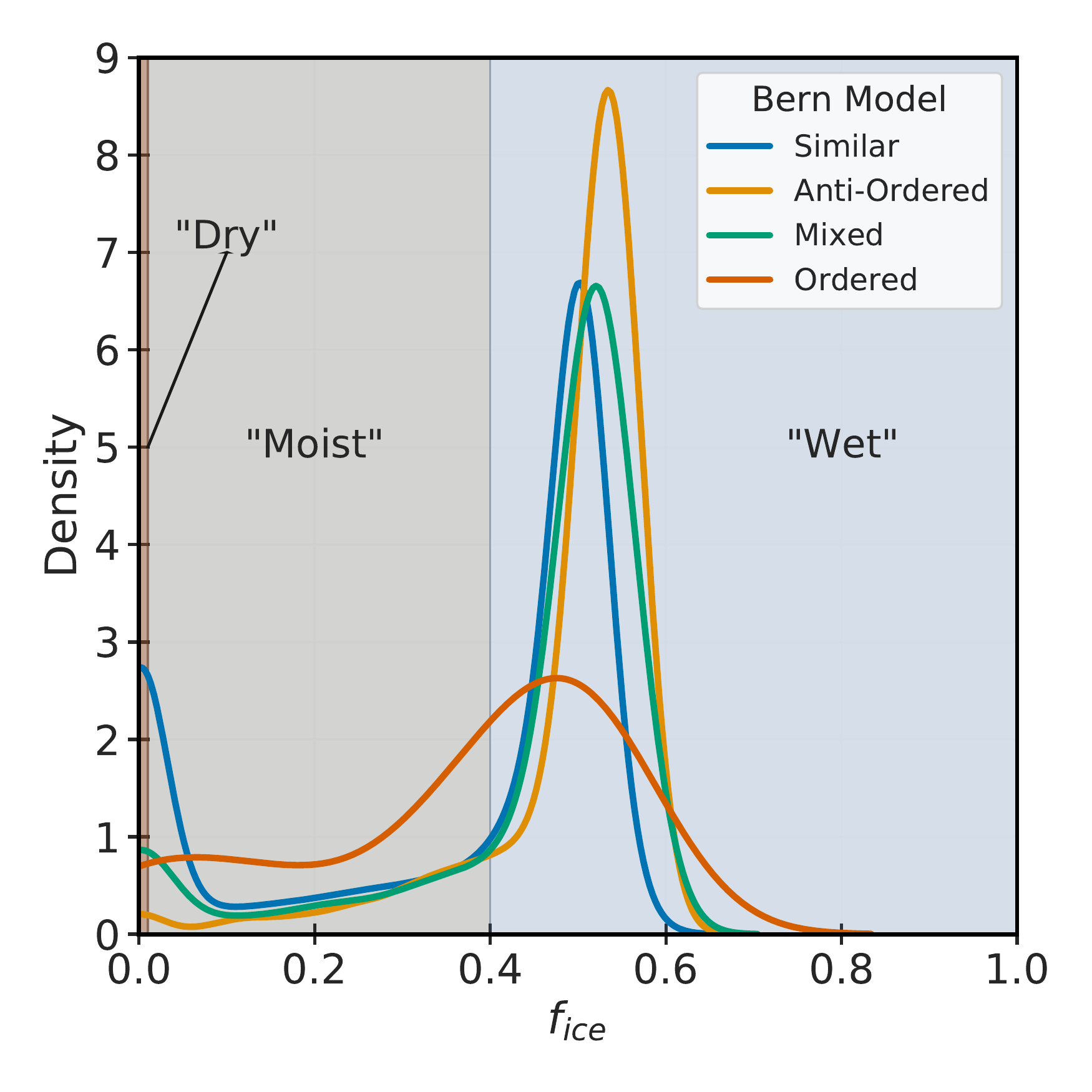}
        \caption{Planetary core water-ice mass fraction. Left: Core water mass fraction of a planet as a function of its starting location. The architecture of the system to which a planet belongs to is shown by marker characteristics. The vertical shaded regions shows the location of the ice line. Middle: Water mass fraction architecture seen through the \cstext/ versus the \cvtext/ plot. The shape of the marker shows a system's mass architecture, and the colour depicts its Aryabhata's number (see \papertwo/ for definition). Right: Distribution of $f_{ice}$ across architecture classes. Depending on $f_{ice}$, planets are labelled as `dry', `moist', or `wet'.}
        \label{fig:icefrac_dist}
        \end{figure*}
        
        Our model calculates the internal structure of a planetary core \cite[for details see][]{Emsenhuber2021A, Mordasini2012(MR)}. We solved 1D differential equations demanding mass conservation and hydrostatic equilibrium, with a modified polytrope equation serving as the equation of state \citep{Seager2007}. The chemical composition of each planetary core is also tracked. This is accomplished by tracking the chemical makeup of the accreted planetesimals and other colliding planets. The underlying chemical models have thirty-two refractory and eight volatile species \citep{Thiabaud2014, Marboeuf2014b, Marboeuf2014a}. These different chemical species are grouped into three different materials which make the planet's core, in our model: (a) iron, (b) silicates, and (c) ice. All refractory species (except iron) make up the silicate mantle and all volatile species contribute to ice. Since $\text{H}_{2}\text{O}$ constitutes $~60\%$ of all ice by mass, we label this latter component as water ice. The water mass fraction ($f_{ice}$)  of each planetary core is computed.  
        
        We assume that inside the $\text{H}_{2}\text{O}$ ice line, only refractory elements contribute to the solid phase of a planetesimal. Outside this evolving ice line, due to their condensation, volatile elements also contribute to the solid phase of a planetesimal. Figure \ref{fig:icefrac_dist} (left) shows the water mass fraction of a planet's core as a function of its initial location. Most planets which start inside the ice line have little to no volatiles in their cores. A jump in $f_{ice}$ is seen around the ice line. Outside the ice line, most planets have at-least $40\% \ f_{ice}$ in their cores. This suggests that the history of formation and evolution of a planet is imprinted on its water mass fraction. 
         
        We are interested in studying the ice mass fraction architecture of a planetary system. However, we cannot directly apply our framework  (Eqs. \ref{eq:cs}, \ref{eq:cv}) because the water mass fraction is a quantity that admits 0 as a value. While this can lead to ill-defined numbers, this issue has a simple remedy. For quantities that can be 0, we propose the following modification to Eq. \ref{eq:cs}: 
        \begin{equation}
        \label{eq:cs_zero}
        \cs (q) = \lim_{\epsilon \rightarrow 0} \ \frac{1}{\nplanet-1}  \ \sum_{i=1}^{i=\nplanet-1} \  \Bigg(log \ \frac{q_{i+1} + \epsilon}{q_i + \epsilon} \ \Bigg).
        \end{equation}
        Numerically, we calculated the \cstext/ with $\epsilon = 10^{-10}$. We verified this step by calculating the \cstext/ for quantities which do not admit zero (such as masses). \change{In a bootstrapped numerical experiment of 10,000 trials, the \cstext/ for mass was calculated using both equations (\ref{eq:cs} and \ref{eq:cs_zero}). The relative difference between the two outcomes ranged between $10^{-12}$ to $10^{-10}$.}
        
        The ice mass fraction architecture of Bern Model systems is shown in Fig. \ref{fig:icefrac_dist} (middle). A prominent feature from this figure is that most systems have $\cs(f_{ice})$ either close to 0 or positive. A system with $\cs(f_{ice}) \approx 0$ and low $\cv(f_{ice})$ will be composed of planets whose core water mass fraction is similar to one another. 
        A system with positive $\cs(f_{ice})$ will be composed of planets whose core water mass fraction increases inside out. Figure \ref{fig:startingdistance} (right) tells us that many planets that started outside the ice line, and are water rich have not suffered any major radial displacement. Thus, a positive $\cs(f_{ice})$ should be a default scenario for most planetary systems. About $74\%$ systems in the Bern model have $\cs(f_{ice}) > 0.1$. Almost $97\%$ of systems have $\cs(f_{ice}) > 0$. We propose the `Aryabhata formation scenario' to explain the `non-default' systems. This scenario and the related quantity `Aryabhata's Number' are described in \papertwo/. 
        
        \subsection{Frequency of dry, moist, and wet planets}
                \def\figwidth{9cm}
        \begin{figure*}
                \centering
                \includegraphics[trim=0.5cm 3.2cm 0.5cm 0.5cm,clip,width=\figwidth]{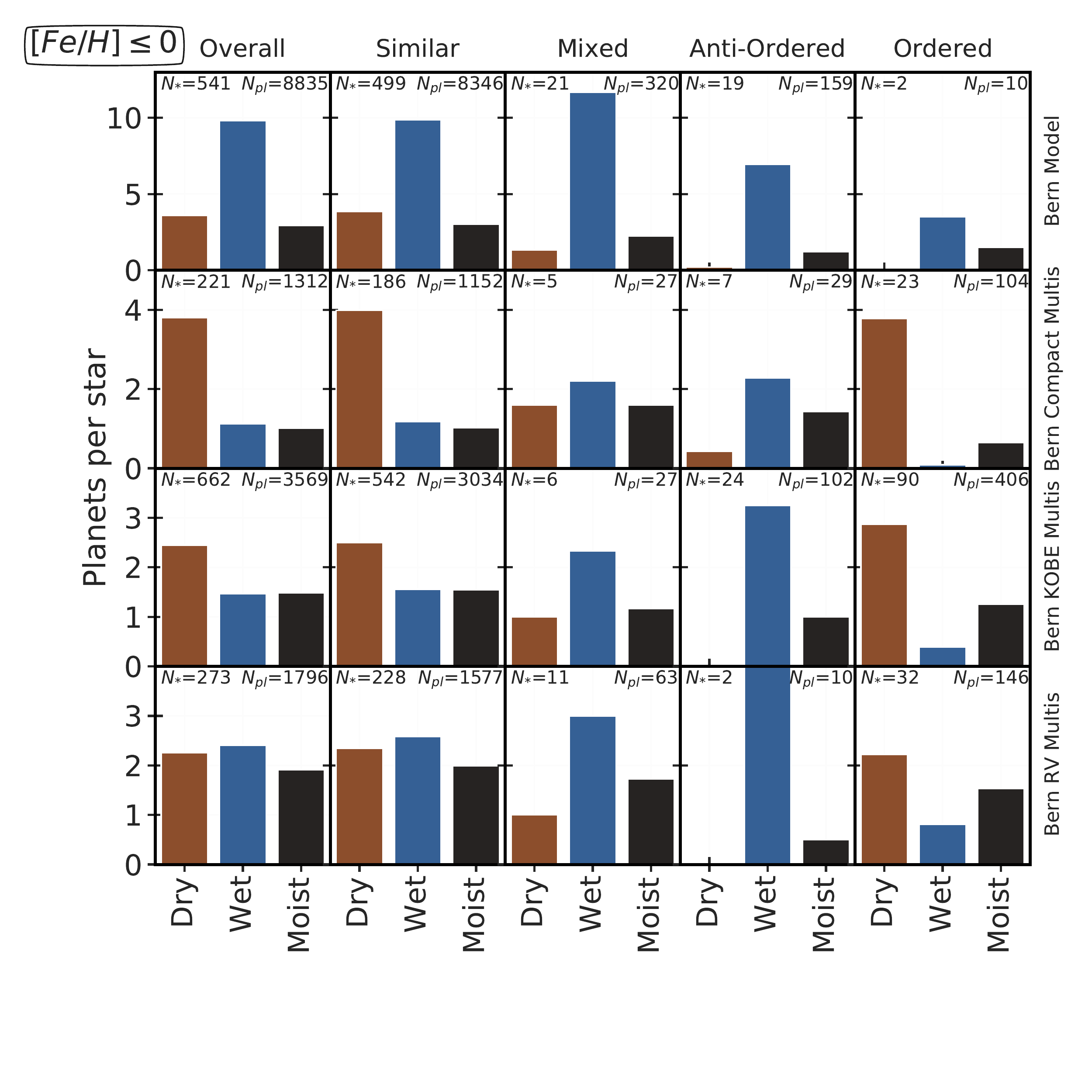}
                \includegraphics[trim=0.5cm 3.2cm 0.5cm 0.5cm,clip,width=\figwidth]{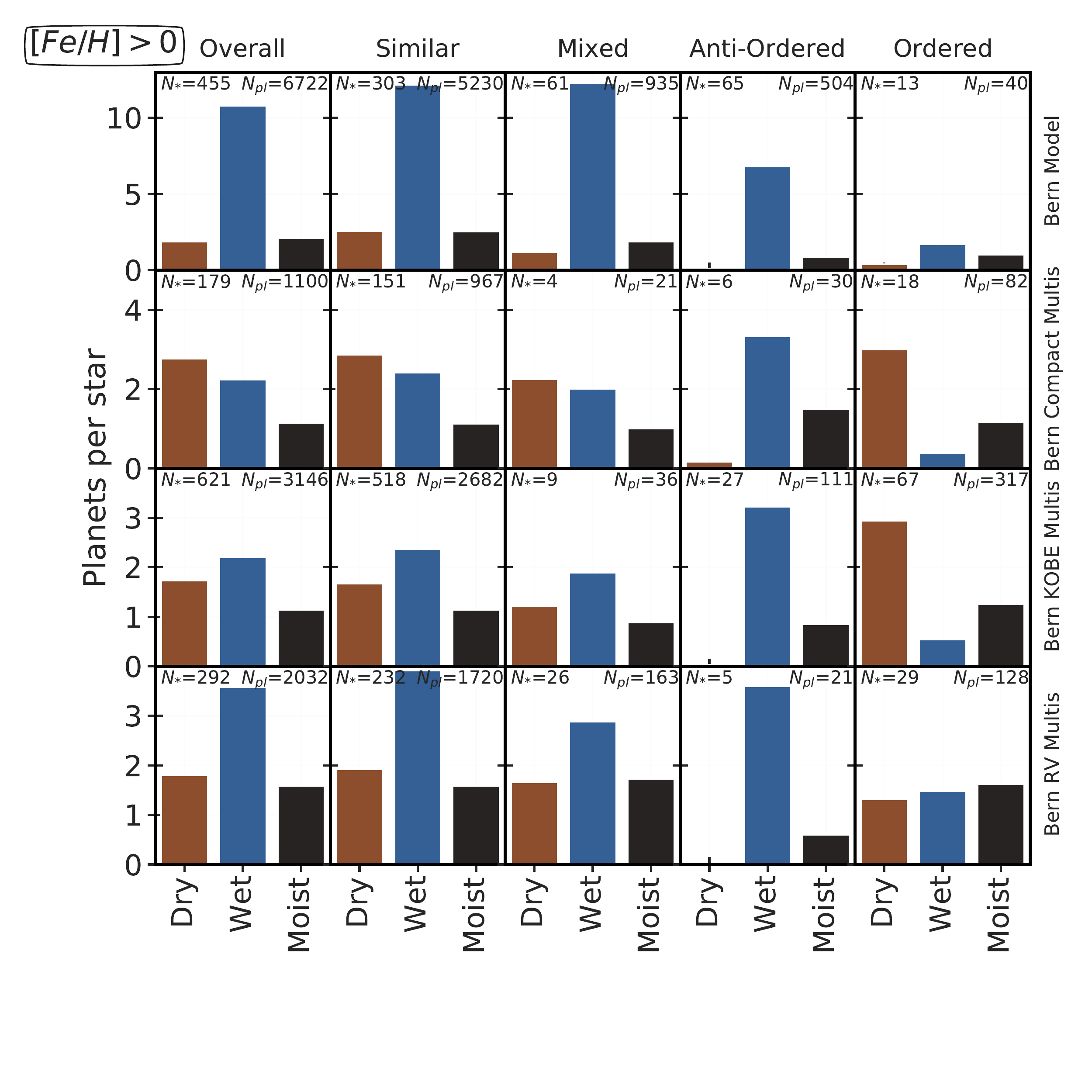}
                \caption{Frequency of planets. This diagram shows the average planet per star for dry, wet, and moist planets in several catalogues (rows), across several architecture classes (columns), and around low (left) and high (right) metallicity stars. The planet per star is simply the total number of planets divided by the total number of stars, after appropriate filters for metallicity, catalogue, or architecture.}
                \label{fig:icefrac_counts}
        \end{figure*}
        
        We are interested in exploring the link between the water mass fraction architecture and the mass architecture of a system. To this end, we divide planets into three categories based on their water mass fraction. A planet is called `dry' if  $f_{ice} \leq 1\%$, `moist' if  $f_{ice} \in (1,40]\%$, and `wet' if  $f_{ice} > 40\%$. These labels serve to simplify our analysis and allows us to see general trends between system architecture and planetary composition. The distribution of water mass fraction across systems of different architecture classes is shown in Fig. \ref{fig:icefrac_dist} (right). While all three planet classes are present in all four architecture classes, there are some observable trends. 
        
        \change{Figure \ref{fig:icefrac_dist} (right) shows that \similar/ architectures host many of the dry planets produced in the Bern model and \antiordered/ architectures are mostly composed of wet planets.} This tells us that many of the planets that start inside the ice line become part of \similar/ architecture systems. Conversely, systems with \antiordered/ architecture are mostly composed of planets that started outside the ice line. Mixed architecture systems are generally composed of more planets that started outside the ice line than inside, as compared to \similar/ architecture systems. Moist planets are present in all architecture classes.  
        We quantify the frequency of dry, moist, and wet planets as a function of mass architecture class (\similar/, \mixed/, \ordered/, or \antiordered/), metallicity (low or high), and source catalogues (Bern model, Bern Compact Multis, Bern KOBE Multis, and Bern RV Multis). Figure \ref{fig:icefrac_counts} shows the planets per star (i.e. the number of each planet type divided by the number of stars) across these forty sub-categories. 
                
        Overall, compared to synthetically observed catalogues, Bern model simulations demonstrate more wet planets. This is understandable since we are looking at the entire underlying population, which includes planets from the outer parts of these systems. Likewise, synthetically observed catalogues tend to have more dry planets. Systems around low-metallicity stars (regardless of the catalogue) generally tend to have a higher frequency of dry planets as opposed to systems around high-metallicity stars. The frequency of wet planets shows a noticeable increase for systems around high-metallicity stars. Amongst the different catalogues, Bern Compact Multis have the highest frequency of dry planets, followed by Bern KOBE Multis, and Bern RV Multis. Low-metallicity environments have a slightly higher average planet per star ($8835/541 \approx 16.3$) than high-metallicity environments ($6722/455 \approx 14.8$).
                
        \textit{Similar systems:}   
        Systems in the underlying Bern model that are characterised by a \similar/ architecture tend to have many wet planets ($\sim 10$ per star) and few dry or moist planets ($\sim 3-4$ per star). However, synthetically observed catalogues seem to have a bias against the discovery of many wet planets. For the \similar/ class of compact multi-planetary systems, dry planets are more common around a low-metallicity star. However, for a high-metallicity star, the frequency of dry and wet planets is roughly the same. For transiting systems, in the Bern KOBE Multis, low-metallicity environments favour more dry planets and equal proportions of wet and moist planets. Conversely, in high-metallicity environments, wet planets occur more frequently than dry or moist planets. For RV systems, the frequency of each planet class is approximately the same in a low-metallicity environment. High-metallicity environments almost double the frequency of wet planets. The average planet per star is similar around both low metallic ($\approx 16.8$) and high metallic environments ($\approx 17.3$).
        
        \textit{Mixed systems:}
        Mixed class systems generally have many wet planets. It is only for compact systems around high-metallicity stars, the frequency of dry planets is higher than wet planets. In all other cases, the frequency of wet planets is greater than the frequency for dry or moist planets. The average planet per star is similar around both low-metallicity ($\approx 15.2$) and high metallicity environments ($\approx 15.3$).
        
        \textit{Anti-Ordered systems:}
        Systems with \antiordered/ architecture have a distinct core water mass fraction architecture. These systems are rich in wet planets. In fact, about $80\%$ of these systems follow the \edit{Aryabhata formation scenario} described in \papertwo/. Compact \antiordered/ systems may have some dry planets. For transit and RV surveys, the frequency of dry planets is zero in our simulations. The total number of planets per star in \antiordered/ systems is slightly higher around low metallicity stars ($159/19 \approx 8.4$), as compared to high metallicity stars ($504/65 \approx 7.8$). In the future, if an \antiordered/ architecture planetary system is to be discovered, it would be interesting to study its core water mass fraction architecture as well. The current work suggests that the Aryabhata's number for these systems should be close to 0 and, irrespective of the detection technique, the system should would be expected to have many wet planets (see \papertwo/); this is one of the main predictions arising from this work.

        \textit{Ordered systems:}
        Juxtaposed directly to the \antiordered/ systems, \ordered/ systems in synthetically observed catalogues tend to be rich in dry planets. These systems are distinct not only because of their frequent dry planets, but also due to a low frequency of wet planets. For all synthetic catalogues, moist planets occur more frequency than wet planets, which is a unique distinguishing feature for these systems. For the Bern model, these systems have low average planets per star: 5 around low-metallicity stars and 3.1 around high-metallicity stars. 
        
        In summary, we note some salient features of these system architectures. Generally, wet planets  survive more frequently around high-metallicity stars. One detection technique that favours the discovery of close-in planets also favours the detection of dry planets. The comparative frequency of planet (dry, wet, or moist) per star seems to be intimately connected with the mass architecture of the system. Similar and \mixed/ systems can host lots of dry or wet planets, depending on the metallicity of the systems and detection technique. Anti-ordered systems, forming prominently via the \edit{Aryabhata formation scenario}, are rich in wet planets. Ordered systems, in simulated observations, are rich in dry planets and have more moist planets than wet planets. The physical connection between the average planet per star and the star's metallicity is sensitive to the formation pathways that a system undergoes.

\section{Habitability as a function of system architecture}
\label{sec:discussion}

In this paper thus far, we have described a new framework for studying the architecture of planetary systems (Sect. \ref{sec:framework}), the characteristics of the four classes of system architecture (Sect. \ref{sec:architecturetypes}), and the relation between the mass architecture of a system and its internal structure and composition architecture (Sect. \ref{sec:internalcomposition}). In this section, we speculate on the idea of studying habitability as a function of system-level architecture.
        

        \def\figwidthz{4.5cm}
        \begin{figure*}
        \centering
        \includegraphics[height=\figwidthz]{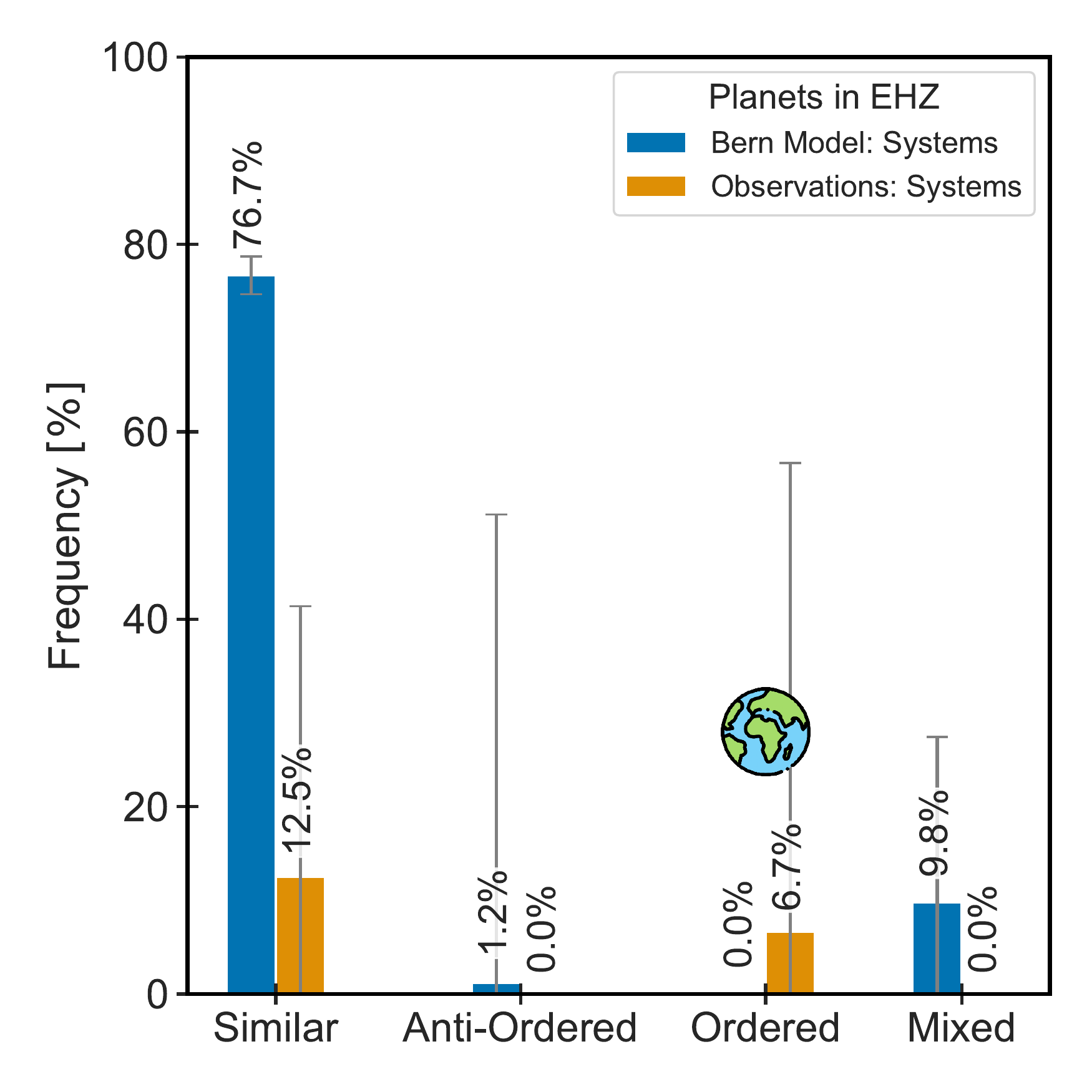}
        \includegraphics[height=\figwidthz]{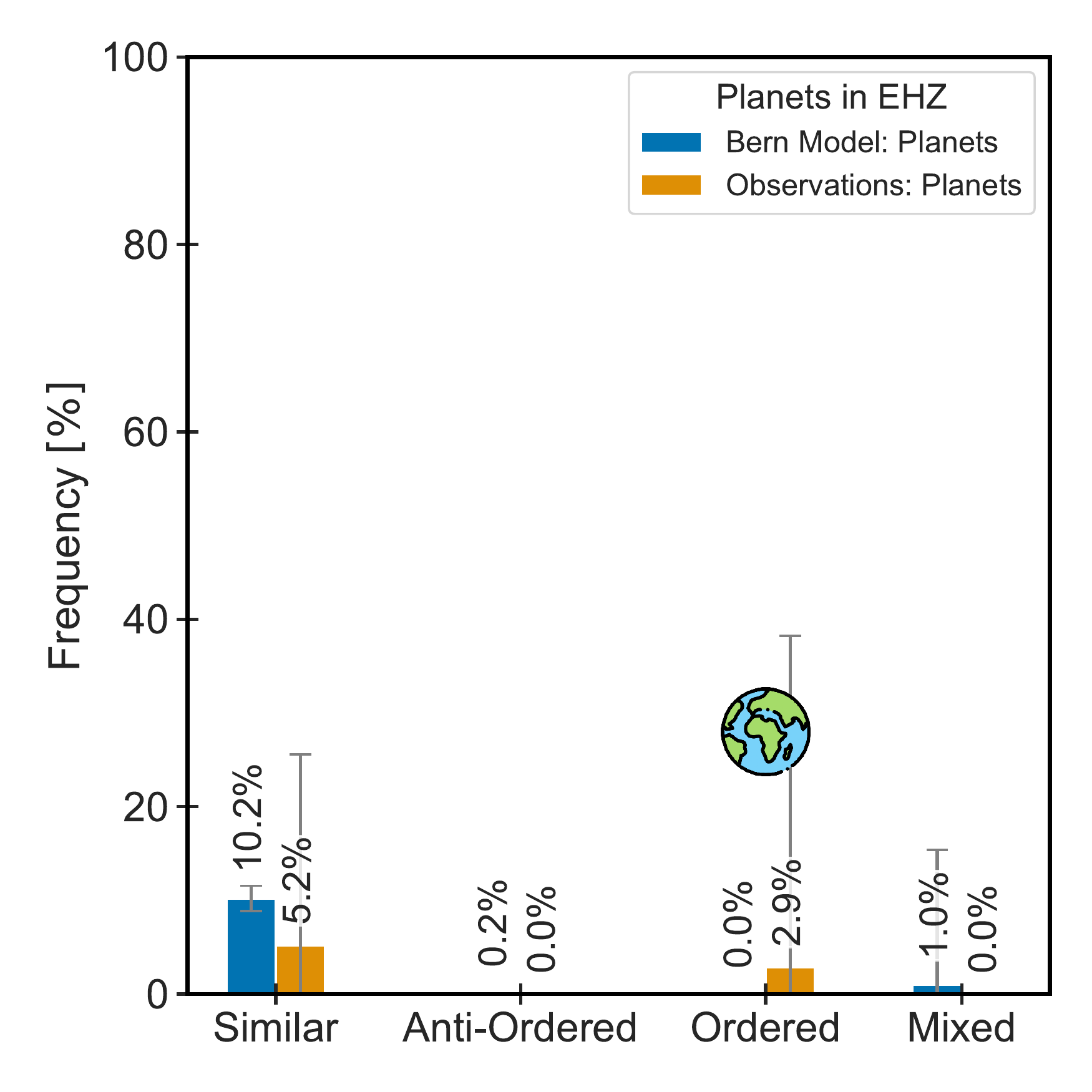}
        \includegraphics[height=\figwidthz]{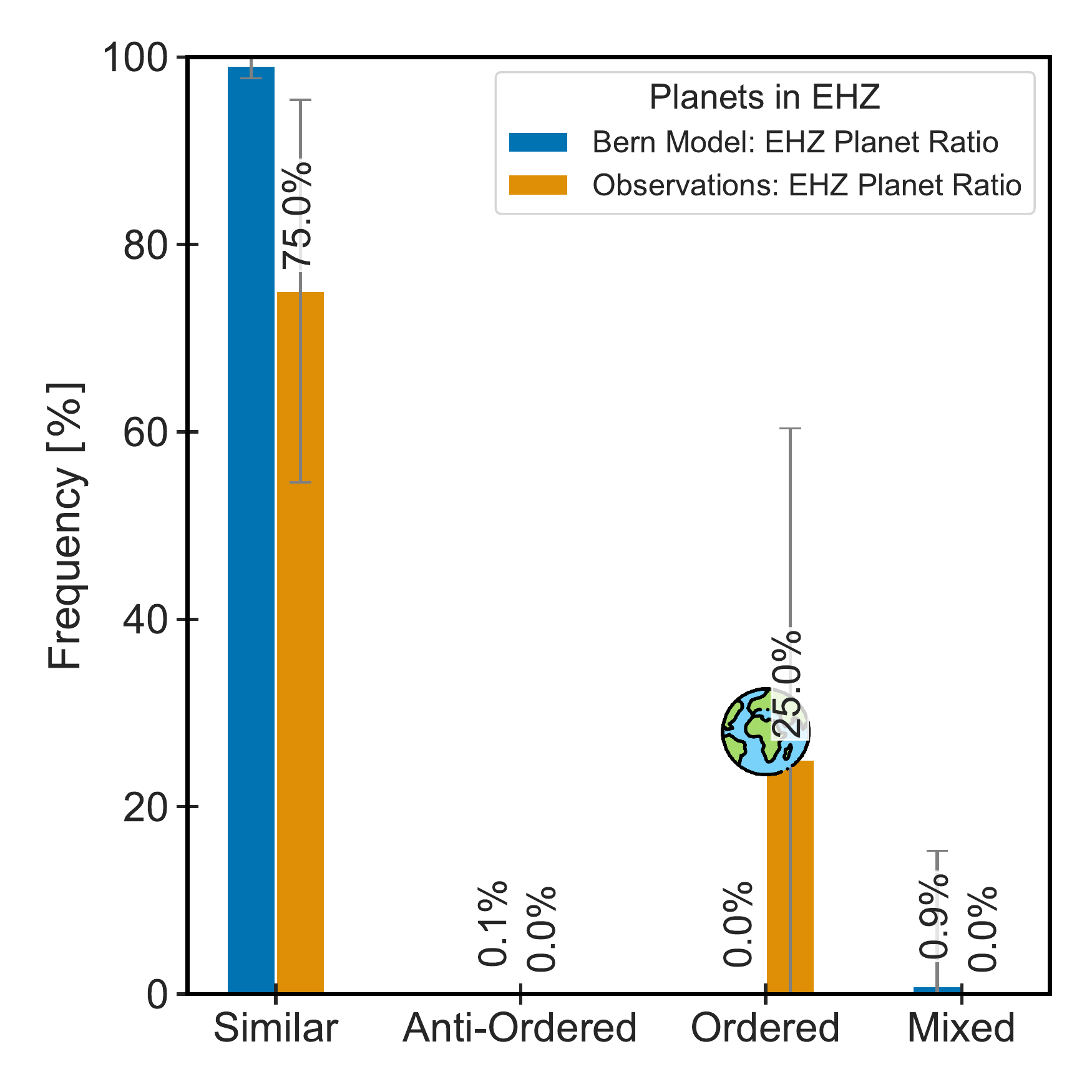}
        \includegraphics[trim=0cm 0.5cm 0cm 0cm,clip,height=\figwidthz]{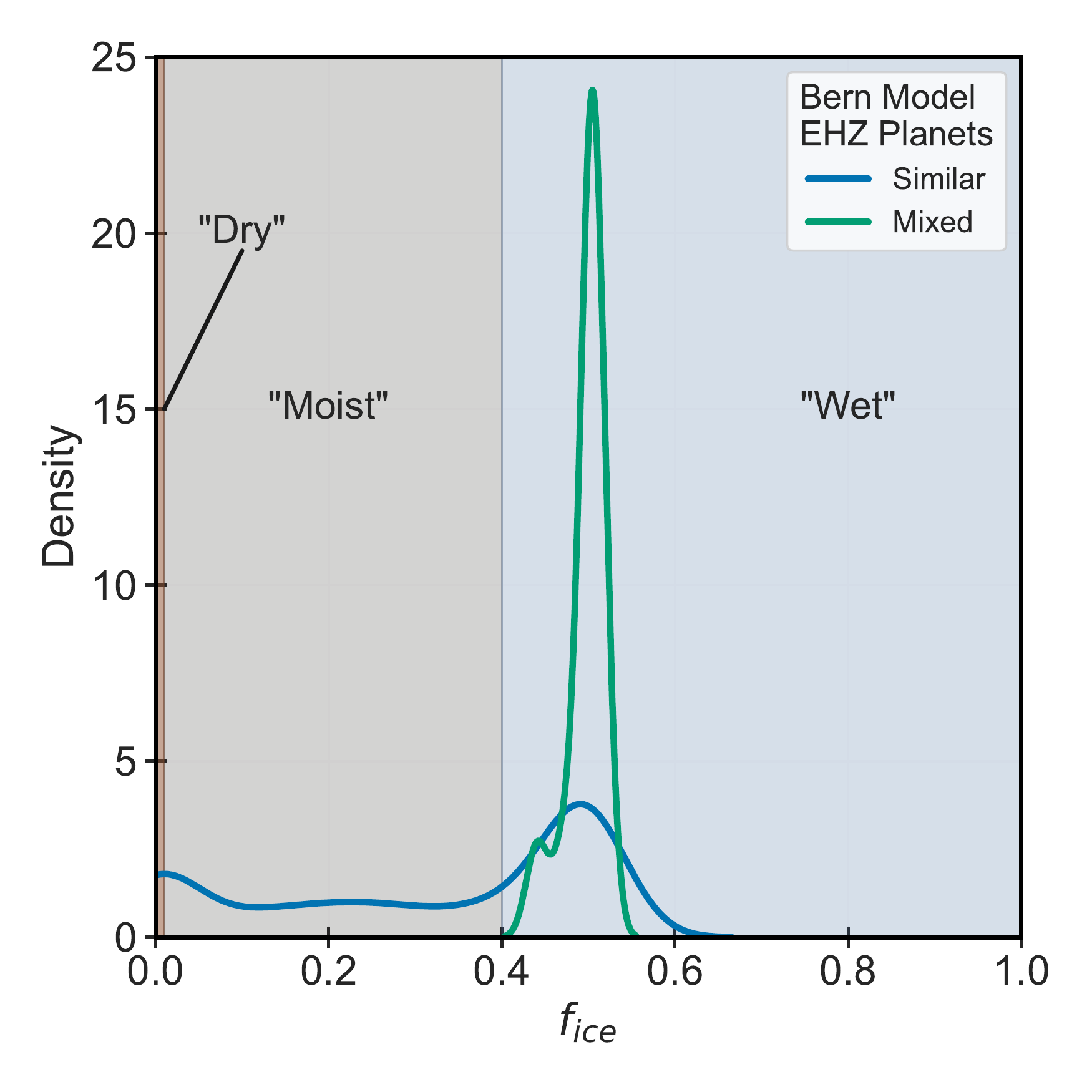}
        \caption{\edit{Planets inside the empirical habitable zone (EHZ). The left-most plot shows the frequency of planetary systems, of a given architecture class, which host at least one planet inside the EHZ. The central-left plot shows the fraction of planets inside a given architecture class which are in the EHZ. The central-right plot shows the fraction of all EHZ planets within a given architecture class. The rightmost plot shows the distribution of $f_\text{ice}$ for EHZ planets across the architecture classes.} The cartoon sketch of Earth emphasises that the only known life-harbouring planet resides in an \ordered/ system. The length of error bars visualises the total number of systems or planets in respective bin as: ${100}/{\sqrt{\mathrm{bin\ counts}}}$. \change{The lengths of the error bars represents the number of planetary systems (left-most panel) and the number of planets (two middle panels) which are inside the bin. Large error bars in the leftmost panel, for example for \antiordered/ architecture emerges from their low count (see Fig. \ref{fig:archfreq})}. \edit{The \change{Gaussian} kernel is estimated using Scott's rule \citep{2015mdet.book.....S}}}
        \label{fig:habitability}        
        \end{figure*}

        Mankind has pondered the existence of other biotic life-forms beyond Earth, as well as outside our own Solar System. Our current understanding of habitability stems from and is focused at an individual planetary level. We consider whether habitability could be correlated with other properties of a planetary system, namely, whether  habitability could be a system-level phenomenon. In this section, we speculate on the role of planetary-system level information on the existence of habitable worlds in such systems.  The framework we present here for studying the system-level architecture
of a planetary system brings to light several novel questions, probing the dependence of habitability and occurrence of habitable worlds (and related concepts) on the architecture of a said system. For example, we wonder how the occurrence rate of habitable planets in the galaxy depends on the occurrence of the four architecture classes. 
        
        In this section, we address this question on \edit{three levels: system, planet, and planet ratio}. We use the concept of empirical Habitable zone (EHZ) planets from \cite{2021arXiv210107500Q, Kopparapu2014}. Planets with masses between $[0.1, 5] \mearth$ and stellar insolation within $[1.776, 0.32] S_\oplus$ are considered to be inside the EHZ. The stellar flux limits correspond to `recent Venus' and `early Mars' scenarios and include the luminosity evolution for a $1 M_\odot$ Solar-twin. At the system level, we note the frequency of systems of a particular architecture to host at least one planet in the EHZ. At the planet level, we count the frequency of planets in the EHZ across each system architecture class. \edit{At the planet ratio level, we show the fraction of all EHZ planets across their architecture class}. Figure \ref{fig:habitability} shows the frequency of EHZ planets\edit{, at all three levels,} as a function of their system architecture for both synthetic and observed exoplanetary systems. 

        Out of all synthetic systems with a \similar/ class architecture, $\approx 77\%$ host at least one EHZ planet. This is remarkably higher than any other architecture class. $\approx 10\%$ of systems with \mixed/ architecture host at least one EHZ planet. The frequency drops to $\approx 1\%$ for \antiordered/ architecture systems and $\approx 0\%$ for \ordered/ systems. One way to interpret these numbers could be to look at the multiplicity distribution across each architecture class in Fig. \ref{fig:architecturecharacteristics}. The frequency of at least one EHZ planets across architecture class seems to follow the multiplicity trends. Similar and \mixed/ architectures have comparably high number of planets. The distribution of the Aryabhata's number shows that \similar/ systems usually have higher Aryabhata's number than \mixed/ systems, implying that \similar/ systems tend to host more planets which started from inside the ice line (see \papertwo/ for Aryabhata's number). This may account for the large frequency of \similar/ systems which host at least one EHZ planet. The multiplicity distribution shows that \antiordered/ systems often host less planets than \similar/ and \mixed/ class systems, while \ordered/ systems have the lowest multiplicities. We see in Sect. \ref{subsec:frequency} that the \similar/ class architecture is perhaps the most common architecture for planetary systems in our galaxy. These results \edit{from the Bern model simulations} suggest that observation campaigns to detect habitable planets will find more EHZ planets in \similar/ class architectures.  
        
        For the observed multi-planetary systems in our catalogue, about $\approx 13\%$ of similar class systems have at least one EHZ planet. About  7\% of ordered class exoplanetary systems in our catalogue host at least one EHZ planet. In the mixed class observed systems in our catalogue, none of them have EHZ planets and there are no known anti-ordered class systems in our catalogue. These frequencies are quite different from their theoretical counterparts. While the lack of a complete and reliable observations catalogue may explain the discrepancy for similar class systems -- it does not completely explain the discrepancy for ordered systems. Our own planet resides in the ordered class system of the Solar System, which is not supposed to be influenced by issues such as completeness or detection biases. This reflects the inability of Bern models to simulate a Solar System analogue -- pointing to a gap in our understanding of the physics that goes into planetary formation and evolution. In addition, many observed ordered class systems may have a different architecture when more planets in these systems are detected. 
        
        At the planet level in our simulations, out of all synthetic planets that exist in \similar/ class systems, about 10\% are inside the EHZ. This frequency is, again, remarkably higher for any other architecture class. About 1\% of all simulated planets in a \mixed/ system are inside the EHZ. Close to 0\% of all planets in \antiordered/ and \ordered/ class architectures are inside the EHZ. From our observational catalogue, while 5\% of observed exoplanets in \similar/ class systems are inside the EHZ. About 3\% of observed exoplanets in \ordered/ class systems are inside the EHZ. 
        
        \edit{The planet ratio level shows the fraction of all EHZ planet that belong to a particular architecture class. In the Bern model, we see that out of all EHZ planets, about 99\% are in the \similar/ class. The share of EHZ planets by other architecture classes is negligible. Amongst the observations, three-quarters of EHZ planets are in \similar/ class and the remaining are in \ordered/ class. The observations and theory are quite misaligned in this scenario. We attribute this discrepancy to the absence of a complete and reliable  catalogue of observations.}
        
        Our observations catalogue has only 41 multi-planetary systems, of which only four host planets inside the EHZ. These systems are Trappist-1 (three planets in EHZ), GJ 667 C (two planets in EHZ), Solar System (two planets in EHZ), and Tau Ceti (one planet in EHZ). The occurrence of architecture classes and the frequency with which they host EHZ planets might be better constrained with future observations. This may allow us to have a better estimate of the occurrence rate of EHZ planets as a function of architecture class. 
 
        Simulations suggest that \ordered/ architecture is a rare outcome of planet formation (about 1.5\% of systems out of 1000 were deemed to be \ordered/) and yet, we live in an \ordered/ system. These two statements can shed new light on the rarity of life in the galaxy. We foresee that the famous Drake equation may be suitably modified to take into account the occurrence rate of different architectures and thereby set more optimal constraints on $\eta_\oplus$ \citep{Sarkar2022}.
        
        \edit{Since water plays a fundamental role for life forms on Earth, it is interesting to probe the core water-ice fraction for the EHZ planets. Figure \ref{fig:habitability} also shows the $f_\text{ice}$ distribution for EHZ planets in the Bern model. As we see before, most of the EHZ planets are in the \similar/ class and $\approx 1\%$ of EHZ planets are in the \mixed/ class. EHZ planets in \similar/ systems are `dry', '`moist', and `wet'. In stark contrast, EHZ planets in \mixed/ class are only `wet' planets. We hope these results may be useful in guiding future missions in finding  EHZ planets that have the potential to harbour life.}
        
        \section{Summary, conclusions, and future work}
        \label{sec:conclusion}  
        
        In this paper, we introduce and explore a new framework for studying the architecture of planetary systems. Our new framework allows us to study, quantify, classify, the global architecture of an entire planetary system at the system-level; and compare the architecture of one planetary system with another. In Sect. \ref{sec:framework}, we detailed the new architecture framework and presented an in-depth discussion comparing our framework with other works in the literature. We present the \cstext/ and the \cvtext/ as two quantities that quantify our conceptual ideas. Our framework gives rise to a new parameter space (the $\cs$ versus $\cv$ plane) in which individual planetary systems can be compared with one another. Throughout this paper, we applied this framework to study the distribution and arrangement of several planetary quantities within a planetary system, thereby understanding the system architecture for that quantity. In this manner, we studied the mass architecture, the radius architecture,  the core mass architecture, the core water mass fraction architecture, and the density architecture of synthetic and observed planetary systems. 
        
        To study some consequences of this framework, we applied our method to several catalogues of planetary systems (introduced in Sect. \ref{sec:catalogues}). We curated, especially for the purposes of this study, a catalogue of observed multi-planetary systems that have four or more planets and include mass measurements for at least four planets. For engendering further studies, additional stellar and planetary properties were collected and presented in Table \ref{table:1}. We also used synthetic planetary systems simulated via the Bern model. To facilitate a comparison of theory with observations, we prepared three synthetic observed catalogues by applying the detection biases on the simulated planetary systems. This led to the Bern RV Multis, Bern KOBE Multis, and the Bern Compact Multis catalogue. We note that there are caveats present in the datasets we  used. The model-dependent results we present here may be improved upon in future studies using better theoretical models and a more complete observational catalogues (e.g. from PLATO). 
        
        \edit{Summary of architecture framework:}
        \begin{enumerate}
            \item \change{The architecture framework is model-independent and therefore does not suffer from any caveats emerging from planet formation theory or observations.} 
            
            \item \change{The same architecture framework can be used to study the multi-faceted aspects of planetary system architecture. When the framework is applied to study planetary masses, the framework informs us of the mass architecture of the system, namely, the arrangement and distribution of masses in the planetary system. In this way, we can use this framework to study the mass architecture, radii architecture, eccentricity architecture, and so on  for the same system. In this series of work, we identified the architecture of a system with its bulk mass architecture. }
        
                \item Planetary system architecture can be one of four classes that are derived from our framework: \similar/, \mixed/, \ordered/, and \antiordered/.
                
                \item A planetary system's architecture is of \similar/ class when the masses of all the planets within such a system are similar to each other. This architecture class corresponds to the `peas in a pod' architecture trend reported in the literature.
                
                \item The architecture class of a planetary system is \ordered/ (or \antiordered/) when the planetary masses in such systems tend to increase or decrease from inside-out. 
                
                \item Planetary systems of \mixed/ class architecture host planets whose masses show broad increasing and decreasing variations.
        \end{enumerate}

        \edit{Our key model-dependent findings are as follows:}
        \begin{enumerate}
                \item \textbf{Frequency of architecture class:} \change{Systems with \similar/ bulk mass architecture are the most common outcome of simulations}, followed by the other three architecture classes. \edit{Our model} suggests that \similar/ architecture should be the most common exoplanetary system architecture in our Galaxy and beyond. This explains why radius similarity in exoplanets was already detected from the first four months of Kepler data \citep{Lissauer2011}.
                
                \item \textbf{Distance bi-modality:} We found hints of a bi-modality in the exoplanetary distance distribution arising from the two different modes of orbital migrations. This bi-modality is \change{readily visible (see Fig. \ref{fig:architecturecharacteristics})} for \similar/ and \mixed/ mass architecture exoplanetary systems observed via RV.
                
                \item \textbf{Core mass architectures:} We found that for most systems, the \change{bulk} mass architecture is inherited from the core mass architecture. In addition, the accretion of gases tends to highlight the underlying core mass architecture by amplifying it. In this way, the observed mass architecture of a system could serve as a gateway for studying the distribution and arrangement of the planetary core masses, which tends to be simpler for theoretical modelling. 
                
                \item \edit{\textbf{In situ formation:} We found that most planets belonging to the \similar/ \change{bulk mass architecture} class form in situ inside the ice line. In contrast, planets inside the ice line belonging to \mixed/, \antiordered/, and \ordered/ show large inward migrations.}
                
                \item \textbf{Core water-ice mass fraction architectures:} Synthetic planetary systems were found to have two scenarios for their core water mass fraction architecture. The default scenario consists of relatively more dry planets in the inner parts of a system and more wet planets in the outer parts of the system. This is probably a direct consequence of the starting location of planets: planets starting inside (or outside) the ice line tend to be dry (or wet). About one-fifths of simulates systems do not follow the default scenario described above. We propose the `Aryabhata formation scenario' to explain their core-water mass fraction architecture (see \papertwo/).
        
                \item \textbf{Linking architecture and internal composition:} We found that wet planets are more likely to survive around high-metallicity stars. Among other predictions, we showed that \antiordered/ observed systems should be rich in wet worlds, while \ordered/ observed systems are expected to have many dry planets (based on the core-accretion planet formation paradigm).
                
                \item \textbf{Density Architectures:} Synthetic systems that are \ordered/ (or \antiordered/) in mass tend to also be \ordered/ (or \antiordered/) in their bulk densities. Some mass \similar/ systems may also have low dispersion in their planetary bulk densities. The density architecture is sensitive to the Aryabhata's number (i.e. the starting location of various surviving planets; see \papertwo/). The density architecture of observed systems is in good agreement with the density architecture of synthetically observed simulated systems. Detection biases seem to favour the discovery of planetary systems where the densities show anti-ordering, mixing, or similarity. 
                
                \item \textbf{Radius architectures:} The radius architecture of most planetary systems closely follows their mass architecture. Therefore, most mass similar systems also show similarity in radius (also for mass mixed, ordered, or anti-ordered systems). However, this is not always true. Future studies can calibrate a classification scheme based on planetary radii.  
                
                \item \textbf{Habitability as a system-level phenomenon:} We reflected on the prospect of studying habitability as a function of system-level properties such as system architecture. Similar architecture systems represent an excellent observation target for finding life outside the solar systems because these systems tend to host many more planets inside the empirical habitable zone that other architecture classes.
                
                \item The current version of the Bern model seems to have difficulty in producing planets inside the EHZ of an \ordered/ architecture system. Nevertheless, more data is required to conclude whether the existence of Earth, an inhabited planet in an \ordered/ system, is an exception or whether there are additional gaps in our understanding of planet formation. 
        \end{enumerate}
        
        This paper is the first in a series. The current work presents a new testing ground, the architecture space, for theoretical models and for comparing observations with theory. We can now constrain our understanding of planet formation not only on the level of an individual planet -- but at the global systemic level. This is a multi-faceted approach, since the system architecture of several quantities can now be uniformly assessed and compared with observations. In our next paper (\papertwo/), we show another important aspect emerging from this architecture framework which asserts that systems with comparable architecture often have the same formation pathways. We present ideas to further the nature versus nurture debate around planet formation. While \similar/ architectures are usually a product of their starting conditions, stochastic multi-body effects are responsible for shaping the other three architecture classes. This work leads to several future studies which will be presented in other papers in this series. Davoult et al. (in prep.) explore how the present architecture framework can be employed for an efficient usage of telescope time to hunt for habitable worlds. Other possible explorations that emerge from this work include: (a) a data-driven approach to classifying planetary architecture based on radii and (b) a suitable modification to Drake's equation that accounts for the empirical occurrence rate of system architectures.

        \begin{acknowledgements}
                The authors thank the anonymous referee for their careful reading, constructive suggestions, and insightful questions, which has allowed the quality of this manuscript to be improved.
                This work has been carried out within the framework of the National Centre of Competence in Research PlanetS supported by the Swiss National Science Foundation under grants $51NF40\_182901$ and $51NF40\_205606$. The authors acknowledge the financial support of the SNSF.
                \textit{Data:} The synthetic planetary populations (NGPPS) used in this work are available online at \url{http://dace.unige.ch} under section ``Formation \& evolution''. This research has made use of the NASA Exoplanet Archive, which is operated by the California Institute of Technology, under contract with the National Aeronautics and Space Administration under the Exoplanet Exploration Program: \url{https://exoplanetarchive.ipac.caltech.edu} (DOI: 10.26133/NEA6). The artwork used to depict Earth in Fig. \ref{fig:habitability} is taken from flaticon.com.
                \textit{Software:} \kobe/ \citep{Mishra2021, 2021ascl.soft06001M}, Python \citep{python3}, NumPy \citep{numpy}, SciPy \citep{scipy}, Seaborn \citep{seaborn}, Pandas \citep{pandas}, Matplotlib \citep{matplotlib}.
                
        \end{acknowledgements}
        
\bibliographystyle{aa}
\bibliography{exoplanets,mps_bibliography}
        
        \begin{appendix}
        
        \section{Bern Model: Additional details}
                \label{sec:bernmodel}
                
                In this section, we provide some additional details on the physics included in the Bern model and how it is utilised to simulate synthetic planetary systems. Finally, we give an overview on comparisons between the output of the Bern Model and observed planetary systems. \change{For the historic development, we refer to  \cite{Alibert2004, Alibert2005, Mordasini2009, 2011A&A...526A..63A, Mordasini2012(MR),Mordasini2012(models), Alibert2013, Fortier2013, Marboeuf2014a, Thiabaud2014, Dittkrist2014, 2014ApJ...795...65J} and reviews in \cite{Benz2014, Mordasini2018}.}
                
                \change{The Bern model is based on the core accretion paradigm of planetary formation \citep{Pollack1996}}. The model includes stellar evolution for a solar-mass star, using evolution tracks from \cite{Baraffe2015}. The star interacts with the protoplanetary disk and influences its thermodynamical properties. The protoplanetary disk has two phases: gas and solid. We model this disk using the approaches of viscous angular momentum transport \citep{Lynden-Bell1974, Veras2004, Hueso2005}. Turbulence is characterised by the \citet{Shakura1973} approach, with $\alpha = 2 \times 10^-3$. Gas from the disk is accreted by planets, host star, and lost via photo-evaporation. The 1D geometrically thin disk evolution is studied up to 1000 au. The initial mass of this gas disk and its lifetime are randomly drawn for each run of the simulation.        The solid phase of the disk is composed of a swarm of planetesimals. The solid disk is modelled as a fluid which evolves via (a) accretion by growing planets; (b) interaction with the gaseous disk; (c) dynamical stirring from planets and other planetesimals; and so on \citep{Fortier2013}. The initial mass of the solid disk depends on the metallicity of the star and also on the condensation state of the molecules in the disk \citep{Thiabaud2014}. The host star metallicity is randomly drawn for each run of the simulation. 
                
                We added 100 protoplanetary embryos to the protoplanetary disk. The initial location of each embryo was varied from one simulation to another. It was also ensured that no two embryos start within 10 hill radii of each other \citep{Kokubo1998, Kokubo2002}. Embryos accrete from their feeding zones and any overlap may lead to competition \citep{Alibert2013}. The accretion rate depends on the collision probability between a protoplanet and a planetesimal, which in turn is influenced by the dynamical state of the solid disk. 
                
                Gas accretion occurs in several phases \citep{Mordasini2012(models)}. Initially, the gas disk smoothly transitions as a gaseous envelope around all planets -- the attached phase. For planets that are massive enough to undergo runaway gas accretion, the rate of gas supply from the disk may not be enough. In these scenarios, the planet detaches from the gas disk and rapidly contracts to $R_J$. After the gas disk dissipates, all planets are in the isolated phase. Gas accretion from the disk is no longer possible and in this phase, the planets contract and cool. For all the planets, their internal structure is modelled at each time step. We assume planets are spherically symmetric and composed of accreted materials that arranges itself in layers: iron code, silicate mantle, water ice, and H/He gaseous envelope (if accreted). 

        Next, we use these recipes to simulate several thousands of planetary systems in an approach called population synthesis \citep{Emsenhuber2021B}. We start 1000 star-disk-embryo systems with some fixed as well as some randomly drawn properties. The initial properties are inspired by observations of disks \cite{Tychoniec2018}. The then numerically modelled these systems, endowing them with additional physics at the same time. Numerically, we incorporated multi-body dynamical interactions via \textit{N}-body simulations. Planet-disk interactions leading to orbital migration and eccentricity and inclination damping were also incorporated in the \textit{N}-body \cite{Coleman2014, Paardekooper2011, Dittkrist2014}. We followed these numerically intensive steps for 20 Myrs and then stopped the \textit{N}-body calculations. The model then continued to evaluate the internal structure of all planets in the system for 10 Gyrs. 
        
        The recent version of these simulations has been published in the New Generation Planetary Population Synthesis (NGPPS) series of papers \citep{Emsenhuber2021A, Emsenhuber2021B, Schlecker2021a, Schlecker2021b, Burn2021, Mishra2021}.  The output of these models have been compared with observations in several works. \cite{2022arXiv220309759D} compares the occurrence rates of synthetic systems with observations. \cite{Schlecker2021a} studies the warm Super Earth and cold Jupiter correlation in the synthetic systems. \cite{Mishra2021} analyse the 'peas in a pod' architecture and compare synthetic systems with observations from \cite{Weiss2018}. \cite{Mulders2018} present a detailed comparison of the synthetic models with Kepler observations. 
        
        \section{Stellar and planetary data references}
                \label{sec:starplanetreferences}
        
                \begin{enumerate}
\item \object{Sun}: \cite{Archinal2018, Standish_JPL,2018Icar..299..460W,2017PEPS....4...24H,2006AJ....132.2520J,2014AJ....148...76J,2009AJ....137.4322J}
\item \object{Trappist-1}: \cite{2021PSJ.....2....1A, 2017Natur.542..456G, 2017ApJ...845..110B, 2018A&A...613A..68G}
\item \object{TOI-178}: \cite{2021A&A...649A..26L}
\item \object{HD 10180}: \cite{2011A&A...528A.112L, 2014ApJ...792..111K}
\item \object{HD 219134}: \cite{2021AJ....161..117S, 2020A&A...635A...6B,2015ApJ...814...12V}
\item \object{HD 34445}: \cite{Vogt2017}
\item \object{Kepler-11}: \cite{Berger2020, Lissauer2013}
\item \object{Kepler-20}: \cite{Fressin2011, Buchhave2016}
\item \object{Kepler-80}: \cite{MacDonald2016, Shallue2017}
\item \object{K2-138}: \cite{2019A&A...631A..90L}
\item \object{55 Cnc}: \cite{Bourrier2018}
\item \object{GJ 667 C}: \cite{Anglada-Escude2013}
\item \object{HD 158259}: \cite{Hara2020, Gaspar2016}
\item \object{HD 40307}: \cite{2016A&A...585A.134D, 2019AJ....158..138S}
\item \object{Kepler-102}: \cite{Berger2020, 2014ApJS..210...20M}
\item \object{Kepler-33}: \cite{Berger2020, 2012ApJ...750..112L, 2017AJ....154....5H}
\item \object{Kepler-62}: \cite{Berger2020, 2013Sci...340..587B}
\item \object{HD 20781}: \cite{2019A&A...622A..37U}
\item \object{TOI-561}: \cite{2021MNRAS.501.4148L, 2021AJ....161...56W}
\item \object{DMPP-1}: \cite{2020NatAs...4..399S}
\item \object{GJ 3293}: \cite{2017A&A...602A..88A}
\item \object{GJ 676 A}: \cite{2016A&A...595A..77S,2017AJ....153..136S}
\item \object{GJ 876}: \cite{2018A&A...609A.117T, 2012ApJ...748...93R}
\item \object{HD 141399}: \cite{2016A&A...588A.145H}
\item \object{HD 160691}: \cite{2007ApJ...657..546G, 2007A&A...462..769P}
\item \object{HD 20794}: \cite{2007ApJ...657..546G, 2007A&A...462..769P}
\item \object{HD 215152}: \cite{2007ApJ...657..546G, 2007A&A...462..769P}
\item \object{HR 8799}: \cite{2008Sci...322.1348M, 2019A&A...623L..11G, 2021AJ....161..114S}
\item \object{K2-266}: \cite{2018AJ....156..245R}
\item \object{K2-285}: \cite{2018AJ....156..245R}
\item \object{Kepler-89}: \cite{Berger2020, 2013ApJ...768...14W}
\item \object{Kepler-106}: \cite{Berger2020, 2014ApJS..210...20M}
\item \object{Kepler-107}: \cite{Berger2020, 2019NatAs...3..416B}
\item \object{Kepler-223}: \cite{Berger2020, 2016Natur.533..509M}
\item \object{Kepler-411}: \cite{Berger2020, 2019A&A...624A..15S}
\item \object{Kepler-48}: \cite{Berger2020, 2014ApJS..210...20M}
\item \object{Kepler-65}: \cite{Berger2020, 2019AJ....157..145M}
\item \object{Kepler-79}: \cite{Berger2020, 2021ApJ...908..114Y}
\item \object{WASP-47}: \cite{2017AJ....154..237V}
\item \object{tau Cet}: \cite{2017AJ....154..237V}
\item \object{HD 164922}: \cite{2020A&A...639A..50B, 2021arXiv210511583R}
                \end{enumerate}
                
        \section{Derivation of limits}
                \label{sec:limits}
                \begin{figure}
                \resizebox{\hsize}{!}{\includegraphics{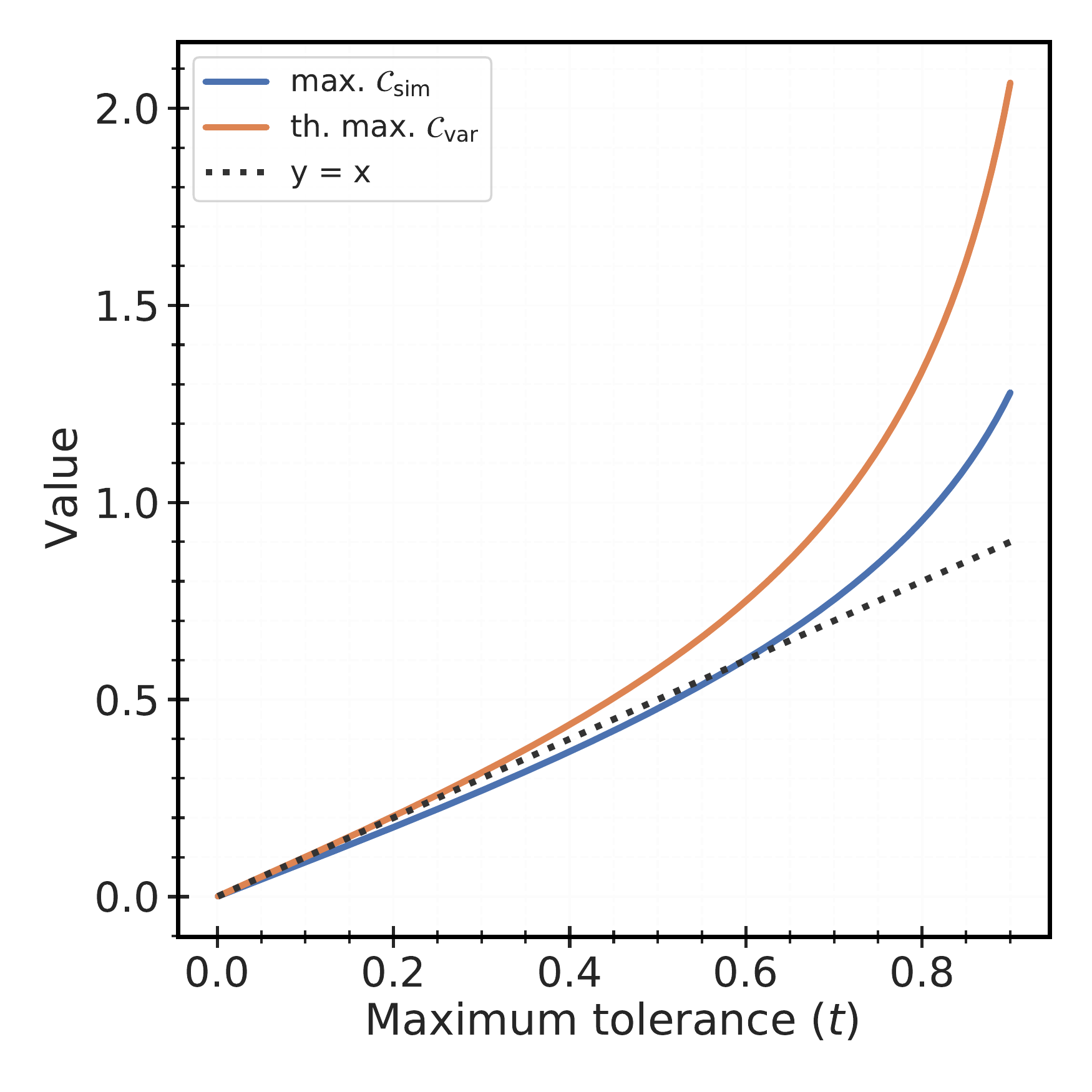}}
                \caption{Maximum value of the \cstext/ (blue) and the theoretical maximum value of the \cvtext/ (orange) is plotted against the maximum tolerance, $t$.}
                \label{fig:cscvlimits}
                \end{figure}
We consider a set $\mathcal{Q}$ of quantities $q$, namely, $\mathcal{Q} = \{q_i\}$ where $q_i$ could be the mass, radius or other parameter of a planet, and the index, $i \in [1,\nplanet],$ identifies a planet (with 1 being the innermost planet). We assume that all $q_i \ge 0$. The quantities $q_i$ are expressed as:
\begin{equation}
\label{eq:qi_decomposition_appendix}
        q_i = q' \ (1 \pm t_i).
\end{equation}
The quantities, $q_i$, are decomposed around some value $q'$ such that all $t_i$ are minimised; $t_i$ is a measurement of the fractional difference (or tolerance) between $q'$ and $q_i$. Since all individual tolerances are a positive quantity, they will satisfy the following relation:
\begin{equation}
        0 \le t_i \le t.
\end{equation}

\subsection{Mean}
Let us consider the mean of the quantities, $\bar{q_i}$:

\begin{equation}
\begin{split}
        \bar{q_i}       &= \frac{Q}{\nplanet}  \ = \frac{\sum_i q_i}{\nplanet}\\
                                &= \frac{q'}{\nplanet} \big(\nplanet \pm t_1 \pm t_2 \pm \dots  \pm t_\nplanet \big)
\end{split}
.\end{equation}
The mean takes its maximum value only when all individual $t_i$ values take their maximum  and are added up. This gives:
\begin{equation}
\begin{split}
        \max \bar{q_i} = q' \ \big(1 + t\big)
\end{split}
.\end{equation}
Similarly, the minimum value of the mean is:
\begin{equation}
\begin{split}
\min \bar{q_i} = q' \ \big(1 - t\big).
\end{split}
\end{equation}

The extreme value of the mean occurs when all the individual quantities are extremised. However, in this scenario, since all quantities are equal, the \cvtext/ is identically 0.

\subsection{Coefficient of similarity}

We start with the definition of the \cstext/,
        \begin{equation}
                \cs (q) = \frac{1}{\nplanet - 1}  \ \sum_{i=1}^{i=\nplanet - 1} \  \Bigg(log \ \frac{q_{i+1}}{q_i} \ \Bigg).
        \end{equation}
Inserting Eq. \ref{eq:qi_decomposition_appendix}, in the definition, we get:
\begin{equation}
        \cs (q) = \frac{1}{\nplanet - 1}  \ \sum_{i=1}^{i=\nplanet - 1} \  \Bigg(log \ \frac{1 \pm t_{i+1}}{1 \pm t_i} \ \Bigg).
\end{equation}
This formulation shows that the \cstext/ depends only on the fractional differences (tolerances) between $q_i$ values -- and not on their actual values. This is a desirable property. 
 
Next, we evaluate the $\max \cs$ as, 
\begin{equation}
\begin{split}
 \max \cs (q) &= \max \Bigg[ \frac{1}{\nplanet - 1}  \ \sum_{i=1}^{i=\nplanet - 1} \  \Bigg(log \ \frac{1 \pm t_{i+1}}{1 \pm t_i} \ \Bigg)  \Bigg],
                        \\&= \frac{1}{\nplanet - 1} \max \Bigg[ \sum_{i=1}^{i=\nplanet - 1} \  \Bigg(log \ \frac{1 \pm t_{i+1}}{1 \pm t_i} \ \Bigg) \Bigg],
                        \\&= \frac{1}{\nplanet - 1} \sum_{i=1}^{i=\nplanet - 1} \ log \   \max \Bigg[ \frac{1 \pm t_{i+1}}{1 \pm t_i} \Bigg].
\end{split}
\end{equation}
In the first step, we commuted the $\max$ operator with the fraction $(\nplanet - 1)^{-1}$ because we are interested in the maximum for a constant $\nplanet$.  Next, knowing that the maximum of a sum occurs at the sum of maximum summands and that $log$ is a monotonically increasing function, we further commute the $\max$ operator.

We observe the following:\ 
\begin{equation}
        \max \Bigg[ \frac{1 \pm t_{i+1}}{1 \pm t_i} \Bigg] \quad \quad \text{when} \quad 
        \left\{
        \begin{array}{lr}
        \pm t_{i+1} &\rightarrow + t
        \\      
        \pm t_{i}       &\rightarrow - t
        \end{array}
        \right\}.
\end{equation}

This implies that
\begin{equation}
\label{eq:cslimit}
\begin{split}
        \max \ \cs (q) &= log \ \frac{1 + t}{1 - t},
        \\      
        \min \ \cs (q) &= - \max \ \cs = log \ \frac{1 - t}{1 + t},
\end{split}
\end{equation}
where the second equality can be similarly derived. Fig. \ref{fig:cscvlimits} shows the variation of $\max \ \cs$ as a function of tolerance $t$. We note that the limits of the \cstext/ do not depend on $\nplanet$, and we verified our results with numerical simulations. $\blacksquare$

\subsection{Coefficient of Variation}

We start with the definition of the \cvtext/,
\begin{equation}
        \cv (q) = \frac{\sigma(q)}{\bar{q}}
,\end{equation}
and we note that the minimum value of the \cvtext/ is zero and it occurs when all $q_i$ values are equal, thereby giving no variance. 

In the literature, we can find some derivations for the maximum value of the \cvtext/ \citep{Katsnelson1957,Sharma2010}. \cite{Katsnelson1957} show that the upper limit of the \cvtext/ is $\sqrt{\nplanet - 1}$ when all but one $q_i$ is zero. However, this limit is only a particular case of our formulation (specifically, $q_1 = q'$ and $q_{i \ne 1} = 0$). Here, we derive the limits for a more general scenario.

We consider that: 
\begin{equation}
        \cv^2  = \frac{1}{\nplanet} \ \sum_{i=1}^{i=\nplanet} \ \bigg(\underbrace{1 - \frac{q_i}{\bar{q}}}_{= A} \bigg)^2.
\end{equation}
Here, we have squared the definition of $\cv$ and used the definition of the standard deviation $\sigma(q)$. As an aside, we note that the equation above shows that the \cvtext/ is zero when all $q_i = \bar{q}$, as noted before. We note that the maximum value of $\cv^2$ occurs when the term $A$ (in parenthesis) is maximised. Denoting $\sum_{i=1} q_i$ by $Q$, we consider the term in the parenthesis,
\begin{equation}
        A = 1 - \frac{\nplanet q_i}{Q} = \frac{Q - \nplanet q_i}{Q}.
\end{equation}

The condition for the general maxima of the \cvtext/, in our formulation, is when one of the quantity (say $q_1$ takes the largest possible value, while all others take the smallest possible value):
\begin{equation}
\begin{split}
        q_1 &= q' \ (1+t) \\
        q_{i \ne 1} &= q' \ (1-t).
\end{split}
\end{equation}
The mean in this scenario becomes (marked with $''$):
\begin{equation}
        \bar{q}'' = \frac{q'(1+t) + (n-1) \times q(1-t)}{\nplanet} = \frac{q'}{\nplanet} \bigg[ n(1-t) + 2t\bigg].
\end{equation}
The variance in this scenario becomes (marked with $''$):
\begin{equation}
        \sigma''^{2}  (q) = \frac{1}{\nplanet} \Biggl\{ \bigg[\underbrace{q'(1+t) - \bar{q}''}_{= \ 2q't \big(\frac{\nplanet-1}{\nplanet}\big)}\bigg]^2  + \ (\nplanet-1) \bigg[ \underbrace{q'(1-t) - \bar{q}''}_{= \ \frac{- 2 q' t}{\nplanet}} \bigg]^2     \Biggr\}.
\end{equation}
This gives: 
\begin{equation}
        \sigma'' (q) = \bigg(\frac{2 q' t}{\nplanet}\bigg) \sqrt{\nplanet - 1}.
\end{equation}

Finally, the general expression for the maximum value of the \cvtext/ becomes:
\begin{equation}
        \max \cv (q) = \frac{\sigma''(q)}{\bar{q}''} =  \frac{2t \ \sqrt{\nplanet - 1}}{n (1-t) + 2t}.
\end{equation}

\change{This expression recovers the particular case derived in literature when we set $t=1$. From this expression, we note that the upper limit of the \cvtext/ does not depend on the actual values of the quantities, but it depends on the number of quantities in the set, $\mathcal{Q,}$ and the maximum tolerance, $t$. This new formulation allows us to extract the upper limit of the \cvtext/ for any set whose maximum tolerance, $t,$ is known. Interestingly, the above expression gives appropriate result when absurd inputs are considered. For example, when there are no planets in a system, $\max \cv \big|_{\nplanet = 0} = \sqrt{-1}$, and when there is only one planet in a system, $\max \cv \big|_{\nplanet = 1} = 0$. For a system of two planets, the upper limit is exactly the fractional difference (or tolerance), that is, $\max \cv \big|_{\nplanet = 2} = t$. }
        
Furthermore, varying over $\nplanet$, and assuming $t \in [0,1)$, allows us to derive the theoretical maximum possible value for the \cvtext/. This occurs at $\nplanet = \frac{2}{1 - t}$ and gives:
        \begin{equation}
        \label{eq:cvlimit}
                \max \ \cv (q) \bigg|_{\nplanet = \frac{2}{1-t} } (q) = \frac{t}{\sqrt{1 - t^2}}.
        \end{equation}
Figure \ref{fig:cscvlimits} shows the variation of the theoretical $\max, \ \cv$, as a function of tolerance $t$.$\blacksquare$

        \section{Classification boundaries for architectures classes}
                \label{sec:classificationboundaries}
                
                In this section, we present some considerations that motivate the boundaries between the four architecture classes for planetary masses. In the current formulation (Eq. \ref{eq:classify}), there are two boundaries that need to be identified. We deal with the distinction between \similar/ and \mixed/ class first, and then distinguish \ordered//\antiordered/ architecture classes.
                
                \subsection{Similar versus mixed}
                \label{subsec:boundarymixed}
                \def\figwidth{8cm}
        \begin{figure*}
                \centering
                \includegraphics[width=\figwidth]{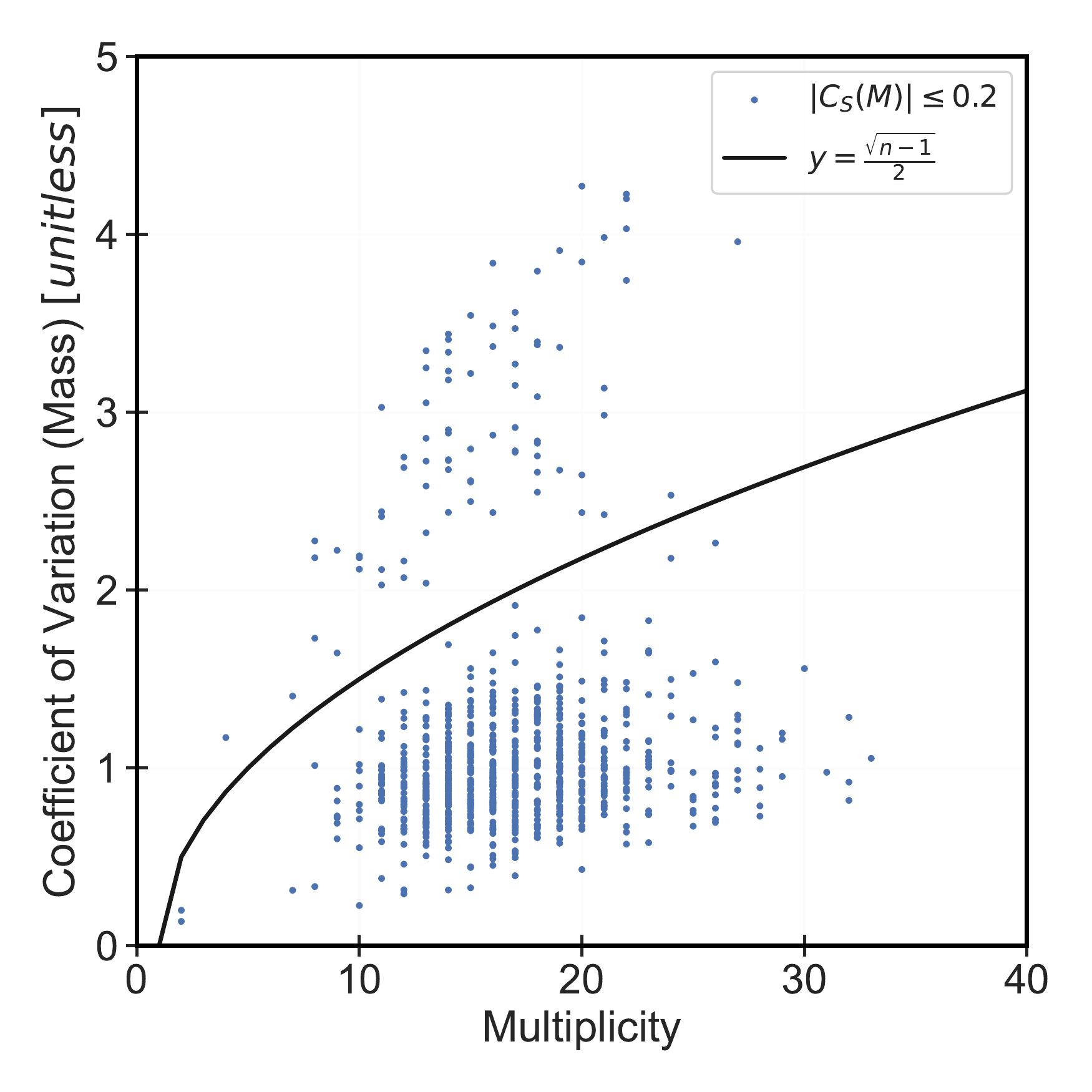}
                \includegraphics[width=\figwidth]{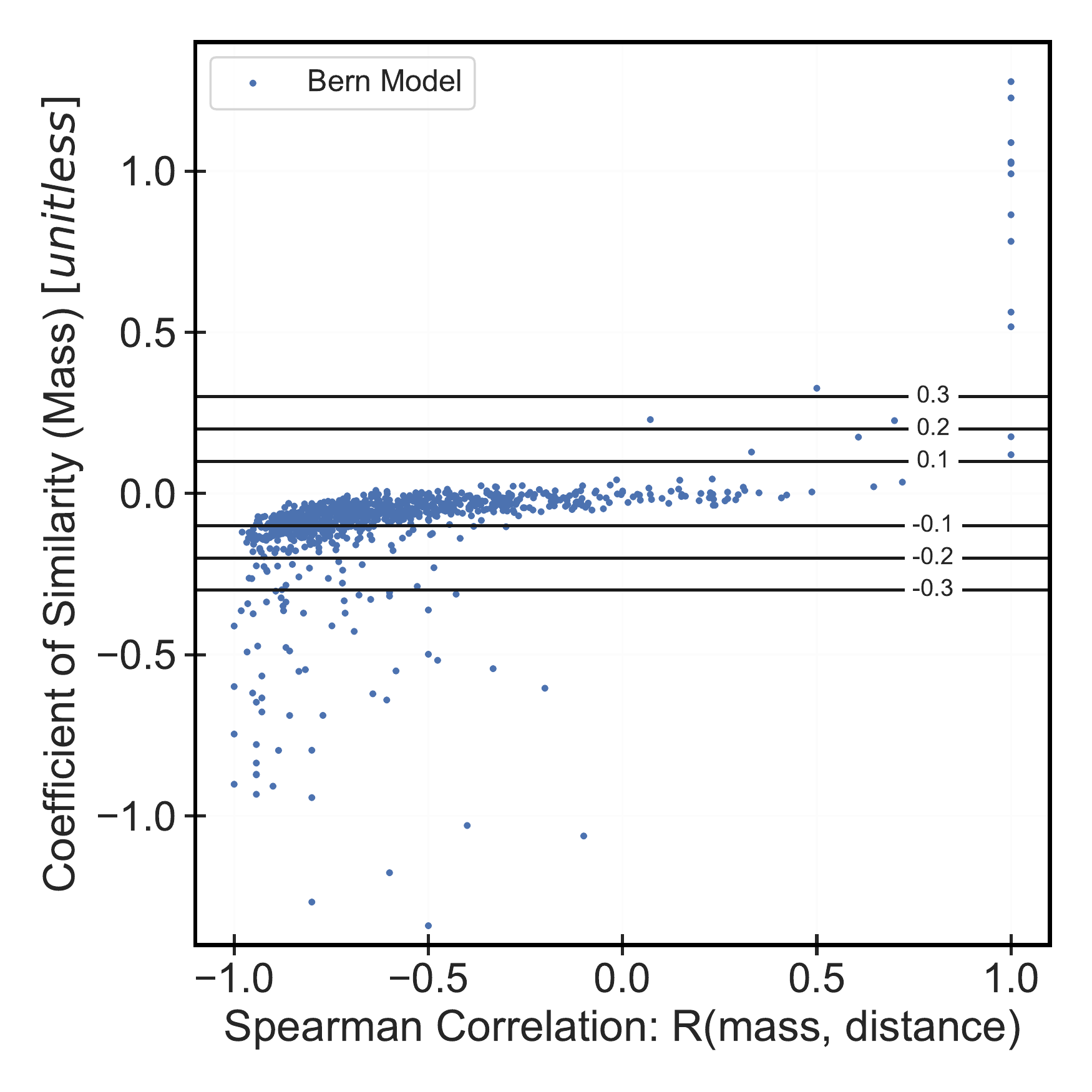}
                \caption Classification boundaries for architecture classes. {\textit{Left: }Boundary between \similar/ and \mixed/ class. The panel show the \cvtext/ for synthetic planetary systems as a function of the number of planets in a system for systems with $|\cs(M)| \leq 0.2$. Two clusters are clearly distinguishable, allowing us to fix the boundary between the \similar/ and \mixed/ architecture classes. 
                \textit{Right: }Boundary between \ordered/ and \antiordered/. This plot shows the \cstext/ of synthetic planetary systems as a function of the Spearman correlation coefficient between the planetary masses and distances of that system. Thick horizontal lines correspond to potential boundaries.}
                \label{fig:cvversusn}
        \end{figure*}

        We saw in Sect. \ref{subsec:concept}, it is difficult to distinguish between \mixed/ and \similar/ architecture classes using the \cstext/ alone. Mixed systems are characterised by large increasing or decreasing variations, which may cancel each other out and lead to small values of $\cs(M)$. Nevertheless, the \cvtext/ can distinguish between large variations in values. Figure \ref{fig:cvversusn} shows the $\cv(M)$ as a function of the number of planets in a planetary system. The left panel shows all synthetic systems from the Bern model, while we only show systems with $|\cs(M)| \leq 0.2$ in the right panel. 
        
        Clearly, there are two clusters of planetary systems. The cluster on the lower right-hand side corresponds to \similar/ class systems. Mixed systems, having large values of $\cv(M)$, are spread over the top left region. It is clear that the boundary between \similar/ and \mixed/ classes depends on the number of planets. The black line (corresponding to $y = \frac{\sqrt{n-1}}{2}$) neatly separates the two clusters. We have chosen this equation to disentangle \similar/ architectures from the \mixed/ class. This equation has, incidentally, two key properties:\ 1) it ensures that no two planet system can be of \mixed/ architecture and 2) it happens to be exactly half of the maximum possible value of the \cvtext/.
        
        \subsection{Ordered and anti-ordered}
                \label{subsec:boundaryordered}
                
                
                \begin{figure}
                \resizebox{\hsize}{!}{\includegraphics[trim=1.5cm 2.5cm 0.5cm 4.5cm,clip,height=5cm]{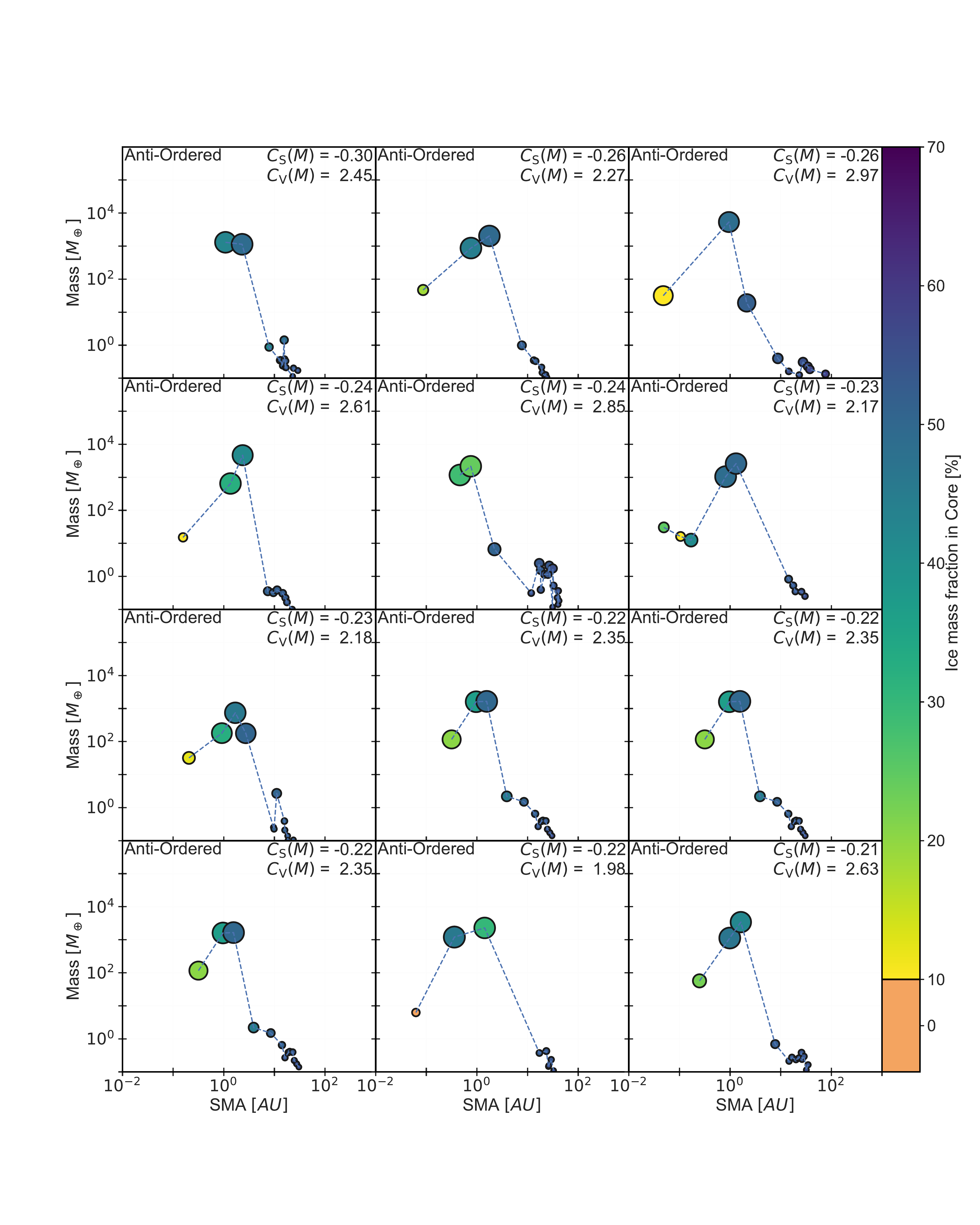}}
                \caption{Mass-distance diagram. This plot shows the planetary masses as a function of distance for some planetary systems with $-0.3 < \cs(M) < -0.2$. The dashed line connects that planets in the system and serves to highlight the arrangement and distribution of masses.
                The size of each circle corresponds to the planet's radius and the colour of each planet also shows its core water mass fraction. }
                \label{fig:ma_cs_limit}
                \end{figure}
                
                Having motivated the boundary between \similar/ and \mixed/ class, we are now left with three groupings of architecture classes. These three groupings correspond to $\cs(M) << 0$ (\antiordered/), $\cs(M) \sim 0$ (\similar//\mixed/), and $\cs(M) >> 0$ (\ordered/). This suggests that we require two boundaries to distinguish these three groups. However, we posit that the boundaries between \ordered/ and \antiordered/ should be symmetric around 0. Thus, we are left with only one boundary.
                
                \ordered/ (or \antiordered/) systems differ in their architecture from \similar//\mixed/ classes in that the quantity (mass here) continues to show an increasing (or decreasing) trend with distance. For all planetary systems in the Bern model, we measure the Spearman correlation coefficient, $R,$ between the planetary masses and their distance from the host star. The Spearman R, measuring the monotonicity between two datasets, varies from -1 to +1, with 0 indicating no correlation. A positive correlation implies that as x increases, so does y. Negative correlations imply that as x increases, y decreases. 
                
                Figure \ref{fig:cvversusn} shows the $\cs(M)$ of synthetic systems as function of their Spearman correlation R (mass and distance). We note that there is a large cluster of points towards $\cs(M) \sim 0$. This group corresponds to the \similar/ and \mixed/ architecture classes. There are some points to the top right (including those with $R=+1$ -- corresponding to planetary systems in which planetary masses are monotonically increasing with distance). There is a scatter of points towards the bottom-left (including some systems with $R=-1$). 
                
                First, we note that the comparison of the \cstext/ with Spearman R fulfils some expectation. For example, there are no points in bottom-right or top-left sections of this plot. Second, our objective is to isolate the central cluster of points from all other scattered points. We note that $|\cs(M)| = 0.1$ fails as a boundary, since it does not include the full central cluster. Both $|\cs(M)| = 0.2 \ \text{and} \ 0.3$ could succeed. Going beyond, a value of 0.3 would add many unnecessary points to the central cluster. 
                
                To further motivate our choice of boundary, namely, $|\cs(M)| = 0.2$, we show the mass-distance diagram of 12 randomly selected systems with $-0.3 < \cs(M) < -0.2$ (out of 19) in Fig. \ref{fig:ma_cs_limit}. We note that all systems show the qualitative features of an \antiordered/ system, namely, massive planets in the inner region and small planets in the outer region. Since all of these planets have their $\cs(M) < -0.2$, we use $|\cs(M)| = 0.2$ as a boundary between \ordered/, \antiordered/, and \similar/+\mixed/ architecture classes. Future works may explore improvements to our selected boundaries using additional ideas from K-means or hierarchical clusterings.
                
        \section{A gallery of architecture types: Mass-distance diagrams}
        \label{sec:gallery}
        
        \def\figwidthd{9cm}
                        \begin{figure*}
                \centering
                \includegraphics[trim=2cm 4cm 0.5cm 4cm,clip,width=\figwidthd]{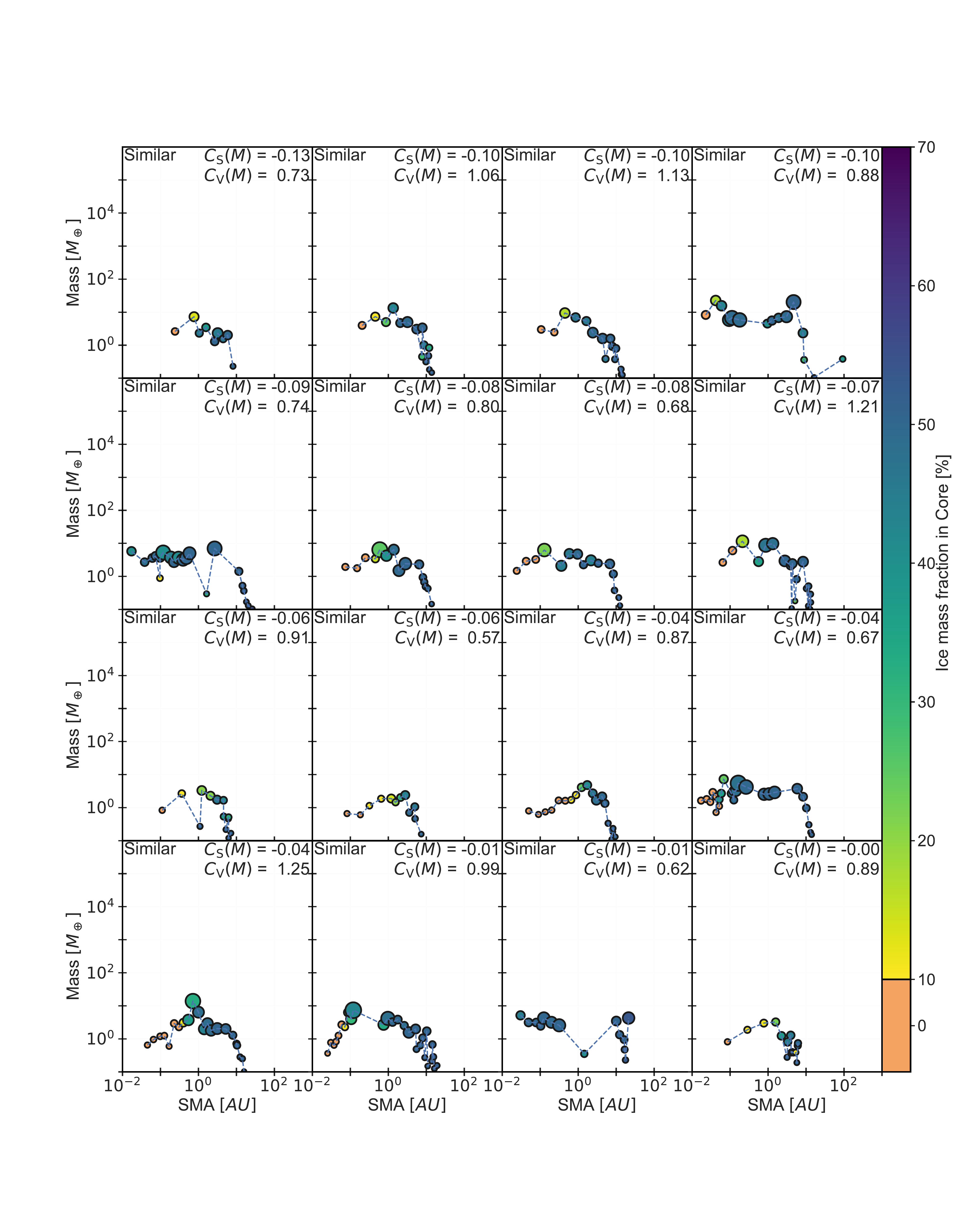}
                \includegraphics[trim=2cm 4cm 0.5cm 4cm,clip,width=\figwidthd]{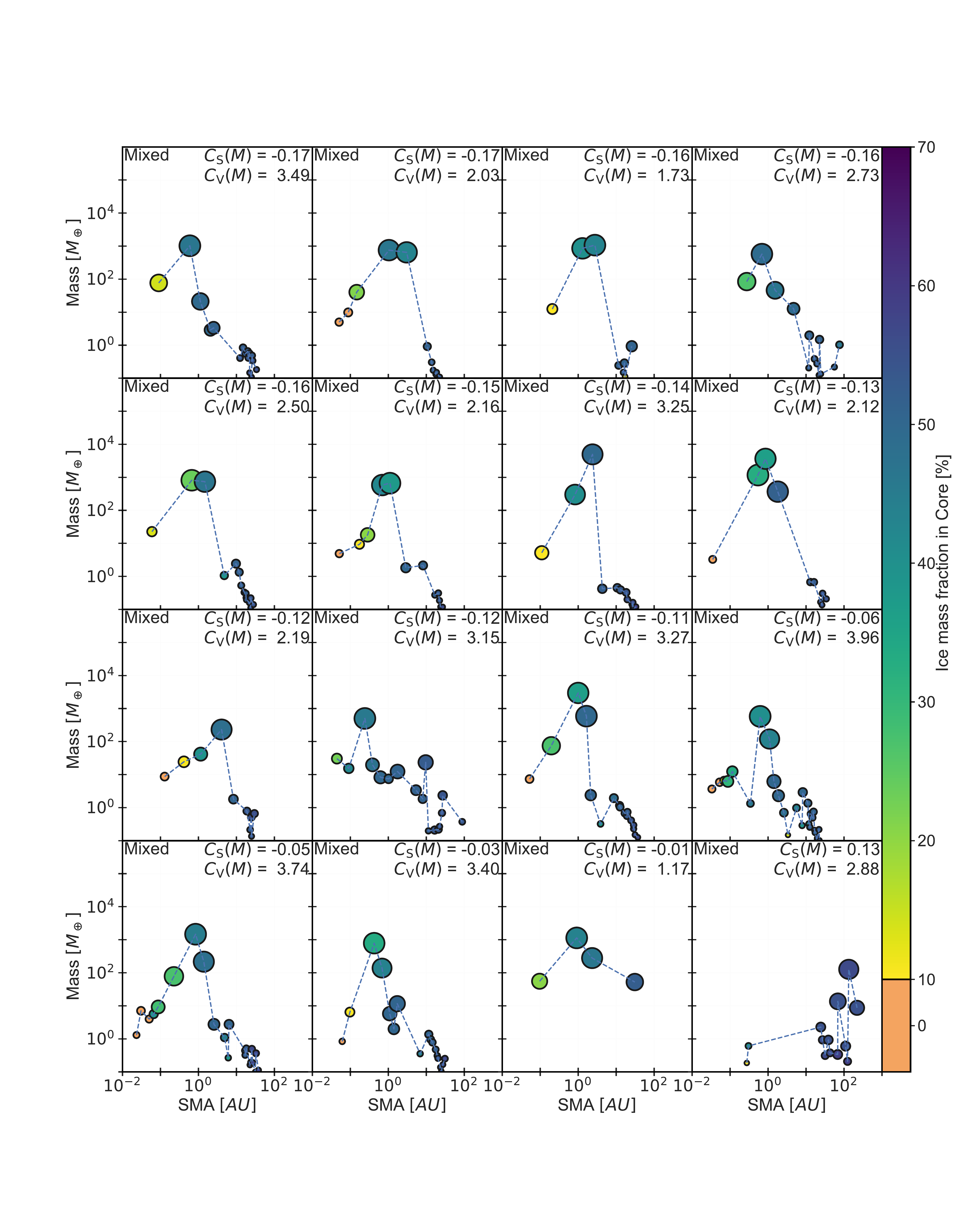}
                \caption{A gallery of planetary system architectures. These plots show the mass-distance diagram for \similar/ (left) and \mixed/ (right) planetary systems from the Bern Model. Each circle represents a planet, its size corresponds to the planetary radius, and its colour represents the fraction of ice in the planetary core. Each panel shows the $\cs(M)$ as well as the $\cv(M)$ of the system.}
                \label{fig:gallerysimilar}
                \end{figure*}
        
                \begin{figure*}
                \centering
                \includegraphics[trim=2cm 4cm 0.5cm 4cm,clip,width=\figwidthd]{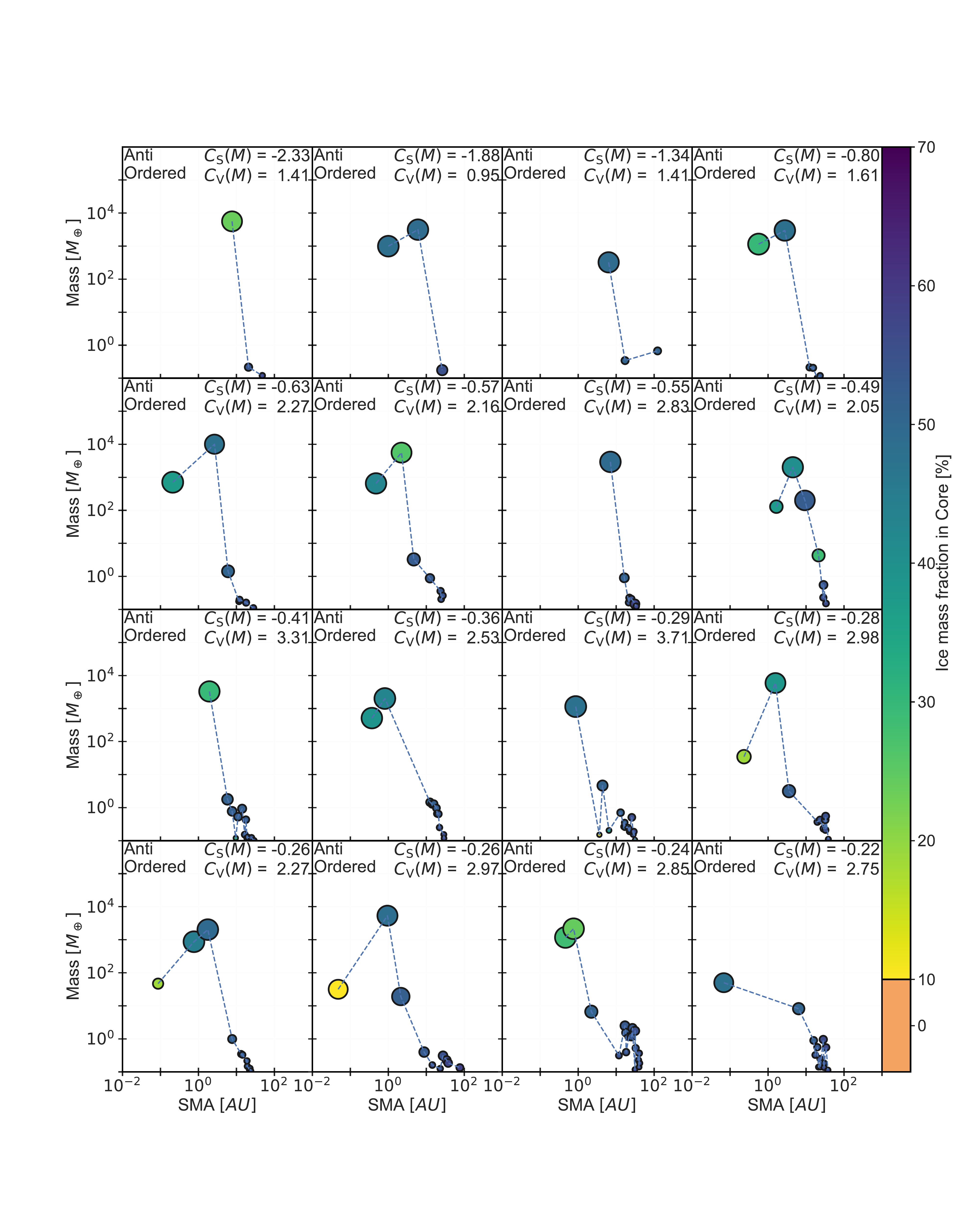}
                \includegraphics[trim=2cm 4cm 0.5cm 4cm,clip,width=\figwidthd]{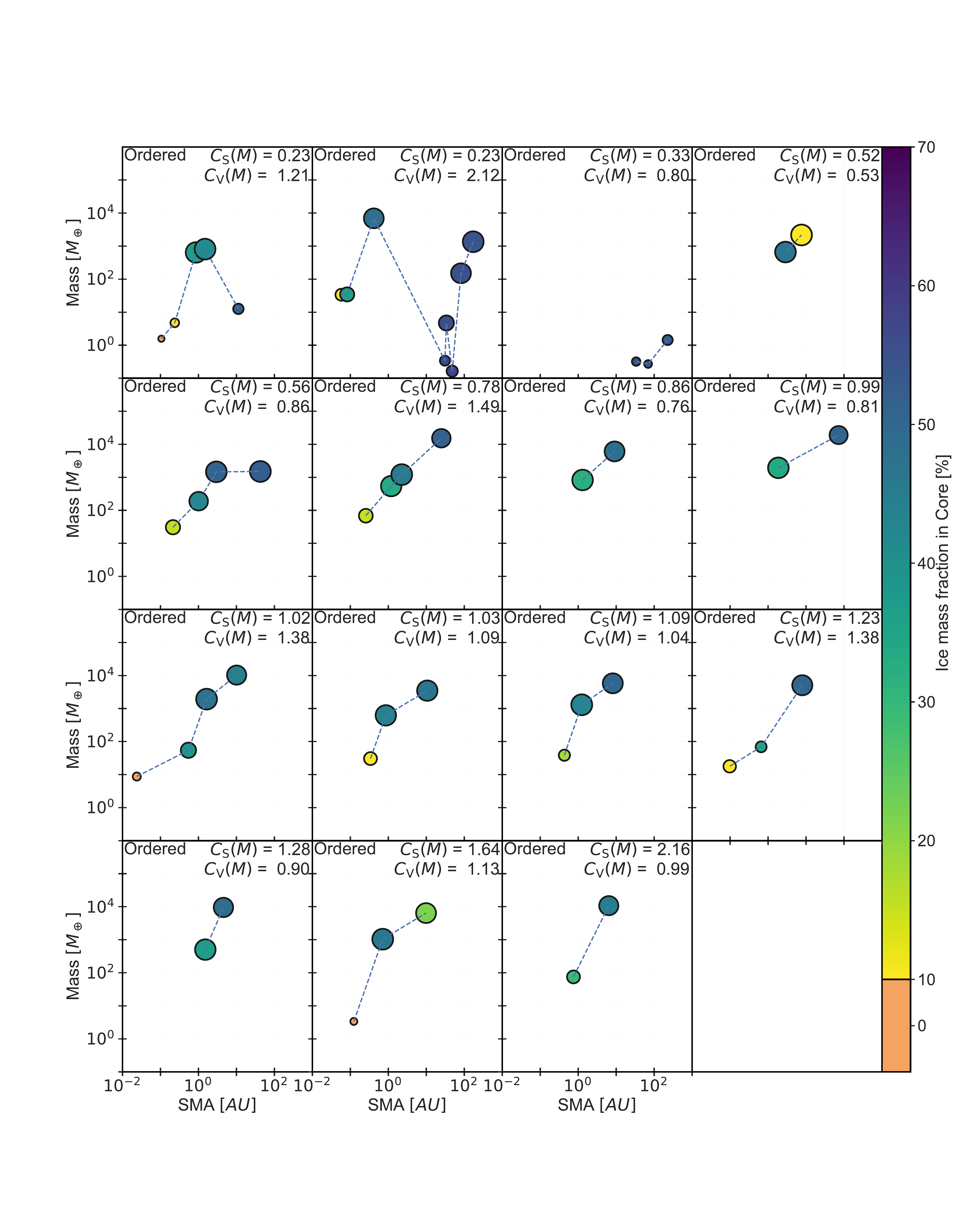}  
                \caption{A gallery of planetary system architectures. These plots show the mass-distance diagram for anti-ordered (left) and ordered (right) planetary systems from the Bern Model. Each circle represents a planet, its size corresponds to the planetary radius, and its colour represents the fraction of ice in the planetary core. Each panel shows the $\cs(M)$ as well as the $\cv(M)$ of the system.}
                \label{fig:galleryantiordered}
                \end{figure*}
        \end{appendix}
        
\end{document}